\pdfoutput=1
\documentclass[epj,nopacs,fleqn]{svjour}
\usepackage{slashed,multirow,arydshln,cite,amsmath,amssymb,url,graphicx,color}
\usepackage[utf8]{inputenc}

\begin{document}

\newcommand{\bea}{\begin{eqnarray}}
\newcommand{\eea}{\end{eqnarray}}
\newcommand{\be}{\begin{equation}}
\newcommand{\ee}{\end{equation}}
\def\bsp#1\esp{\begin{split}#1\end{split}}
\newcommand\sss{\scriptscriptstyle}

\newcommand{\amc}{{\sc Mad\-Graph5\textunderscore}a{\sc MC@NLO}}
\newcommand{\fr}{{\sc Feyn\-Rules}}
\newcommand{\ml}{{\sc MadLoop}}
\newcommand{\mfks}{{\sc MadFKS}}
\newcommand{\nloct}{{\sc NLOCT}}
\newcommand{\py}{{\sc Pythia}~8}
\title{QCD next-to-leading-order predictions matched to parton showers for
  vector-like quark models}

\author{
  Benjamin~Fuks\inst{1,2,3}
  and Hua-Sheng~Shao\inst{4}
}
\institute{
  Sorbonne Universit\'es, UPMC Univ.~Paris 06, UMR 7589, LPTHE, F-75005 Paris, France
  \and
  CNRS, UMR 7589, LPTHE, F-75005 Paris, France
  \and
  Institut Universitaire de France, 103 boulevard Saint-Michel, 75005 Paris, France
  \and
  Theoretical Physics Department, CERN, CH-1211 Geneva 23,Switzerland
}

\date{}

\abstract{Vector-like quarks are featured by a wealth of beyond the Standard
  Model theories and are consequently an important goal of many LHC searches for
  new physics. Those searches, as well as most related phenomenological studies,
  however, rely on predictions evaluated at the leading-order accuracy in QCD and
  consider well-defined simplified benchmark scenarios. Adopting an effective
  bottom-up approach, we compute next-to-leading-order predictions for
  vector-like-quark pair production and single production in association with
  jets, with a weak or with a Higgs boson in a general new physics setup. We
  additionally compute vector-like-quark contributions to the production of a
  pair of Standard Model bosons at the same level of accuracy. For all processes
  under consideration, we focus both on total cross sections and on
  differential distributions, most these calculations being performed for the
  first time in our field. As a result, our work paves the way to precise
  extraction of experimental limits on vector-like quarks thanks to an accurate
  control of the shapes of the relevant observables and emphasise the
  extra handles that could be provided by novel vector-like-quark probes
  never envisaged so far.}

\titlerunning{QCD NLO predictions matched to parton showers for vector-like quark models}
\authorrunning{B.~Fuks and H.-S.~Shao}

\maketitle
\vspace*{-14cm}
\noindent 
  CERN-TH-2016-210, MCNET-16-38
\vspace*{12cm}


\section{Introduction\label{sec:intro}}
The Standard Model of particle physics is a successful theory of nature,
although it exhibits many conceptual issues, like the hierarchy problem or the
strong $CP$ problem, and practical limitations like the absence of a viable
candidate for explaining the dark matter pervading our universe. As a result, it
is commonly acknowledged as an effective theory stemming from a more fundamental
setup that has still to be observed and confirmed experimentally. This effective
description has been recently strengthened with the discovery of a Higgs boson
with properties very similar to those expected from the Standard Model in 2012.
The null results of all collider searches for new particles predicted by most of
beyond
the Standard Model theories are, however, at the same time pushing the limits on
the mass of these potential new particles to higher and higher energy scales.

Among the viable options for new physics, many extensions of the Standard
Model predict the existence of additional quark species that should be
observable during the next runs of the Large Hadron Collider (LHC) at CERN.
One of the common feature of such theories consists of the predicted vector-like
nature of the additional quarks, \textit{i.e.}~their left-handed and
right-handed components lie in the same representation of the electroweak
symmetry group. These quarks appear for instance in models with extra
space-time dimensions that exhibit an extended gauge symmetry or a new strong
dynamics giving rise to massive composite states~\cite{Antoniadis:1990ew,
Kaplan:1991dc,ArkaniHamed:2002qx,Contino:2006qr,Matsedonskyi:2012ym}. As a
result, vector-like quark searches play an important role in the ATLAS and CMS
experimental program.

Current searches, relying on signatures induced by both the vector-like quark
pair-production and single-production modes, impose strong constraints on the
masses of the heavy quarks that are now bounded to be above about
750-1500~GeV~\cite{Aad:2014efa,Aad:2015kqa,Aad:2015tba,ATLAS:2016sno,
CMS:2016ccy,CMS:2015alb,CMS:2016ete}, this wide range reflecting the wealth of
options for describing how the new states decay into a pair of Standard Model
particles. Care must, however, be taken with the
interpretation of these limits as they are extracted once various simplifying
assumptions are accounted for the new physics signal. Most of the bounds
indeed assume that the quark partners decay, with a branching ratio of 100\%,
into third-generation top or bottom quarks, while more general situations where
decays into second or first generation quarks are possible are less explored
yet. Sizeable couplings to light quarks are still allowed by indirect
constraints~\cite{delAguila:2000rc,Cacciapaglia:2015ixa,Ishiwata:2015cga}, which
could have a severe impact on electroweak vector-like quark processes at the
LHC~\cite{Atre:2011ae}. The latter, which induce vector-like quark decays, are
driven by the couplings of the extra quarks to the weak and Higgs bosons. They
admit a simple model-independent parameterisation~\cite{Buchkremer:2013bha}
regardless of the representation of the new physics particles under the
electroweak symmetry group, and this bottom-up approach is now adopted by the
experimental collaborations. Most LHC searches for heavy quarks hence turn out
to be agnostic of the ultraviolet completion of the model and can rather easily
be reinterpreted in any framework, and options for combining different
searches can also be considered.

Current vector-like quark searches, as well as most associated theoretical
works, are, however, based on Monte Carlo simulations of the new physics signals
where hard-scattering matrix-elements are evaluated at the leading-order (LO)
accuracy in QCD. In addition, event samples featuring different final-state jet
multiplicities are sometimes also merged in order to get a better control on the
shapes of the key differential distributions. The formal precision of these
calculations is nonetheless rather limited, which directly impacts the
extraction of any limit on the properties of the new particle properties or the
corresponding measurements in the case of a discovery. In this work, we build up
a new procedure that allows us to make use of the \amc\
framework~\cite{Alwall:2014hca} to compute total rates and differential
distributions at the next-to-leading-order (NLO) accuracy in QCD for processes
involving vector-like quarks. This more precisely concerns vector-like quark
production on the one hand, and the production of Standard Model particles when
vector-like quark diagrams contribute, regardless of the strong or electroweak
nature of the Born process.

Our methodology relies on the joint use of the \fr~\cite{Alloul:2013bka} and
\nloct~\cite{Degrande:2014vpa} packages, the latter making use of
{\sc FeynArts}~\cite{Hahn:2000kx}, for automatically generating a UFO
library~\cite{Degrande:2011ua} that contains all tree-level vertices and
counterterms necessary for NLO QCD computations. This UFO library can then be
further used by \amc\ for event generation, at the LO and NLO accuracies in QCD
as well as for loop-induced processes.
Virtual loop contributions are numerically evaluated through
the {\sc MadLoop} module~\cite{Hirschi:2011pa} and combined with the
real-emission diagrams following the FKS subtraction method as implemented in
{\sc MadFKS}~\cite{Frixione:1995ms,Frederix:2009yq}, and the matching to parton
showers is finally achieved according to the MC@NLO
prescription~\cite{Frixione:2002ik}.

In Section~\ref{sec:model}, we detail how we have modified the model-independent
parameterisation of Ref.~\cite{Buchkremer:2013bha} to make it suitable for NLO
calculations in QCD matched to parton showers for vector-like quark processes.
We also define a series of benchmark scenarios for the phenomenological
studies performed in Section~\ref{sec:LHCpheno}. We investigate vector-like
quark pair production and single production (in association with either a jet
or a weak gauge or Higgs boson), as well as diboson production. We
summarise our work in Section~\ref{sec:conclusions}.

\section{A model-independent parameterisation for vector-like
quark models\label{sec:model}}

\subsection{Model description}
Vector-like quarks appear in many extensions of the Standard
Model. They are usually included as fields lying in the fundamental
representation of $SU(3)_c$ and carry colour charges similar to those of
their Standard Model counterparts. However, they can lie in various
representations of the weak interaction symmetry group $SU(2)_L$ and be assigned
different hypercharge $U(1)_Y$ quantum numbers. Focusing on phenomenologically
viable minimal models that comprise a single Standard Model Higgs field
$\Phi$, only weak triplets, doublets and singlets of vector-like quarks are
allowed~\cite{Cacciapaglia:2011fx}. Consequently, the particle content of the
theory can solely include four species of extra quarks, which we denote by $X$,
$T$, $B$ and $Y$, their respective electric charges being $Q=5/3$,
$2/3$, $-1/3$ and $-4/3$.

Although vector-like and Standard Model quarks with the same electric charge
mix, the mixing pattern and the resulting phenomenology can be simplified when
minimality requirements are imposed. Since we consider that the Higgs sector
contains a single Standard-Model-like scalar Higgs field $\Phi$, quark mixings
are solely generated by its Yukawa interactions. The mass splitting between the
vector-like quarks is consequently also constrained to be small and connected to
the Higgs vacuum expectation value $v$, so that the extra quarks will always
directly decay into a gauge or Higgs boson and one of the Standard Model
quarks~\cite{Cacciapaglia:2011fx}. A model-independent effective
parameterisation apt to describe the phenomenology of this vector-like quark
setup has been recently proposed~\cite{Buchkremer:2013bha}, but it is
not suitable for higher-order QCD calculations.
The reason is that in the latter parameterisation, the strengths of the
interactions of the vector-like and Standard Model quarks with a single
Higgs boson depend on the masses of the model particles. As a result, the
renormalisation of the quark masses and the one of the couplings are
related, which prevents all ultraviolet divergences that arise at the
NLO from cancelling. We therefore modify the modelling
of Ref.~\cite{Buchkremer:2013bha} so that all the couplings of the vector-like
quarks to a gauge or a Higgs boson are free parameters, and we have
the following effective Lagrangian,
\be\bsp
 & {\cal L}_{\rm VLQ} =
    i \bar Y \slashed{D} Y - m_{\sss Y} \bar Y Y
  +  i \bar B \slashed{D} B - m_{\sss B} \bar B B\\
 &\quad +  i \bar T \slashed{D} T - m_{\sss T} \bar T T
  +  i \bar X \slashed{D} X - m_{\sss X} \bar X X\\
 &\quad - h\ \bigg[
    \bar B \Big(\hat\kappa_{\sss L}^{\sss B} P_L + \hat\kappa_{\sss R}^{\sss B} P_R\Big) q_d
   + {\rm h.c.} \bigg] \\
 &\quad - h\ \bigg[
   \bar T \Big(\hat\kappa_{\sss L}^{\sss T} P_L + \hat\kappa_{\sss R}^{\sss T} P_R\Big) q_u
   + {\rm h.c.} \bigg] \\
 &\quad +
  \frac{g}{2 c_{\sss W}}\ \bigg[
    \bar B\slashed{Z} \Big(\tilde\kappa_{\sss L}^{\sss B}P_L +
      \tilde\kappa_{\sss R}^{\sss B} P_R\Big) q_d + {\rm h.c.} \bigg] \\
 &\quad +
  \frac{g}{2 c_{\sss W}}\ \bigg[
   \bar T\slashed{Z} \Big(\tilde\kappa_{\sss L}^{\sss T}P_L +
      \tilde\kappa_{\sss R}^{\sss T} P_R\Big) q_u + {\rm h.c.} \bigg] \\
 &\quad +
  \frac{\sqrt{2} g}{2}\ \bigg[
    \bar Y  \slashed{\bar W} \Big(\kappa_{\sss L}^{\sss Y} P_L \!+\! \kappa_{\sss R}^{\sss Y}
       P_R\Big) q_d + {\rm h.c.} \bigg] \\
 &\quad +
  \frac{\sqrt{2} g}{2}\ \bigg[
   \bar B \slashed{\bar W} \Big(\kappa_{\sss L}^{\sss B} P_L \!+\! \kappa_{\sss R}^{\sss B}
       P_R\Big) q_u + {\rm h.c.} \bigg]\\
 &\quad +
  \frac{\sqrt{2} g}{2}\ \bigg[
   \bar T \slashed{W} \Big(\kappa_{\sss L}^{\sss T} P_L \!+\! \kappa_{\sss R}^{\sss T}
       P_R\Big) q_d + {\rm h.c.} \bigg]\\
 &\quad +
  \frac{\sqrt{2} g}{2}\ \bigg[
   \bar X  \slashed{W} \Big(\kappa_{\sss L}^{\sss X} P_L \!+\! \kappa_{\sss R}^{\sss X}
        P_R\Big) q_u + {\rm h.c.} \bigg]\ ,
\esp\label{eq:LVLQ}\ee
being supplemented to the Standard Model Lagrangian ${\cal L}_{\rm SM}$. The
terms in the first and second lines consist of gauge-invariant kinetic and mass
terms for the vector-like quark fields (taken in the mass eigenbasis)
after restricting the covariant derivatives to their QCD component,
\be
  D_\mu  = \partial_\mu -i g_s T_a G_\mu^a \ .
\ee
The coupling parameter $g_s$ denotes the strong coupling constant, $G_\mu$ the
gluon field and $T$ (and $f$ for further references) the fundamental
representation matrices (the structure constants) of $SU(3)$. Although the
electroweak pieces of the covariant derivative could have been included, they
have been omitted in order to simplify our model description since they are
model-dependent and are expected to yield a negligible effect with respect to
their strong interaction part.

The third and fourth lines of Eq.~\eqref{eq:LVLQ} collects the effective interactions of
the physical Higgs boson $h$ with one Standard Model quark
and its vector-like partner, generation indices being understood for clarity. As
mentioned above, such interactions are yielded by the (flavour-changing) Yukawa
couplings of the Higgs doublet $\Phi$ with the up-type ($q_u$), down-type
($q_d$) and vector-like quarks that induce a mixing of the Standard
Model and the new physics quark sectors. The relevant elements of the mixing
matrices have been included in the strengths of the effective interactions
$\hat\kappa$.

In the last six lines of Eq.~\eqref{eq:LVLQ}, we include the weak interactions
of the $Z$-boson and $W$-boson with one Standard Model quark and one
vector-like quark. In our conventions, we have factorised out the weak coupling
$g$ which represents the overall interaction strength, and the $\kappa$ and
$\tilde\kappa$ parameters include the relevant elements of the
quark mixing matrices, as for the Higgs interactions. Moreover, the $c_{\sss W}$
parameter stands for the cosine of the weak mixing angle.

The ${\cal L}_{\rm VLQ}$ Lagrangian above is equivalent, at the tree level, to
the one of Ref.~\cite{Buchkremer:2013bha} once we impose
\be\bsp
 \big(\hat\kappa^{\sss Q}_{\sss L,R}\big)_f =&\
     \frac{\kappa_{\sss Q}m_{\sss Q}}{v}
    \sqrt{\frac{\zeta_{\sss L,R}^f\ \xi^{\sss Q}_{\sss H}}
     {\Gamma^{\sss Q}_{\sss H}}} \ , \\
 \big(\tilde\kappa^{\sss Q}_{\sss L,R}\big)_f =&\ \kappa_{\sss Q}
    \sqrt{\frac{\zeta_{\sss L,R}^f\ \xi^{\sss Q}_{\sss Z}}
     {\Gamma^{\sss Q}_{\sss Z}}}\ ,\\
 \big(\kappa^{\sss Q}_{\sss L,R}\big)_f =&\ \kappa_{\sss Q}
    \sqrt{\frac{\zeta_{\sss L,R}^f\ \xi^{\sss Q}_{\sss W}}
     {\Gamma^{\sss Q}_{\sss W}}}\ .
\esp\label{eq:conventions}\ee
In this notation, $f$ stands for a generation index and
$\Gamma^{\sss Q}_{\sss X}$ denotes the kinematic factor of the
partial decay width of the extra quark
$Q$ into a final state containing an $X$ boson. The $\kappa_{\sss Q}$ parameter
encodes the magnitude of the coupling of the extra quark $Q$ to the different
electroweak bosons, while the $\zeta$ parameters refer to the mixing of the
vector-like quarks with the Standard Model quarks,
\be
  \zeta_{\sss L,R}^f =
   \frac{|(V_{\sss L,R})_{4f}|^2}{\sum_{i=1}^3 |(V_{\sss L,R})_{4j}|^2}
   \qquad\text{with}\quad \sum_{i=1}^3 \zeta_{\sss L,R}^i = 1 \ .
\label{eq:zeta}\ee
In this expression, we have represented the left-handed and right-handed
$4 \times 4$ mixing matrices between the new quarks and the three Standard Model
quarks by $V_{\sss L,R}$. Finally, the $\xi$ parameters of
Eq.~\eqref{eq:conventions} determine the relative importance of the various
decay modes of the vector-like quarks, their sum being equal to one.

The calculation of differential and total cross sections for LHC
processes at the NLO accuracy in QCD necessitates to evaluate,
on the one hand, real-emission squared amplitudes and on the other hand,
interferences of tree-level with virtual one-loop diagrams. The ultraviolet
divergences that arise in the latter case are absorbed through the
renormalisation of the fields and parameters appearing in ${\cal L}_{\rm VLQ}$.
This is achieved by replacing all fermionic and non-fermionic bare
fields $\Psi$ and $\Phi$ and bare parameters $y$ by the corresponding
renormalised quantities,
\be\bsp
  \Phi \to & \ \Big[1 + \frac12 \delta Z_\Phi\Big] \Phi\ ,\\
  \Psi \to & \ \Big[ 1 + \frac12 \delta Z^L_\Psi P_L +
      \frac12 \delta Z^R_\Psi P_R \Big]\Psi\ , \\
  y \to &\ y + \delta y \ ,
\esp\ee
where we truncate the renormalisation constants $\delta Z$ and $\delta y$ at the
first order in the strong coupling $\alpha_s=g_s^2/(4 \pi)$. While the
wave-function renormalisation constants of the Standard Model quarks are not
modified by the presence of the vector-like quarks, the one of the gluon field
is given, when the on-shell renormalisation scheme is adopted and when we
include $n_f=5$ massless flavours of quarks, by
\be\bsp
  \delta Z_g = & -\frac{g_s^2}{24 \pi^2} \sum_{q=t,T,B,X,Y}\bigg\{
    - \frac13 + B_0(0,m_q^2,m_q^2) \\ &\quad
      + 2 m_q^2 B_0'(0,m_q^2,m_q^2) \bigg\}\ ,
\esp\ee
where the $B_{0,1}$ and $B_{0,1}'$
functions stand for standard two-point Passarino-Veltman integrals and their
derivatives~\cite{Passarino:1978jh}. The left-handed and right-handed wave
function renormalisation constants $\delta Z_{\sss Q}^{L,R}$ and the mass
renormalisation constants $\delta m_{\sss Q}$ of a vector-like quark $Q$ (with
$Q=T$, $B$, $X$, $Y$) are similar to the top-quark ones and read
\be\label{eq:ZQ}\bsp
  & \delta Z_{\sss Q}^{L,R} = \frac{g_s^2 C_F}{16 \pi^2} \bigg[
        1 + 2 B_1(m_{\sss Q}^2;m_{\sss Q}^2,0)
    \\& \quad
       + 8 m_{\sss Q}^2 B_0'(m_{\sss Q}^2; m_{\sss Q}^2,0)
         + 4 m_{\sss Q}^2 B_1'(m_{\sss Q}^2; m_{\sss Q}^2,0) \bigg] \ , \\
  & \delta m_{\sss Q} =  \frac{g_s^2 C_F\ m_{\sss Q}}{16 \pi^2} \bigg[
        1 - 4 B_0(m_{\sss Q}^2; m_{\sss Q}^2,0) \\ &\quad
       - 2 B_1(m_{\sss Q}^2; m_{\sss Q}^2,0) \bigg] \ .
\esp\ee
In these two expressions, $C_F=(n_c^2-1)/(2 n_c)$ is the quadratic Casimir
invariant associated with the fundamental representation of $SU(3)$, with
$n_c=3$. Finally, in order to fix the renormalisation group
running of $\alpha_s$ so that it
originates from gluons and the $n_f=5$ active light quark flavours, we
renormalise $\alpha_s$ by subtracting, from the gluon self-energy, the
contributions of all massive particles evaluated at zero-momentum transfer,
\be
  \frac{\delta\alpha_s}{\alpha_s} =
      \frac{\alpha_s}{2\pi\bar\epsilon}
        \bigg[\frac{n_f}{3} \!-\! \frac{11 n_c}{6}\bigg]
    \!+\! \frac{\alpha_s}{6\pi}\sum_{q=t,T,B,X,Y}
        \bigg[\frac{1}{\bar\epsilon} \!-\! \log\frac{m_q^2}{\mu_R^2}\bigg] \ ,
\label{eq:das}\ee
where the ultraviolet-divergent pieces are written in terms of
\mbox{$\frac{1}{\bar\epsilon}=\frac{1}{\epsilon} - \gamma_E + \log{4\pi}$} with
$\gamma_E$ being the Euler-Mascheroni constant and $\epsilon$ being connected to
the number of space-time dimensions $D=4-2\epsilon$. The first term in the
right-hand side of Eq.\eqref{eq:das} results from the Standard Model massless
parton contributions, while the second term is connected to the massive states,
namely the top quark and the four considered vector-like quark species.

In Section~\ref{sec:LHCpheno}, we will compute predictions at the
NLO accuracy in QCD for processes involving vector-like
quarks. We will rely on a numerical evaluation of the loop integrals in four
dimensions, which necessitates the
calculation of rational terms associated with the $\epsilon$-dimensional pieces
of the loop integrals. There exist two sets of such rational
terms that are, respectively,
connected to the loop-integral denominators ($R_1$) and numerators ($R_2$).
While the former are universal, the latter are model-dependent and can be seen
as a finite number of counterterm Feynman rules derived from
the bare Lagrangian~\cite{Ossola:2008xq}. Starting from the ${\cal L}_{\rm VLQ}$
Lagrangian of Eq.~\eqref{eq:LVLQ}, several $R_2$ counterterms with external
gauge bosons are modified with respect to the Standard Model case,
\be\bsp
 & R_2^{G G} = \frac{i g_s^2}{96 \pi^2}\bigg[
      \Big( 9 p_2^2 \eta^{\mu_1\mu_2} - 6 p_2^{\mu_1} p_2^{\mu_2} \Big)\\
  &\qquad
    + 2 \eta^{\mu_1\mu_2}\sum_q\Big\{p_2^2-6m_q^2 \Big\}
    \bigg] \delta_{c_1c_2}\ , \\
 & R_2^{G G G} = -\frac{g_s^3  f_{c_1c_2c_3}}{192 \pi^2}\Big( 33+\sum_q8 \Big)
    V^{\mu_1\mu_2\mu_3}\ ,\\
 & R_2^{G G G G} = \frac{i g_s^4}{48 \pi^2}\Big(
   C^{(1)}_{c_1c_2c_3c_4} \eta^{\mu_1\mu_2} \eta^{\mu_3\mu_4} \\&\qquad
   + C^{(2)}_{c_1c_2c_3c_4} \eta^{\mu_1\mu_3} \eta^{\mu_2\mu_4}+
   C^{(3)}_{c_1c_2c_3c_4} \eta^{\mu_1\mu_4} \eta^{\mu_2\mu_3}
    \Big) \ ,\\
 & R_2^{WWGG} = \frac{i g^2 g_s^2}{96 \pi^2}
    V^{\mu_1\mu_2\mu_3\mu_4}\\ &\qquad \times 
     \Big(3+\sum_{Q,f} \Big\{ (\kappa^{\sss Q}_{\sss L})_f^2+ 
       (\kappa^{\sss Q}_{\sss R})_f^2\Big\} \Big) \delta_{c_3c_4}\ ,\\
 & R_2^{ZZGG} =\frac{i g^2 g_s^2}{288 c_{\sss W}^2 \pi^2}
    V^{\mu_1\mu_2\mu_3\mu_4}
     \Big(8-18s_{\sss W}^2 + 20 s_{\sss W}^4 \\ &\qquad +3 \sum_{Q,f}
    \Big\{ (\tilde\kappa^{\sss Q}_{\sss L})_f^2+ 
       (\tilde\kappa^{\sss Q}_{\sss R})_f^2\Big\} \Big) \delta_{c_3c_4}\ ,\\
 & R_2^{GGhh} = \frac{i g_s^2}{16 \pi^2}\\ &\qquad\times 
    \eta^{\mu_1\mu_2} \Big(-y_t^2 -2 \sum_{Q,f}
    \Big\{ (\hat\kappa^{\sss Q}_{\sss L})_f^2+ 
       (\hat\kappa^{\sss Q}_{\sss R})_f^2\Big\} \Big) \delta_{c_3c_4}\ ,
\esp\label{eq:r2sm}\ee
with
\be\bsp
& V^{\mu_1\mu_2\mu_3} =  (p_1-p_2)^{\mu_3}\eta^{\mu_1\mu_2}
      + (p_3-p_1)^{\mu_2}\eta^{\mu_3\mu_1}\\ &\quad
      + (p_2-p_3)^{\mu_1}\eta^{\mu_2\mu_3}\ , \\
& V^{\mu_1\mu_2\mu_3\mu_4} = \eta^{\mu_1\mu_2} \eta^{\mu_3\mu_4} \!+\!
     \eta^{\mu_1\mu_3} \eta^{\mu_2\mu_4} \!+\!
      \eta^{\mu_1\mu_4} \eta^{\mu_2\mu_3} \ , \\
&  C_{abcd}^{(1)} =
    \delta_{ad}\delta_{bc}+\delta_{ac}\delta_{bd}+\delta_{ab}\delta_{cd}\\ &\quad
   + 2 \Big({\rm Tr}\big[T_aT_cT_bT_d\big] \!+\! {\rm Tr}\big[T_aT_dT_bT_c\big]\Big)
      \Big(45 \!+\! \sum_q11\Big)\\ &\quad
   - 6 \Big({\rm Tr}\big[T_aT_bT_cT_d\big] + {\rm Tr}\big[T_aT_bT_dT_c\big]\Big)
      \Big(7+\sum_q2\Big)
  \\ &\quad
   - 6 \Big({\rm Tr}\big[T_aT_cT_dT_b\big] + {\rm Tr}\big[T_aT_dT_cT_b\big]\Big)
      \Big(7+\sum_q2\Big) \ , \\
&  C_{abcd}^{(2)} =
   \delta_{ad}\delta_{bc}+\delta_{ac}\delta_{bd}+\delta_{ab}\delta_{cd}
  \\ &\quad
   - 6 \Big({\rm Tr}\big[T_aT_bT_dT_c\big] + {\rm Tr}\big[T_aT_cT_dT_b\big]\Big)
      \Big(7+\sum_q2\Big)\\ &\quad
   - 2 \Big({\rm Tr}\big[T_aT_cT_bT_d\big] + {\rm Tr}\big[T_aT_dT_bT_c\big]\Big)
      \Big(24+\sum_q5\Big)
  \\ &\quad
   +4\Big({\rm Tr}\big[T_aT_bT_cT_d\big] + {\rm Tr}\big[T_aT_dT_cT_b\big]\Big)
      \Big(24+\sum_q5\Big) \ , \\
&  C_{abcd}^{(3)} = 
    \delta_{ad}\delta_{bc}+\delta_{ac}\delta_{bd}+\delta_{ab}\delta_{cd}
  \\ &\quad
   - 6 \Big({\rm Tr}\big[T_aT_bT_cT_d\big] + {\rm Tr}\big[T_aT_dT_cT_b\big]\Big)
      \Big(7+\sum_q2\Big)\\ &\quad
   - 2 \Big({\rm Tr}\big[T_aT_cT_bT_d\big] + {\rm Tr}\big[T_aT_dT_bT_c\big]\Big)
      \Big(24+\sum_q5\Big)\\&\quad 
   +4\Big({\rm Tr}\big[T_aT_bT_dT_c\big] + {\rm Tr}\big[T_aT_cT_dT_b\big]\Big)
      \Big(24+\sum_q5\Big) \ .
\esp\ee
In addition, the new $R_2$ counterterms involving external vector-like quarks
are given by
\be\bsp
 R_2^{\bar Q Q} = &\ -\frac{i g_s^2}{12 \pi^2}
     \Big( \slashed{p}_2 -2 m_{\sss Q} \Big) \delta_{c_1c_2}\ , \\
 R_2^{\bar Q Q G} = &\
    -\frac{i g_s^2}{6 \pi^2}\gamma^{\mu_3}\ T^{c_3}_{c_1c_2}\ , \\
 R_2^{\bar q_f Q h} = &\ -\frac{i g_s^2}{3 \pi^2}\
     \Big((\hat\kappa^{\sss Q}_{\sss L})_f P_L +
          (\hat\kappa^{\sss Q}_{\sss R})_f P_R \Big) \delta_{c_1c_2}\ , \\
 R_2^{\bar Q q_f h} = &\ -\frac{i g_s^2}{3 \pi^2}\
     \Big((\hat\kappa^{\sss Q}_{\sss L})_f P_R +
          (\hat\kappa^{\sss Q}_{\sss R})_f P_L \Big) \delta_{c_1c_2}\ , \\
 R_2^{\bar q_f Q Z} = &\ R_2^{\bar Q q_f Z} = 
    -\frac{i g g_s^2}{12 c_{\sss W} \pi^2}\gamma^{\mu_3}\\ &\quad
     \times \Big((\tilde\kappa^{\sss Q}_{\sss L})_f P_L +
          (\tilde\kappa^{\sss Q}_{\sss R})_f P_R \Big) \delta_{c_1c_2}\ , \\
 R_2^{\bar q_f Q W} = &\ R_2^{\bar Q q_f W} =
    -\frac{i g g_s^2}{6\sqrt{2} \pi^2}\gamma^{\mu_3}\\ &\quad
     \times \Big((\kappa^{\sss Q}_{\sss L})_f P_L +
          (\kappa^{\sss Q}_{\sss R})_f P_R \Big) \delta_{c_1c_2}\ .
\esp\label{eq:r2vlq}\ee
In the conventions of Eq.~\eqref{eq:r2sm} and Eq.~\eqref{eq:r2vlq}, $c_i$,
$\mu_i$, and $p_i$ indicate the colour index (which can be associated either with
the adjoint or the fundamental representation of $SU(3)$), the Lorentz index,
and the four-momentum of the $i^{\rm th}$ particle incoming to the
$R_2^{\ldots i\ldots}$ vertex, respectively. Moreover, an explicit summation upon
$q$, $Q$
and $f$ implies a summation over all quark species, the extra quark species and
the Standard Model quark species, respectively.

In the phenomenological study undertaken in Section~\ref{sec:LHCpheno}, the
virtual one-loop contributions to the NLO predictions are evaluated with the
\ml\ module~\cite{Hirschi:2011pa}, and then combined with the real contributions
by means of the FKS subtraction method~\cite{Frixione:1995ms} as implemented in
the \mfks\ package~\cite{Frederix:2009yq}. Both \ml\ and \mfks\ being part of
\amc~\cite{Alwall:2014hca}, the entire calculation is entirely automated from
the knowledge of the bare Lagrangian of Eq.~\eqref{eq:LVLQ} and the
specification of the process of interest~\cite{Christensen:2009jx}. Technically,
the translation of the
model Lagrangian into a UFO library~\cite{Degrande:2011ua} that contains
ultraviolet and $R_2$ counterterms and that could be used by \amc\ is
automatically performed with the \fr~\cite{Alloul:2013bka} and
\nloct~\cite{Degrande:2014vpa} packages, the latter program taking care of the
calculation of the one-loop ingredients of the model files. The corresponding
\fr\ and UFO models have been made publicly available and can be downloaded from
the webpage\\
\hspace*{.25cm}\url{http://feynrules.irmp.ucl.ac.be/wiki/NLOModels}.

\subsection{Benchmark scenarios}
\label{sec:benchmarks}
Throughout our phenomenological analysis, we adopt several series of benchmark
scenarios in which one single vector-like quark is light enough so that it could
be reachable at the LHC. Moreover, for the sake of simplicity, we enforce its
decay to proceed via a single
channel. We denote each class of scenarios by the acronym {\bf QVi} where the
symbol {\bf Q} can be either $T$, $B$, $X$ or $Y$ and refers to the nature of
the relevant extra quark, the symbol {\bf V} refers to the nature of the gauge
boson which the vector-like
quark $Q$ decays into and {\bf i} is a generation number related
to the family which the quark $Q$ mixes with. For instance, the scenario
{\bf TW2} would correspond to a setup in which the Standard
Model is supplemented by an extra up-type quark $T$ that decays into a
final state made of a $W$-boson and a strange quark with a branching ratio equal
to 1.

These types of scenarios are motivated by several considerations. The
mixings of the extra quark with the Standard Model sector are severely
constrained by flavour-changing neutral current probes~\cite{Buchkremer:2013bha,Alok:2015iha,Alok:2014yua},
LEP data~\cite{ALEPH:2005ab,Cacciapaglia:2010vn} and atomic parity violation
measurements~\cite{Deandrea:1997wk}. Taking the parameterisation of the
$\kappa$, $\tilde\kappa$ and $\hat\kappa$ parameters of
Eq.~\eqref{eq:conventions}, sizeable mixings with all three generations are only
allowed when the $\kappa_{\sss Q}$ parameters are below
$10^{-2} - 10^{-3}$~\cite{Buchkremer:2013bha}. Those bounds can, however, be
relaxed when the mixing pattern is restricted to involve one or two
quark generations. In our study, we enforce the vector-like quark mixing to
only deal with one specific generation of Standard Model quarks, and we fix the
values of the $\kappa_{\sss Q}$ parameters to their current experimental limits
of 0.07, 0.2 and 0.1 for mixings involving the first, second and third
generation, respectively.

\section{LHC phenomenology}
\label{sec:LHCpheno}
In this section, we compute total cross sections and differential distributions
both at the LO and NLO accuracy for several processes involving vector-like
quarks. We study the genuine effects of the NLO corrections, as well as the
induced reduction of the theoretical uncertainties. We then investigate
the effects of matching fixed-order calculations to parton showers, both at
LO and NLO. In Sections~\ref{sec:VLQpair} and \ref{sec:singleVLQ}, we,
respectively, focus on vector-like quark pair and single production. For these
processes, we consider QCD and electroweak diagrams that arise from the
Lagrangian of Eq.~\eqref{eq:LVLQ}, and we calculate the NLO
effects for both contributions and their possible interferences. In
Section~\ref{sec:others}, we focus on a set of additional processes where
vector-like quark effects could be relevant. We consider the
production of a single vector-like quark in association with a Standard Model
weak or Higgs boson, and of a pair of Standard Model Higgs or weak bosons in the
presence of vector-like quarks.

For the considered processes, the central (total and differential) cross-section
values are computed after setting the renormalisation and factorisation
scales to the average transverse mass of the final-state particles and by using
the NLO set of the NNPDF 3.0 parton density functions (PDF)~\cite{Ball:2014uwa}
accessed via the LHAPDF~6 library~\cite{Buckley:2014ana}. Scale uncertainties
are derived by varying both scales independently by a factor of two up and down,
and the PDF uncertainties are extracted following the recommendations of
Ref.~\cite{Demartin:2010er}, and both contributions to the theoretical
uncertainties are added in quadrature.

\subsection{Vector-like quark pair production at the LHC}
\label{sec:VLQpair}
\begin{figure*}
\centering
 \includegraphics[width=0.25\textwidth]{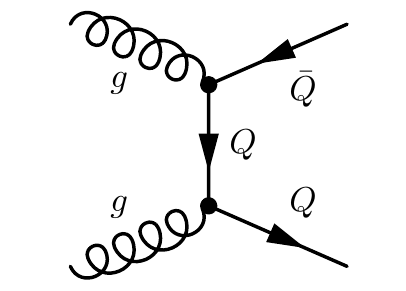}\quad
 \includegraphics[width=0.25\textwidth]{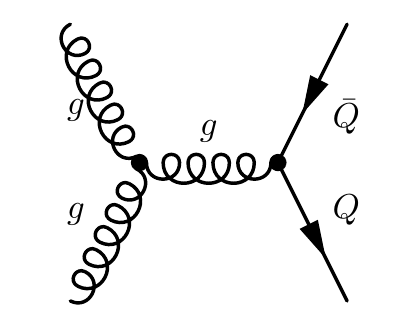}\quad
 \includegraphics[width=0.25\textwidth]{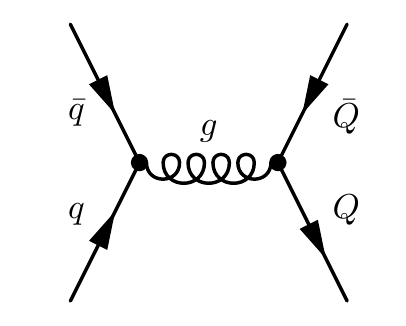}\\
 \includegraphics[width=0.25\textwidth]{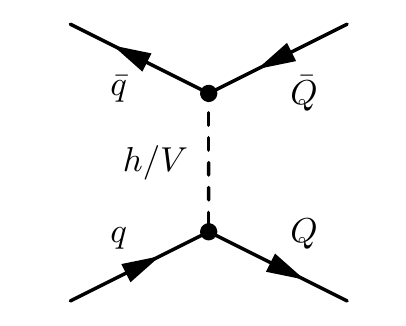}\quad
 \includegraphics[width=0.25\textwidth]{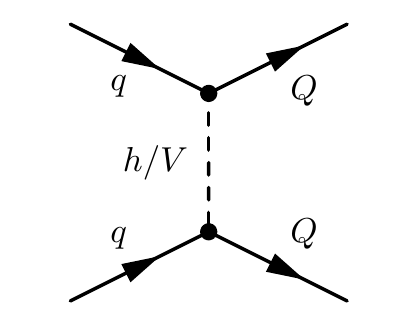}\quad
 \includegraphics[width=0.25\textwidth]{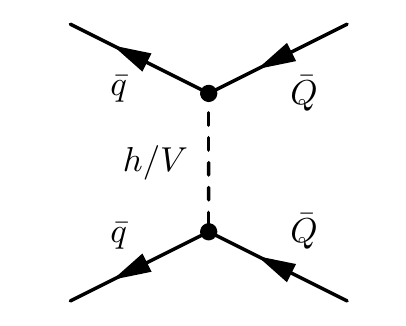}
 \caption{Strong (upper) and electroweak (lower) tree-level diagram
  contributions to the production of a pair of vector-like quarks, possibly
  carrying the same electric charge. The corresponding loop and real-emission
  diagrams are automatically generated by \amc\ and contains one extra power of
  the strong coupling.}
\label{fig:graph_vlqpair}
\end{figure*}

Vector-like quark pair production is in general dominated by QCD contributions,
which has the advantage to be independent of the model details. Model-dependent
electroweak diagrams induced by the last four lines of the Lagrangian of
Eq.~\eqref{eq:LVLQ} may, however, be non-negligible, in particular when
the final state of interest can be produced from the scattering of one or two
valence quarks. We focus on the production of a pair of vector-like quarks
including the cases where they have the same electric charge,
\be
  p p \to Q \bar Q\ , \qquad
  p p \to Q Q\qquad\text{and}\qquad
  p p \to \bar Q \bar Q\ ,
\label{eq:process_vlqpair}\ee
with $Q$ being either $T$, $B$, $X$ or $Y$. While the first of these three
subprocesses receives both strong (diagrams of the first line of
Figure~\ref{fig:graph_vlqpair}) and electroweak (first diagram of the second line of
Figure~\ref{fig:graph_vlqpair}) contributions, the latter two subprocesses
can only be mediated by the $t$-channel exchange of a weak or Higgs boson (last
two diagrams of the second line of Figure~\ref{fig:graph_vlqpair}).

NLO corrections to the strong contributions to the production
of a pair of vector-like quarks (the diagrams of the first line of
Figure~\ref{fig:graph_vlqpair}) can be automatically calculated within the
\amc\ framework. Thanks to the upgrading of the model to NLO as explained in the
previous section, it is now sufficient to type in the program shell,
taking the example of $T\bar T$ production,
\begin{verbatim}
  import model VLQ_NLO_UFO
  generate p p > tp tp~ [QCD]
  output
  launch
\end{verbatim}
With this set of instructions, we first import the UFO model associated
with the Lagrangian of Eq.~\eqref{eq:LVLQ} and then start the calculation of the
cross section and the generation of Monte Carlo events, at the NLO accuracy in
QCD, for the production
of a pair of $T\bar T$ quark and antiquark
(whose UFO names are \verb+tp+ and \verb+tp~+). Other vector-like quark
processes can be obtained by replacing the \verb+tp+ symbol by \verb+bp+ (for a
$B$ quark), \verb+x+ (for an $X$ quark) and \verb+y+ (for a $Y$ quark). LO event
generation can be achieved in the same way once the \verb+[QCD]+ tag is omitted.
We recall that the syntax is case insensitive.

For the electroweak channels (the diagrams of the second line of
Figure~\ref{fig:graph_vlqpair}), the command to be typed in reads
\begin{verbatim}
  generate p p > tp tp~ QCD=0 [QCD]
\end{verbatim}
still for the same example of $T\bar T$ production. Other channels (including
the production of a pair of heavy quarks carrying the same electric charge) can
be addressed similarly.
Care must, however, be taken as mixed electroweak and QCD loops appear at
NLO. In our approach, we focus on NLO calculations in QCD, and not
in QED or in the context of the electroweak theory. As a result, \amc\
automatically discards Feynman diagrams tag\-ged as an electroweak correction to a QCD
graph, although in our case all diagrams should be kept for a proper
cancellation of all divergences. One possible way to
cure this issue would be to include in the UFO model library all
ultraviolet and $R_2$ counterterms that would be necessary for undertaking mixed
NLO calculations in QCD and QED and to implement in \mfks\ the necessary
subtraction terms. This, however, goes beyond the scope of this work, so that in
order to maintain automation from the user standpoint, we have instead released
a public script that should be called for diagram generation in order to prevent
\amc\ from discarding any loop diagram that would be tagged as an electroweak
correction to a QCD Born diagram. More precisely, the script allows for the
inclusion of all box diagrams containing at least two strong interaction
vertices, but it removes weak-boson or Higgs-boson loop contributions that consist
of an electroweak correction to a QCD Born process. Additionally, diagrams
exhibiting the $t$-channel exchange of a weak or Higgs boson but with an
additional gluon are kept. Not using the script instead yields the removal of
several necessary box diagram contributions.

Finally, QCD and electroweak diagrams can interfere. Because of the mixing of
QCD and electroweak interaction orders at the one-loop level and the missing
counterterms and subtraction terms that
have been mentioned above, \amc\ cannot currently be used for the calculation of
these interferences beyond the LO accuracy. We therefore rely on LO simulations
and reweight instead the results to include a $K$-factor approximating the
effect of the QCD corrections, denoted by
$K_{\rm NLOQCD}^{(\rm int)}$, taken as the ratio of the NLO to the LO
(differential) results. We choose this $K_{\rm NLOQCD}^{(\rm int)}$ factor to be
the geometrical average of the $K_{\rm NLOQCD}^{(\rm QCD)}$ and
$K_{\rm NLOQCD}^{(\rm EW)}$ $K$-factors obtained in the context of the
QCD and electroweak diagrams taken independently,
\be
  K_{\rm NLOQCD}^{(\rm int)} =
     \sqrt{K_{\rm NLOQCD}^{(\rm QCD)}\ K_{\rm NLOQCD}^{(\rm EW)}}\ ,
\ee
where our notations explicitly indicate the nature of the corresponding
underlying Born process. We have additionally checked that, for the central
value, our procedure yields numerical differences that are of at most
2\%.
The related \amc\ command allowing for generating interference events is, again
for the example of $T\bar T$ production,
\begin{verbatim}
  generate p p > tp tp~ QCD^2=2 QED^2=2
\end{verbatim}
which is a standard command for LO event generation in \amc.

All \amc\ scripts necessary for differential and total cross-section calculations and
event generation are available from the webpage\\
\hspace*{.25cm}\url{http://feynrules.irmp.ucl.ac.be/wiki/NLOModels},\\ together
with the UFO and \fr\ models.

\begin{figure*}
\centering
 \includegraphics[width=0.80\textwidth]{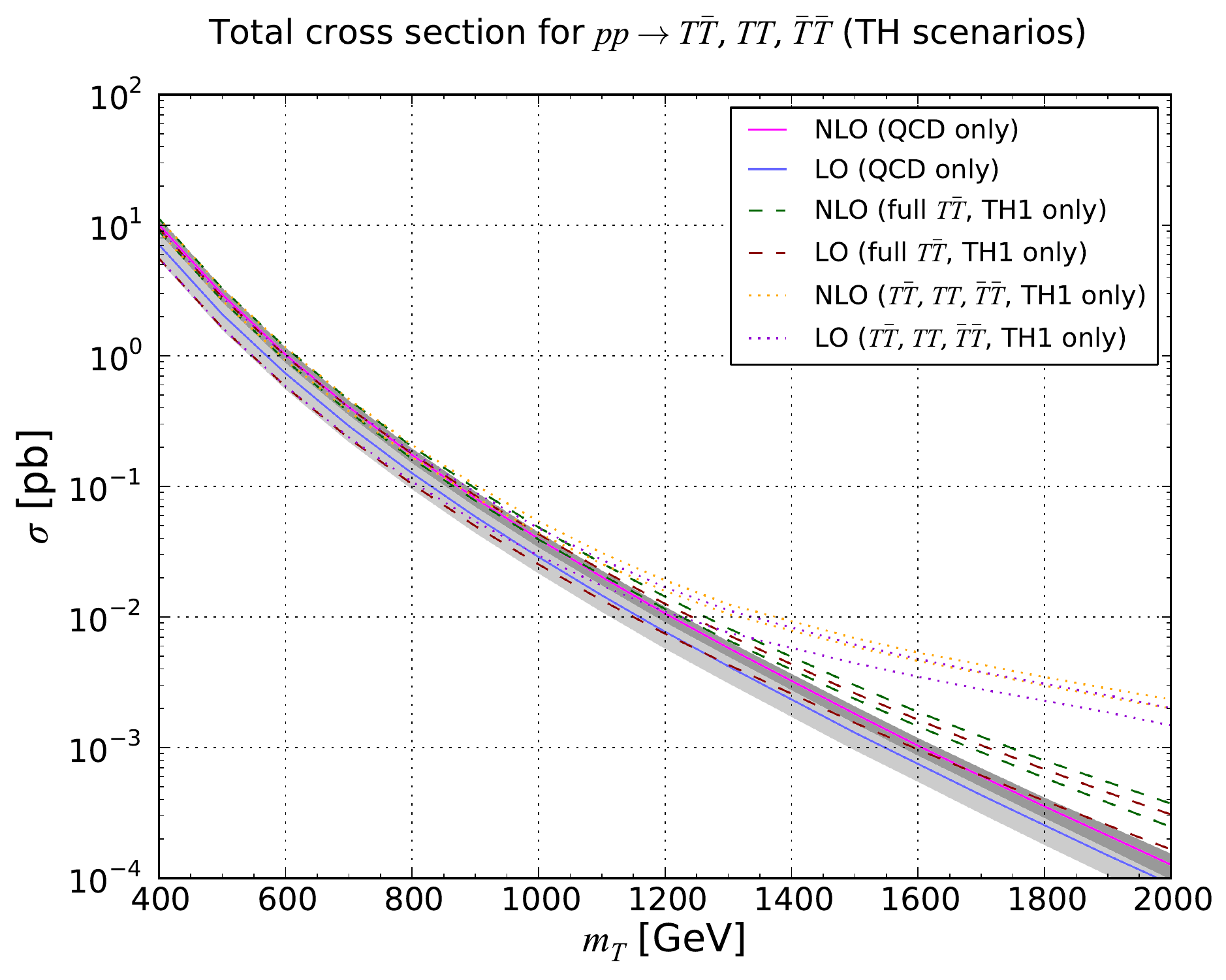}
 \caption{LO and NLO QCD inclusive cross sections for
   $T\bar T$/$TT$/$\bar T \bar T$ pair production at the LHC with \mbox{$\sqrt{s}=13$~TeV}. The QCD
   contribution results are presented together with the associated theoretical
   uncertainty bands, and we indicate the shifts in the bands that are
   induced by including weak or Higgs-boson exchange diagram contributions when
   they are non-negligible, which is only the case for the scenario {\bf TH1}.}
\label{fig:xsec_Tpair}
\end{figure*}

In Figure~\ref{fig:xsec_Tpair} and Table~\ref{tab:Txs}, we present LO and NLO
total cross sections for the production of a pair of vector-like $T$
quarks for the different scenarios introduced in Section~\ref{sec:benchmarks},
and we depict the dependence of the cross sections on the vector-like quark
mass. We first focus on the pure QCD contribution that is independent of the
vector-like quark nature (diagrams of the upper line of
Figure~\ref{fig:graph_vlqpair}). The genuine NLO contributions are found to be
important as they first induce a shift in the cross section of about 50\% within
the entire probed $m_{\sss Q}$ range and next reduce the dependence of the
rate on the unphysical factorisation and renormalisation scales. The uncertainty
band is indeed significantly reduced when NLO effects are accounted for, the
scale dependence being reduced to the level of about $\pm10\%$ over the entire
mass range. At the NLO, the scale dependence induced by the
virtual contributions indeed partially compensates the one stemming from the
Born and real-emission diagrams. Our results for the pure QCD case (strong
production Born diagrams only) agree with the literature, and we recall that
this corresponds to the production of a top-quark pair with a different
top-quark mass~\cite{Aliev:2010zk}. It is in contrast the first calculation
including the impact of the electroweak diagrams at the NLO accuracy in QCD.

\begin{table*}
 \centering
 \renewcommand{\arraystretch}{1.40}
 \setlength{\tabcolsep}{12pt}
 \begin{tabular}{cc|cc}
   $m_{\sss T}$~[GeV] & Scenario & $\sigma_{\rm LO}$ [pb] & $\sigma_{\rm NLO}$ [pb]\\
   \hline
   \multirow{2}{*}{$400$} & {\footnotesize QCD} & $(7.069\ 10^{0}){}^{+32.0\%}_{-22.6\%}{}^{+2.7\%}_{-2.7\%}$ &
     $(1.004\ 10^{1}){}^{+9.4\%}_{-11.3\%}{}^{+2.5\%}_{-2.5\%}$\\
   & {\footnotesize \bf TH1 } & $(7.022\ 10^{0}){}^{+30.2\%}_{-23.8\%}{}^{+1.2\%}_{-4.1\%}$ &
     $(9.980\ 10^{0}){}^{+8.0\%}_{-12.5\%}{}^{+1.2\%}_{-3.8\%}$\\
   \hdashline
   \multirow{2}{*}{$800$} & {\footnotesize QCD} & $(1.261\ 10^{-1}){}^{+33.2\%}_{-23.2\%}{}^{+3.8\%}_{-3.8\%}$ &
     $(1.733\ 10^{-1}){}^{+8.5\%}_{-11.1\%}{}^{+4.4\%}_{-4.4\%}$\\
   & {\footnotesize \bf TH1 } & $(1.244\ 10^{-1}){}^{+18.8\%}_{-31.2\%}{}^{+-7.3\%}_{-14.0\%}$ &
     $(1.702\ 10^{-1}){}^{+-2.3\%}_{-20.0\%}{}^{+-6.0\%}_{-13.9\%}$\\
   \hdashline
   \multirow{2}{*}{$1200$} & {\footnotesize QCD} & $(7.685\ 10^{-3}){}^{+34.0\%}_{-23.7\%}{}^{+5.8\%}_{-5.8\%}$ &
     $(1.061\ 10^{-2}){}^{+8.8\%}_{-11.4\%}{}^{+5.8\%}_{-5.8\%}$\\
   & {\footnotesize \bf TH1 } & $(1.053\ 10^{-2}){}^{+-1.7\%}_{-36.7\%}{}^{+-18.4\%}_{-25.8\%}$ &
     $(1.372\ 10^{-2}){}^{+-16.6\%}_{-29.0\%}{}^{+-18.2\%}_{-25.8\%}$\\
   \hdashline
   \multirow{2}{*}{$1600$} & {\footnotesize QCD} & $(7.477\ 10^{-4}){}^{+34.9\%}_{-24.2\%}{}^{+8.5\%}_{-8.5\%}$ &
     $(1.030\ 10^{-3}){}^{+9.0\%}_{-11.6\%}{}^{+8.6\%}_{-8.6\%}$\\
   & {\footnotesize \bf TH1 } & $(3.395\ 10^{-3}){}^{+-3.3\%}_{-27.0\%}{}^{+-13.3\%}_{-19.9\%}$ &
     $(4.117\ 10^{-3}){}^{+-14.6\%}_{-21.8\%}{}^{+-14.4\%}_{-20.9\%}$\\
   \hdashline
   \multirow{2}{*}{$2000$} & {\footnotesize QCD} & $(8.980\ 10^{-5}){}^{+35.5\%}_{-24.5\%}{}^{+18.3\%}_{-18.3\%}$ &
     $(1.260\ 10^{-4}){}^{+8.7\%}_{-11.7\%}{}^{+17.8\%}_{-17.8\%}$\\
   & {\footnotesize \bf TH1 } & $(1.563\ 10^{-3}){}^{+4.2\%}_{-20.0\%}{}^{+-5.4\%}_{-13.0\%}$ &
     $(1.960\ 10^{-3}){}^{+-6.3\%}_{-14.0\%}{}^{+-6.0\%}_{-13.6\%}$\\
  \end{tabular}
  \renewcommand{\arraystretch}{1.0}
  \caption{\small \label{tab:Txs}LO and NLO QCD inclusive cross sections for
   $T\bar T$/$TT$/$\bar T\bar T$ production at the LHC, running at a center-of-mass energy of
   $\sqrt{s}=13$~TeV. The results are shown together with the associated scale
   and PDF relative uncertainties. For all scenarios but the {\bf TH1} one, the
   predictions match the pure QCD results.}
\end{table*}

In the considered scenarios, electroweak contributions to the
production of a pair of vector-like quarks possibly carrying the same
electric charge are found to be important
only for the {\bf TH1} scenario, the associated results being shown on
Figure~\ref{fig:xsec_Tpair} as dashed and dotted bands. These bands are,
respectively,
related to the production of a pair of vector-like quark-antiquark (dashed) and
to the production of a pair of heavy quark regardless of their electric charge
(dotted). In this last case, the contributions from the three
processes of Eq.~\eqref{eq:process_vlqpair} are summed over. The {\bf TH1}
scenario features an extra quark that mixes with the
first generation of Standard Model quarks so that parton density effects could
lead to an enhancement of the production rate
due to quark-antiquark, quark-quark and antiquark-antiquark
(electroweak) scattering diagrams involving one or two initial
valence quarks. This is particularly pronounced for setups featuring heavy
vector-like quarks that require to probe large Bjorken-$x$ phase-space
regions. As a
consequence, the central cross-section values and the scale and parton density
uncertainties are different from the pure QCD context since non-QCD
diagrams (featuring a different initial state) dominate. This is
illustrated in Table~\ref{tab:Txs} for a few mass choices. Whereas the
production rates are always larger, the
uncertainties can be either smaller or larger than in the QCD case.

We observe a huge gain in cross section for {\bf TH1} scenarios with a
very heavy extra quark. This stems from Eq.~\eqref{eq:conventions} that
shows that the coupling of the extra quark to the Higgs boson and a lighter
quark has been taken proportional to the vector-like quark mass, and
is thus enhanced for large values of $m_{\sss Q}$. In principle, such a coupling
should also be proportional to the related mixing matrix element $\zeta$
that compensates this enhancement, as shown in Eq.~\eqref{eq:conventions} and in
Eq.~\eqref{eq:zeta}. Setting $\zeta=1$, this feature is translated into the
adopted value for the $\kappa_Q$ parameters.

Similar properties can be found
in the context of the production of $B$, $X$ and $Y$ vector-like quarks, as
illustrated in Appendix~\ref{app:vlqpair}.

\begin{figure*}
\centering
 \includegraphics[width=0.48\textwidth]{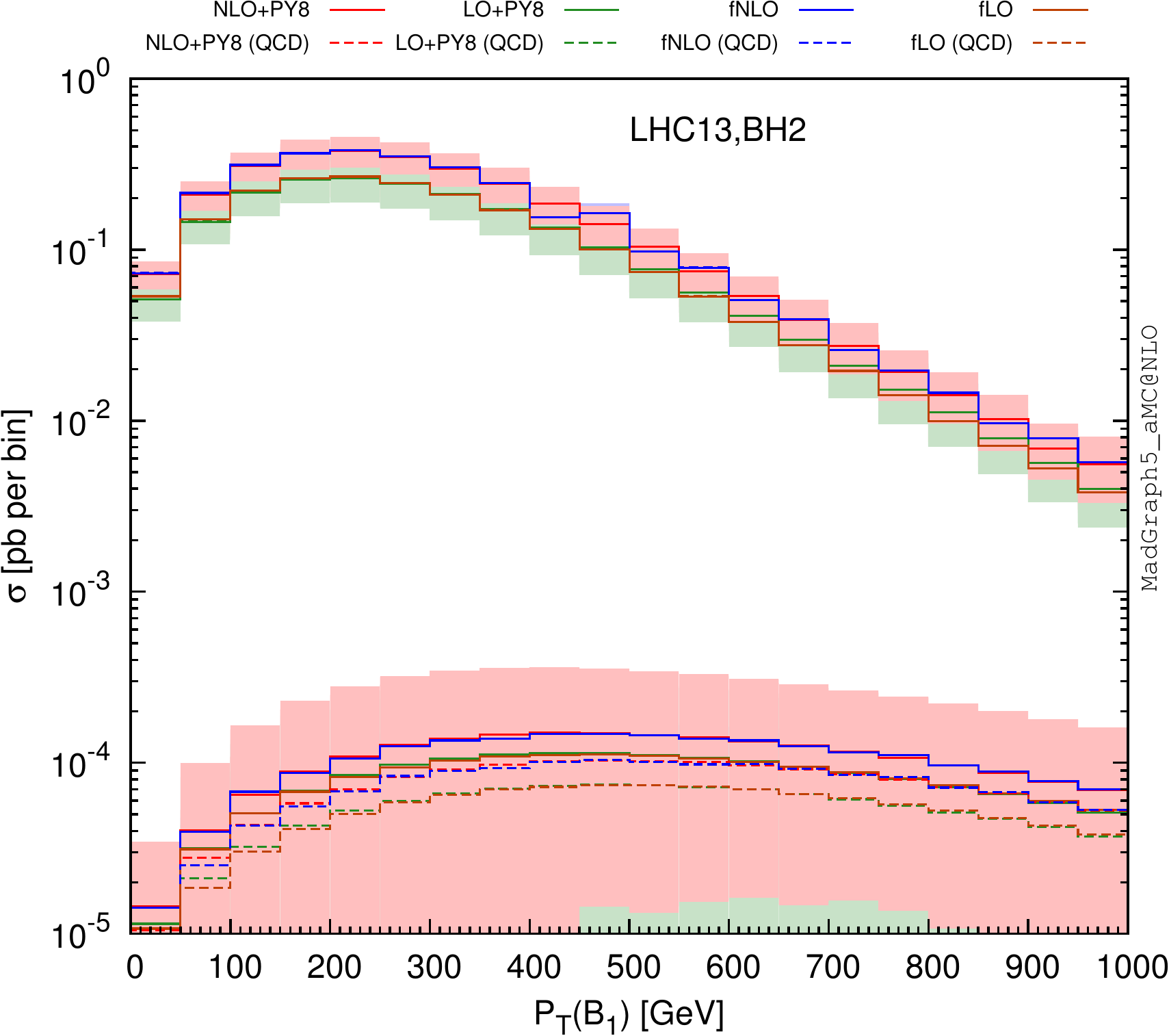}\quad
 \includegraphics[width=0.48\textwidth]{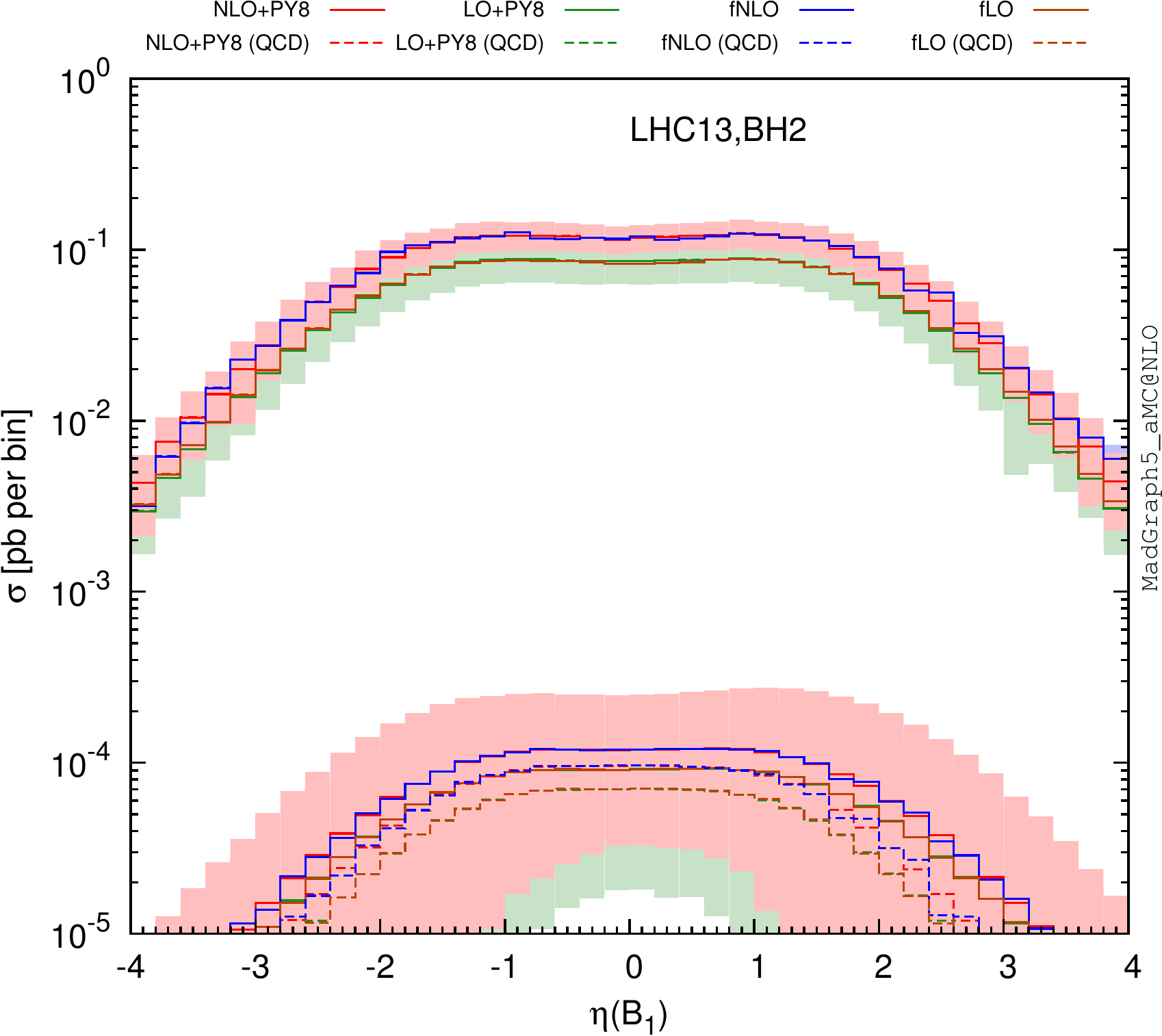}\\
 \includegraphics[width=0.48\textwidth]{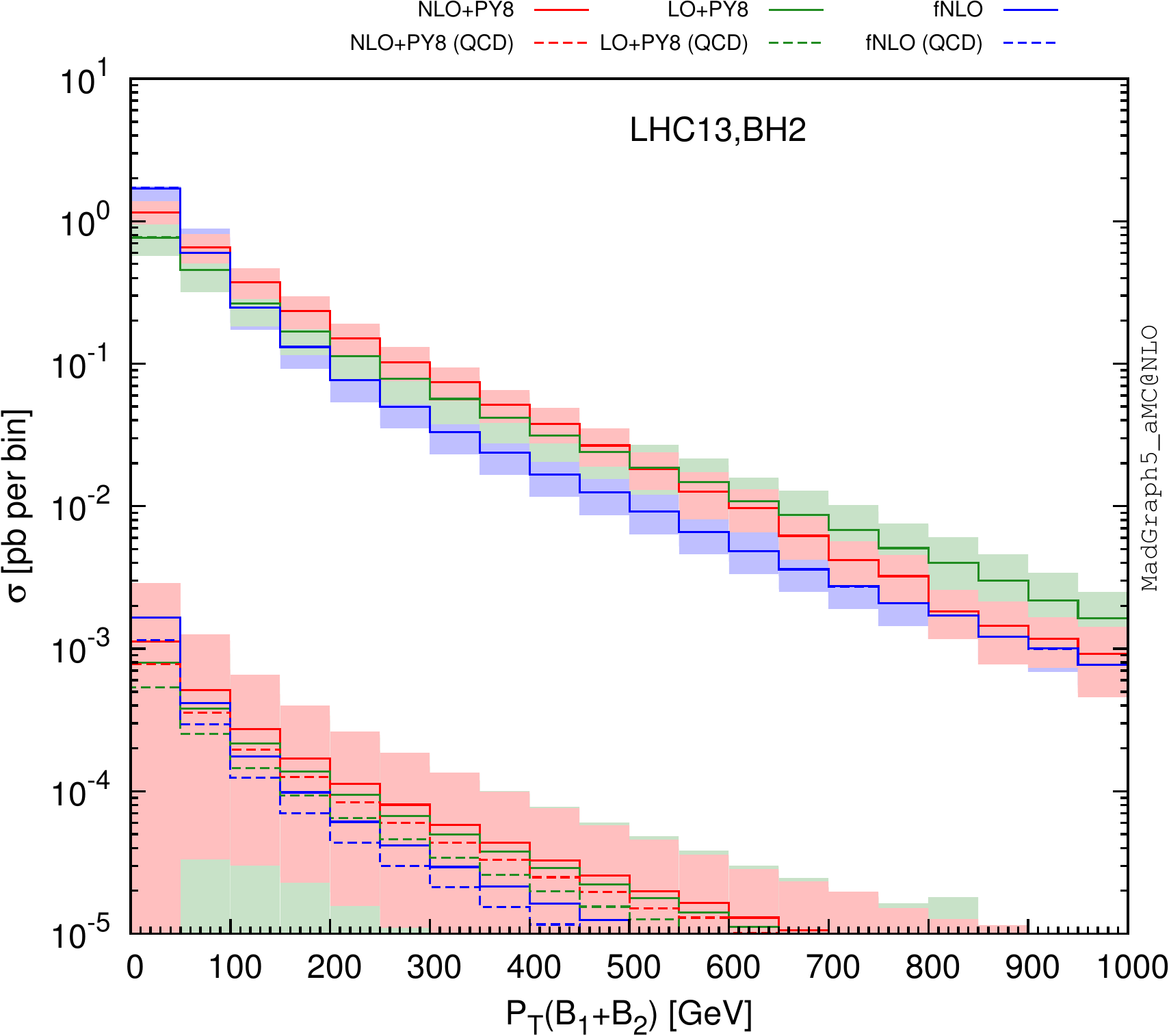}\quad
 \includegraphics[width=0.48\textwidth]{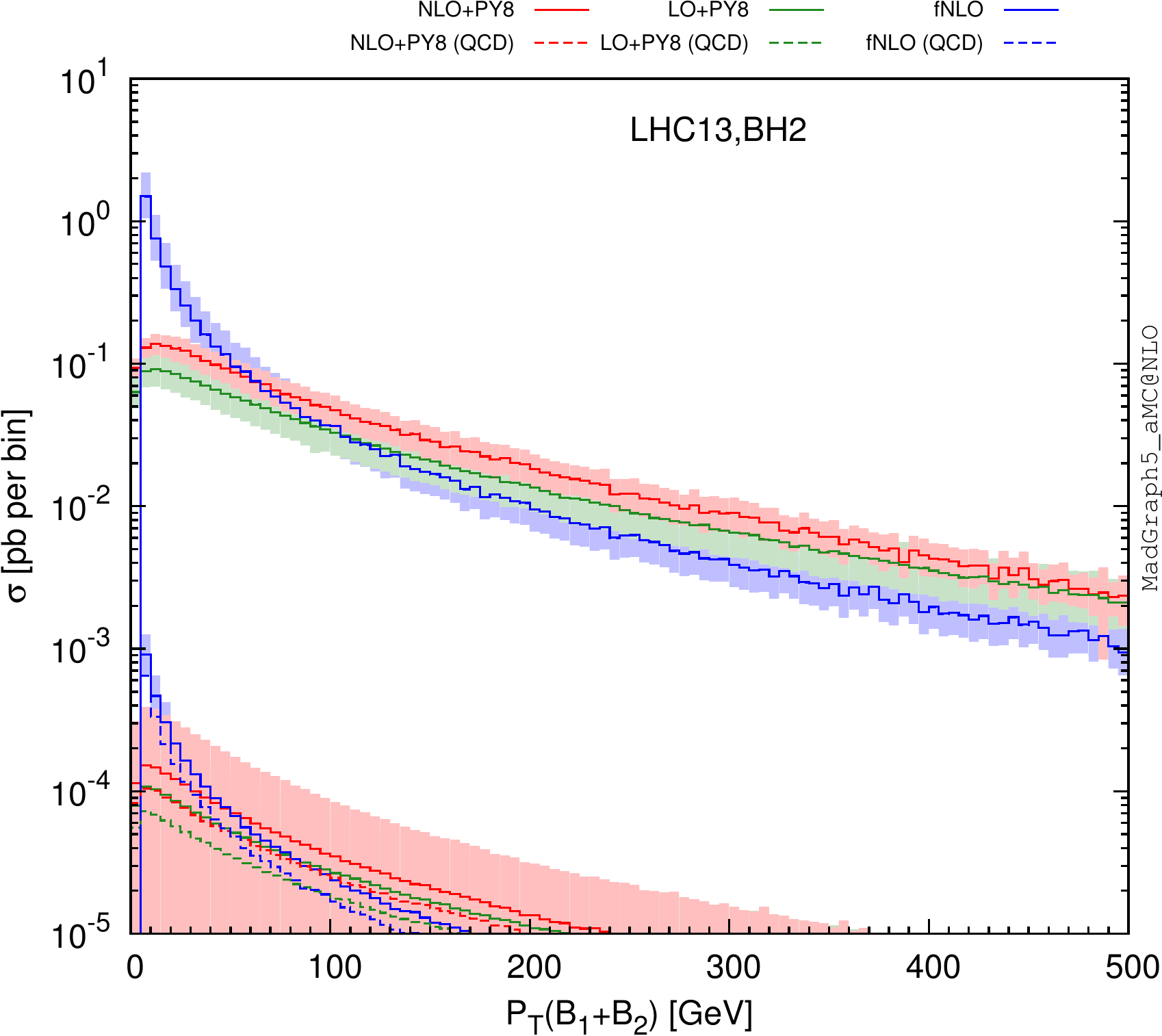}\\
 \caption{Properties of the pair-produced vector-like $B$-quarks. We present
  the transverse-momentum (upper left) and pseudorapidity (upper right)
  distribution for the first $B$-quark, as well as the transverse-momentum
  spectrum of the $B$-quark pair (lower left for the $[0,1000]$~GeV
  transverse-momentum region and lower
  right for a zoom in the $[0,500]$~GeV region). We compare fixed-order QCD
  predictions at the LO accuracy (purple dashed curve),
  NLO accuracy (blue dashed curve), and after matching these two calculations to
  parton showers (green and red dashed curves for the LO and NLO cases,
  respectively), and the solid lines depict the results once the electroweak
  diagram contributions are included. We have fixed the $B$-quark mass either to
  500~GeV (upper series of curves) or to 1500~GeV (lower series of curves).}
\label{fig:diff_vlqpair}
\end{figure*}

Accurate differential distributions are often helpful for setting more precise
exclusion limits and refine the experimental search strategies. Our
implementation in the \amc\ platform can be used to
this aim, and we present in Figure~\ref{fig:diff_vlqpair} several differential
distributions in several observables, including NLO and parton-shower effects.
We have chosen the {\bf BH2} class of benchmark
scenarios with a vector-like quark mass set either to 500~GeV (upper series of
curves on the figure) or 1500~GeV (lower series of curves on the figure). In our
calculations, we have made use of the {\sc MadSpin}~\cite{Artoisenet:2012st} and
{\sc MadWidth}~\cite{Alwall:2014bza} programs for automatically
taking care of the heavy quark
decays in a way in which both off-shell and spin correlation effects are
retained, matched the fixed-order calculcation with the parton-shower and
hadronisation infrastructure as modelled by the \py\
package~\cite{Sjostrand:2014zea}, and we have reconstructed all final-state jets
by means of the anti-$k_T$ algorithm~\cite{Cacciari:2008gp} with a radius
parameter set to 0.5 as implemented in {\sc FastJet}~\cite{Cacciari:2011ma}.
As shown on Figure~\ref{fig:diff_vlqpair} and Figure~\ref{fig:diff_vlqpair2},
our predictions confirm the total cross-section findings of Table~\ref{tab:Bxs}
(see Appendix~\ref{app:vlqpair}). The contributions of the electroweak diagrams
are, respectively, negligible and significant for light and heavy
vector-like quarks. Moreover, the parton density uncertainties dominate for
setups exhibiting a
large $M_{\sss B}$ value, rendering the theoretical predictions barely reliable.

\begin{figure*}
\centering
 \includegraphics[width=0.48\textwidth]{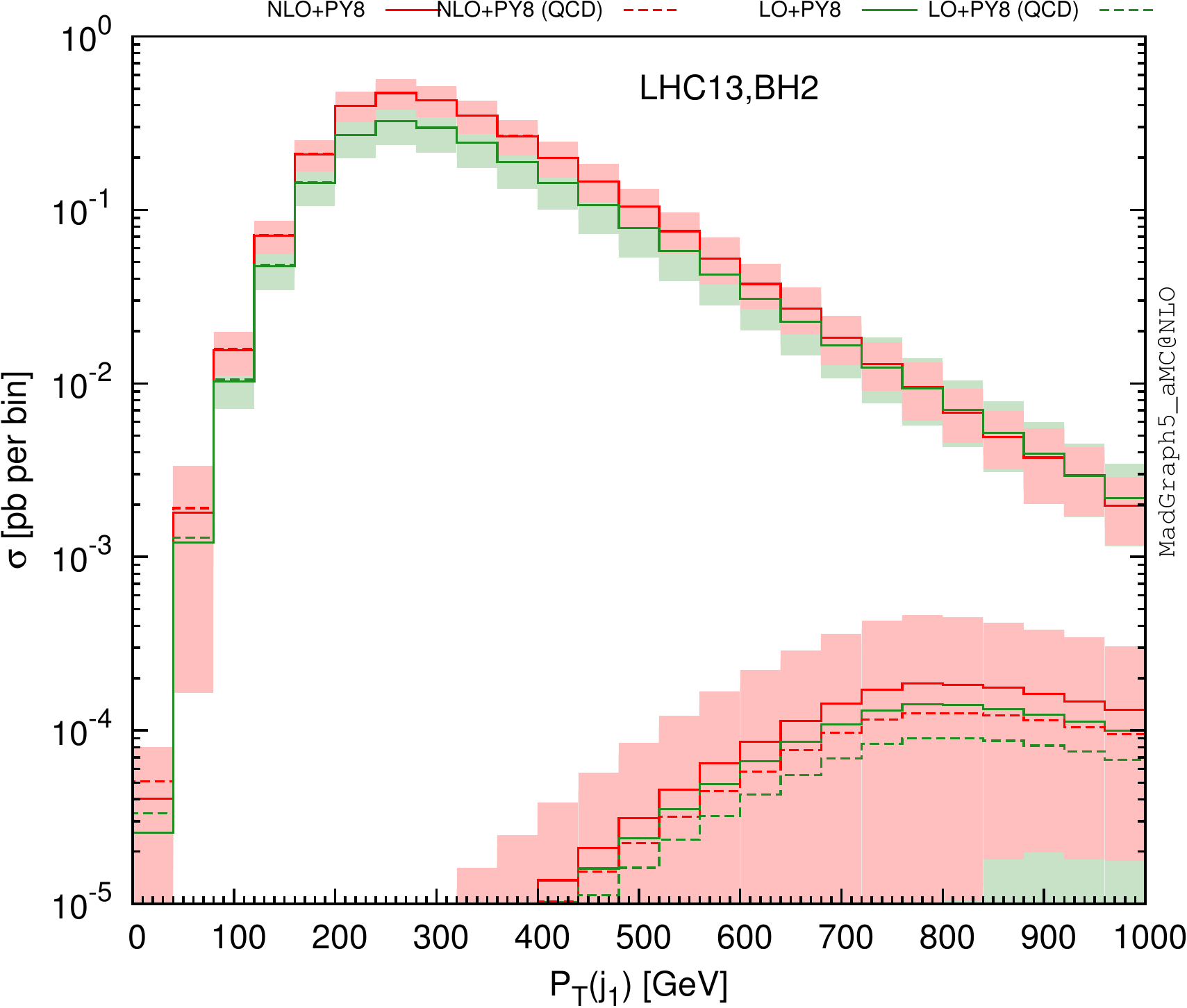}\quad
 \includegraphics[width=0.48\textwidth]{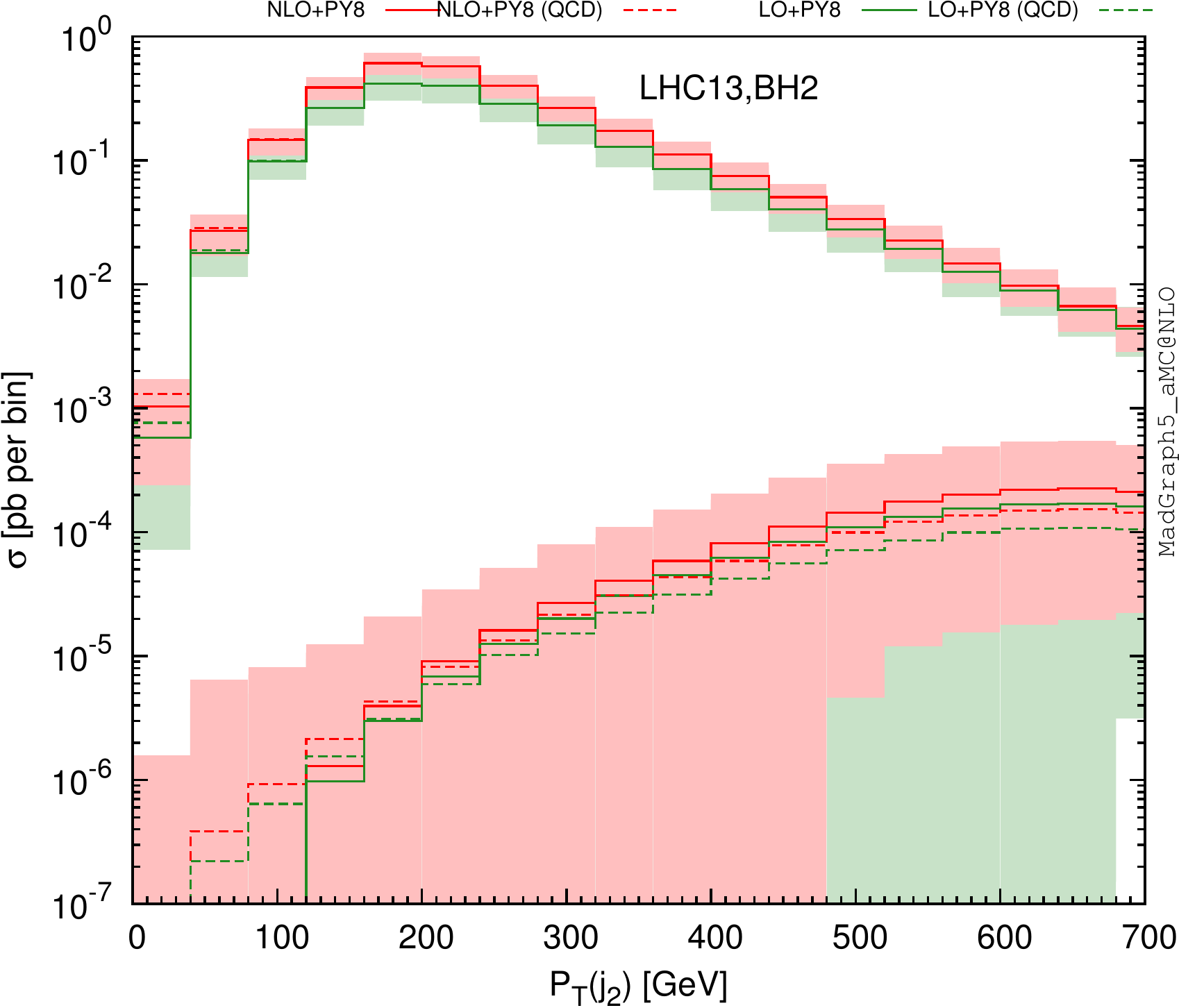}\\
 \includegraphics[width=0.48\textwidth]{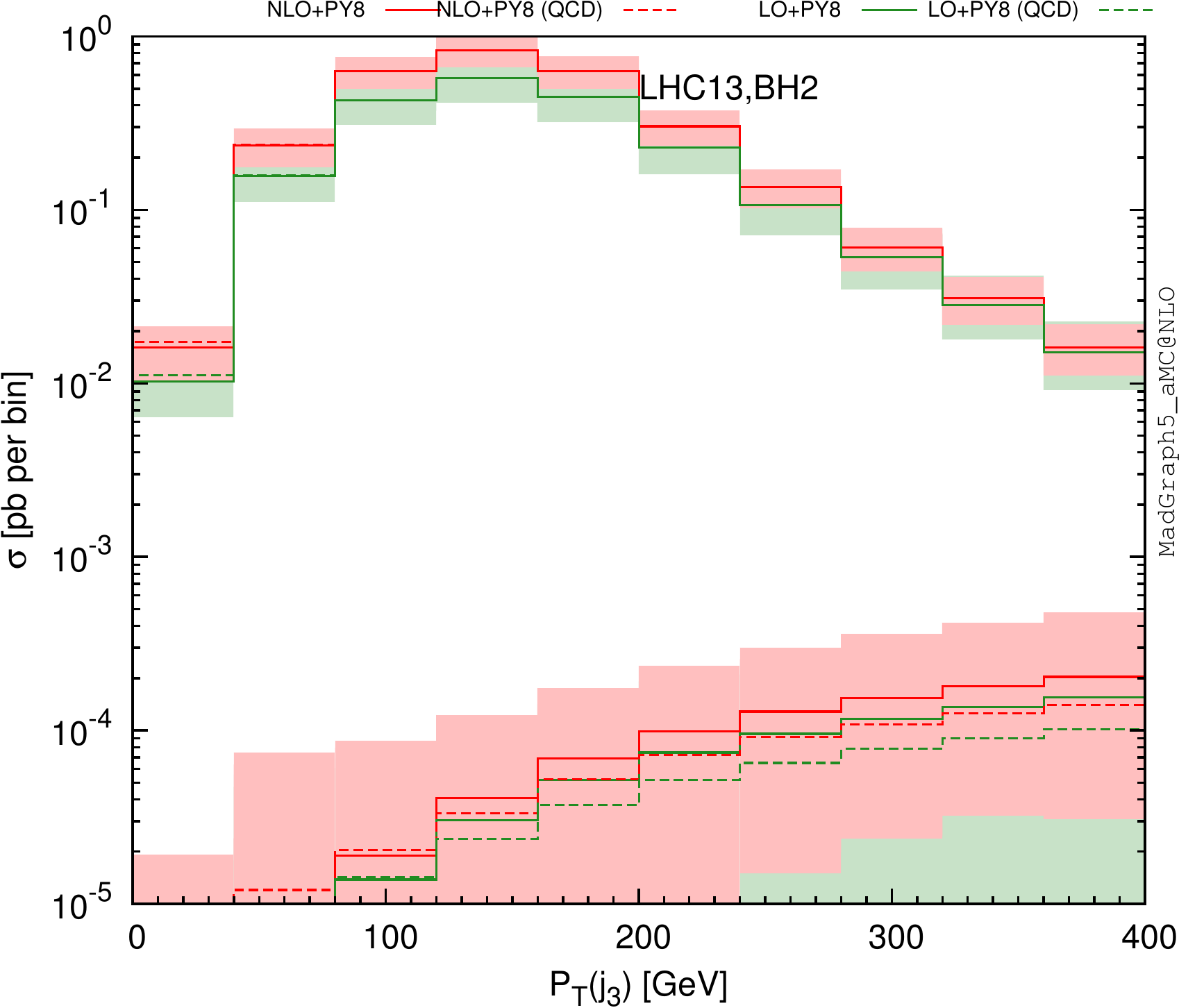}\quad
 \includegraphics[width=0.48\textwidth]{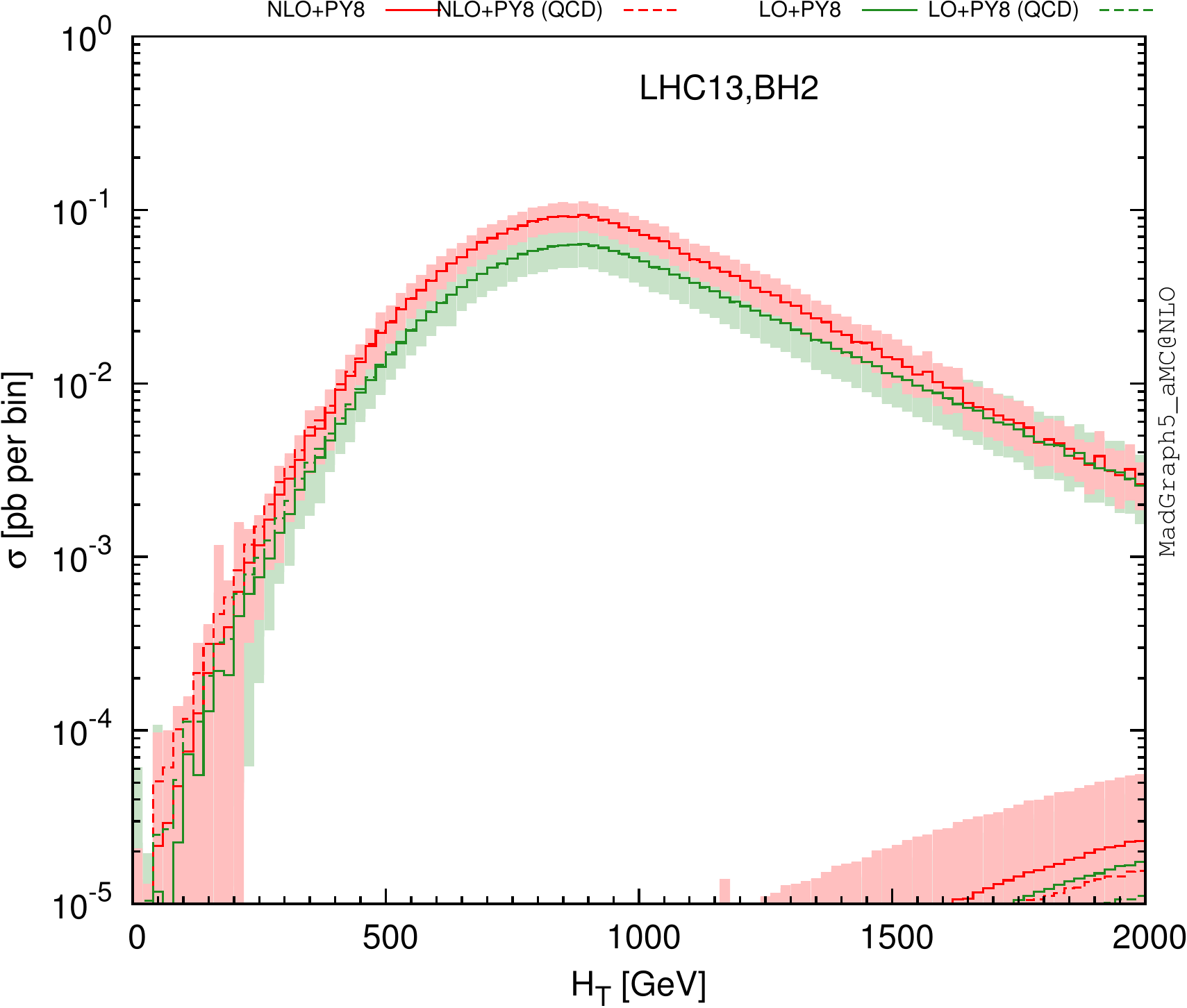}\\
 \caption{Same as in Figure~\ref{fig:diff_vlqpair} but for the distribution in
  the transverse momentum of the first three leading jets and the $H_T$
  variable defined as the sum of the transverse momentum of all the
  final-state jets and leptons.}
\label{fig:diff_vlqpair2}
\end{figure*}

In the two upper panels of Figure~\ref{fig:diff_vlqpair}, we study the
properties of the first produced $B$-quark and show its transverse momentum
$p_T(B_1)$ and pseudorapidity $\eta(B_1)$ distributions. In the case of a
light $B$-quark, the parton-shower effects (green and red solid lines) slightly
affect the shapes of the spectra predicted by the fixed-order calculations (blue
and brown solid lines), both at the LO and NLO accuracies. For
heavier vector-like quarks, slight modifications can be observed in the small
$p_T$ and large $|\eta|$ region although the accurate modelling of the first
extra jet does not yield any impact at the level of the individual $B$-quarks.
These differences are largely covered by the theoretical uncertainties
stemming from the poor knowledge of the parton densities in the relevant
phase-space regions.  These PDF uncertainties are dominant so that the reduction at the level of the scale uncertainties has a small impact. However, PDF uncertainties are expected to improve in the coming years thanks to new LHC data, so that it is mandatory to have NLO calculations available to get a better control on the predictions.
In contrast, parton-shower effects are directly visible when
distributions related to the two $B$-quarks considered as a pair are considered
(Figure~\ref{fig:diff_vlqpair}). Focusing on the related
transverse-momentum distributions, the fixed-order predictions (for which only
the $p_T(B_1+B_2)=0$ bin is populated at LO) diverge at small $p_T$ due to
soft and collinear radiation giving rise to large logarithms that must be
resummed to all orders to obtain reliable predictions. This resummation is
effectively achieved by matching the fixed-order calculations to parton
showers, and the resulting distribution exhibits a reliable behaviour
with a peak for $p_T(B_1+B_2)$ of about 10--20~GeV. Uncertainties originating from the choice of the shower algorithm and its inherent free parameters are,
however, not estimated.

The magnitude of the electroweak diagrams is also studied (dashed lines). In the
case of lighter $B$-quarks, the theoretical predictions are essentially driven
by the QCD contributions so that no differences can be noticed. This contrasts
with the heavy $B$-quark case where electroweak diagrams enhance the total
rate by about 30\% (see Appendix~\ref{app:vlqpair}) and also impact the
differential distributions both in terms of normalisation and shape. This
originates from the different topologies of the electroweak diagrams that
feature a $t$-channel colourless boson exchange.

We now include the vector-like quark decays into
a Higgs boson and a strange quark and reconstruct the final-state jets as
detailed above. Considering only hard central jets for which
$|\eta|<2.4$ and $p_T>30$~GeV, we present the transverse-momentum distributions
of the three leading jets as well as the spectrum of the $H_T$ variable defined
as the scalar sum of the transverse momenta of all final-state jets and leptons
in Figure~\ref{fig:diff_vlqpair2}, the generated events being inclusive in the
Higgs-boson decays. Leptons are included only if their transverse
momentum is larger than 30~GeV, their pseudorapidity smaller than 2.4 in
absolute value and if they are isolated from any jet by an angular distance in
the transverse plane $\Delta R$ of at least 0.5.
Focusing on fixed-order predictions matched to parton showers, we observe that
the first two leading jets are in general hard, as they result from the decay of
two heavy coloured particles. In contrast, the structure
of the $p_T$ dependence of the third jet is more representative of the one
expected from a radiation jet, this jet being most of the time arising from
initial-state or final-state radiation. Turning to the $H_T$ distribution (lower
right panel of the figure), one notices a peak for very small $H_T$ values
when the $B$-quark mass is fixed at 500~GeV. This
feature arises from events where the two jets originating from the heavy quark
decays are mis-reconstructed, the leading jet being thus the radiation
jet so that the associated jet activity in the events is not significantly
large.

\subsection{Single vector-like quark production in association with jets}
\label{sec:singleVLQ}

\begin{figure*}
\centering
 \includegraphics[width=0.25\textwidth]{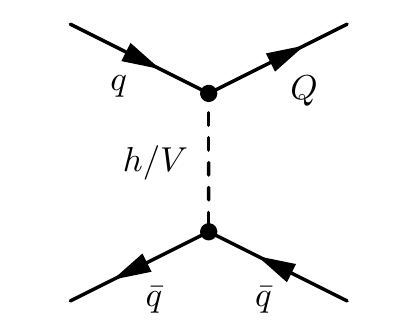}\
 \includegraphics[width=0.25\textwidth]{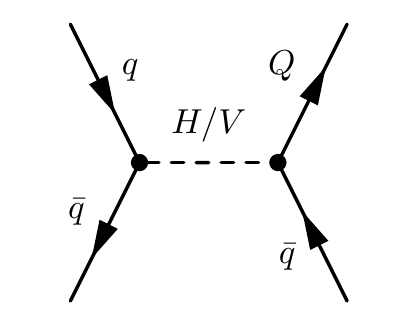}
 \caption{Representative Feynman diagram for single vector-like quark production
  in association with a quark. Other diagrams exist for final-state antiquarks,
  and all flavour assignments are understood. Virtual- and real-emission diagrams
  necessary for QCD correction calculations can be generated by \amc.}
\label{fig:graph_vlqsingle}
\end{figure*}

\begin{table*}
 \centering
 \renewcommand{\arraystretch}{1.40}
 \setlength{\tabcolsep}{12pt}
 \begin{tabular}{cc|cc}
   $m_{\sss B}$~[GeV] & Scenario & $\sigma_{\rm LO}$ [pb] & $\sigma_{\rm NLO}$ [pb]\\
   \hline
   \multirow{4}{*}{$400$} & {\footnotesize \bf BZ1} & $(1.277\ 10^{0}){}^{+2.9\%}_{-3.0\%}{}^{+2.6\%}_{-2.6\%}$ &
     $(1.288\ 10^{0}){}^{+0.9\%}_{-0.4\%}{}^{+2.6\%}_{-2.6\%}$\\
   & {\footnotesize \bf BZ2 } & $(3.573\ 10^{0}){}^{+0.8\%}_{-1.3\%}{}^{+4.8\%}_{-4.8\%}$ &
     $(3.668\ 10^{0}){}^{+1.0\%}_{-0.3\%}{}^{+4.7\%}_{-4.7\%}$\\
   & {\footnotesize \bf BW1 } & $(2.021\ 10^{0}){}^{+2.9\%}_{-3.0\%}{}^{+1.7\%}_{-1.7\%}$ &
     $(2.033\ 10^{0}){}^{+0.9\%}_{-0.6\%}{}^{+1.7\%}_{-1.7\%}$\\
   & {\footnotesize \bf BW2 } & $(3.037\ 10^{0}){}^{+0.0\%}_{-0.7\%}{}^{+1.4\%}_{-1.4\%}$ &
     $(3.161\ 10^{0}){}^{+1.3\%}_{-0.7\%}{}^{+1.4\%}_{-1.4\%}$\\
   \hdashline
   \multirow{4}{*}{$800$} & {\footnotesize \bf BZ1} & $(3.197\ 10^{-1}){}^{+6.6\%}_{-5.9\%}{}^{+3.3\%}_{-3.3\%}$ &
     $(3.461\ 10^{-1}){}^{+1.0\%}_{-1.2\%}{}^{+3.2\%}_{-3.2\%}$\\
   & {\footnotesize \bf BZ2 } & $(6.777\ 10^{-1}){}^{+4.9\%}_{-4.6\%}{}^{+8.0\%}_{-8.0\%}$ &
     $(7.671\ 10^{-1}){}^{+1.4\%}_{-1.3\%}{}^{+7.4\%}_{-7.4\%}$\\
   & {\footnotesize \bf BW1 } & $(5.450\ 10^{-1}){}^{+6.4\%}_{-5.8\%}{}^{+1.8\%}_{-1.8\%}$ &
     $(5.858\ 10^{-1}){}^{+1.0\%}_{-1.1\%}{}^{+1.8\%}_{-1.8\%}$\\
   & {\footnotesize \bf BW2 } & $(5.335\ 10^{-1}){}^{+3.4\%}_{-3.5\%}{}^{+2.0\%}_{-2.0\%}$ &
     $(5.938\ 10^{-1}){}^{+1.4\%}_{-0.8\%}{}^{+1.8\%}_{-1.8\%}$\\
   \hdashline
   \multirow{4}{*}{$1200$} & {\footnotesize \bf BZ1} & $(1.129\ 10^{-1}){}^{+8.8\%}_{-7.6\%}{}^{+4.2\%}_{-4.2\%}$ &
     $(1.291\ 10^{-1}){}^{+1.7\%}_{-2.3\%}{}^{+4.0\%}_{-4.0\%}$\\
   & {\footnotesize \bf BZ2 } & $(1.966\ 10^{-1}){}^{+7.2\%}_{-6.5\%}{}^{+12.6\%}_{-12.6\%}$ &
     $(2.268\ 10^{-1}){}^{+1.7\%}_{-2.0\%}{}^{+11.8\%}_{-11.8\%}$\\
   & {\footnotesize \bf BW1 } & $(2.021\ 10^{-1}){}^{+8.5\%}_{-7.3\%}{}^{+2.0\%}_{-2.0\%}$ &
     $(2.298\ 10^{-1}){}^{+1.5\%}_{-2.1\%}{}^{+2.1\%}_{-2.1\%}$\\
   & {\footnotesize \bf BW2 } & $(1.406\ 10^{-1}){}^{+5.7\%}_{-5.4\%}{}^{+3.3\%}_{-3.3\%}$ &
     $(1.645\ 10^{-1}){}^{+1.8\%}_{-1.8\%}{}^{+3.1\%}_{-3.1\%}$\\
   \hdashline
   \multirow{4}{*}{$1600$} & {\footnotesize \bf BZ1} & $(4.607\ 10^{-2}){}^{+10.3\%}_{-8.8\%}{}^{+5.2\%}_{-5.2\%}$ &
     $(5.519\ 10^{-2}){}^{+2.6\%}_{-3.1\%}{}^{+4.9\%}_{-4.9\%}$\\
   & {\footnotesize \bf BZ2 } & $(6.934\ 10^{-2}){}^{+8.9\%}_{-7.8\%}{}^{+18.9\%}_{-18.9\%}$ &
     $(8.316\ 10^{-2}){}^{+2.5\%}_{-2.8\%}{}^{+17.6\%}_{-17.6\%}$\\
   & {\footnotesize \bf BW1 } & $(8.603\ 10^{-2}){}^{+9.9\%}_{-8.5\%}{}^{+2.4\%}_{-2.4\%}$ &
     $(1.022\ 10^{-1}){}^{+2.4\%}_{-3.0\%}{}^{+2.4\%}_{-2.4\%}$\\
   & {\footnotesize \bf BW2 } & $(4.447\ 10^{-2}){}^{+7.4\%}_{-6.7\%}{}^{+5.1\%}_{-5.1\%}$ &
     $(5.423\ 10^{-2}){}^{+2.5\%}_{-2.6\%}{}^{+4.7\%}_{-4.7\%}$\\
   \hdashline
   \multirow{4}{*}{$2000$} & {\footnotesize \bf BZ1} & $(2.022\ 10^{-2}){}^{+11.6\%}_{-9.7\%}{}^{+6.4\%}_{-6.4\%}$ &
     $(2.516\ 10^{-2}){}^{+3.3\%}_{-3.9\%}{}^{+6.1\%}_{-6.1\%}$\\
   & {\footnotesize \bf BZ2 } & $(2.751\ 10^{-2}){}^{+10.2\%}_{-8.7\%}{}^{+26.9\%}_{-26.9\%}$ &
     $(3.412\ 10^{-2}){}^{+3.2\%}_{-3.5\%}{}^{+25.1\%}_{-25.1\%}$\\
   & {\footnotesize \bf BW1 } & $(3.922\ 10^{-2}){}^{+11.2\%}_{-9.4\%}{}^{+2.9\%}_{-2.9\%}$ &
     $(4.873\ 10^{-2}){}^{+3.3\%}_{-3.8\%}{}^{+2.8\%}_{-2.8\%}$\\
   & {\footnotesize \bf BW2 } & $(1.564\ 10^{-2}){}^{+8.8\%}_{-7.8\%}{}^{+7.3\%}_{-7.3\%}$ &
     $(1.981\ 10^{-2}){}^{+3.2\%}_{-3.3\%}{}^{+6.9\%}_{-6.9\%}$\\
  \end{tabular}
  \renewcommand{\arraystretch}{1.0}
  \caption{\small \label{tab:BJxs}LO and NLO QCD inclusive cross sections for
   single $B$ production at the LHC, running at a center-of-mass energy of
   $\sqrt{s}=13$~TeV. The results are shown together with the associated scale
   and PDF relative uncertainties in the context of several benchmark scenarios.}
\end{table*}

Single vector-like quark
production mechanisms are of a pure electroweak nature. The associated
predictions are thus model-dependent as the sizes of the electroweak
vector-like quark couplings are free parameters of the model.
Comparing vector-like quark single and pair production, the latter gets an
enhancement originating from the presence of strong diagram contributions (first
line of Figure~\ref{fig:graph_vlqpair}) together with a phase-space
suppression for large vector-like quark mass values. In contrast, electroweak
single vector-like quark
production is less suppressed for large vector-like quark masses, which could
compensate the weakness of the involved interaction
vertices and make this channel the main LHC discovery mode for a heavy
vector-like quark. As a results, several ATLAS and CMS vector-like quark
searches also
target their single-production mode~\cite{Aad:2015voa,Aad:2016qpo,Aad:2016shx,%
ATLAS:2016ovj,CMS:2016ccy}.

\begin{figure*}
\centering
 \includegraphics[width=0.48\textwidth]{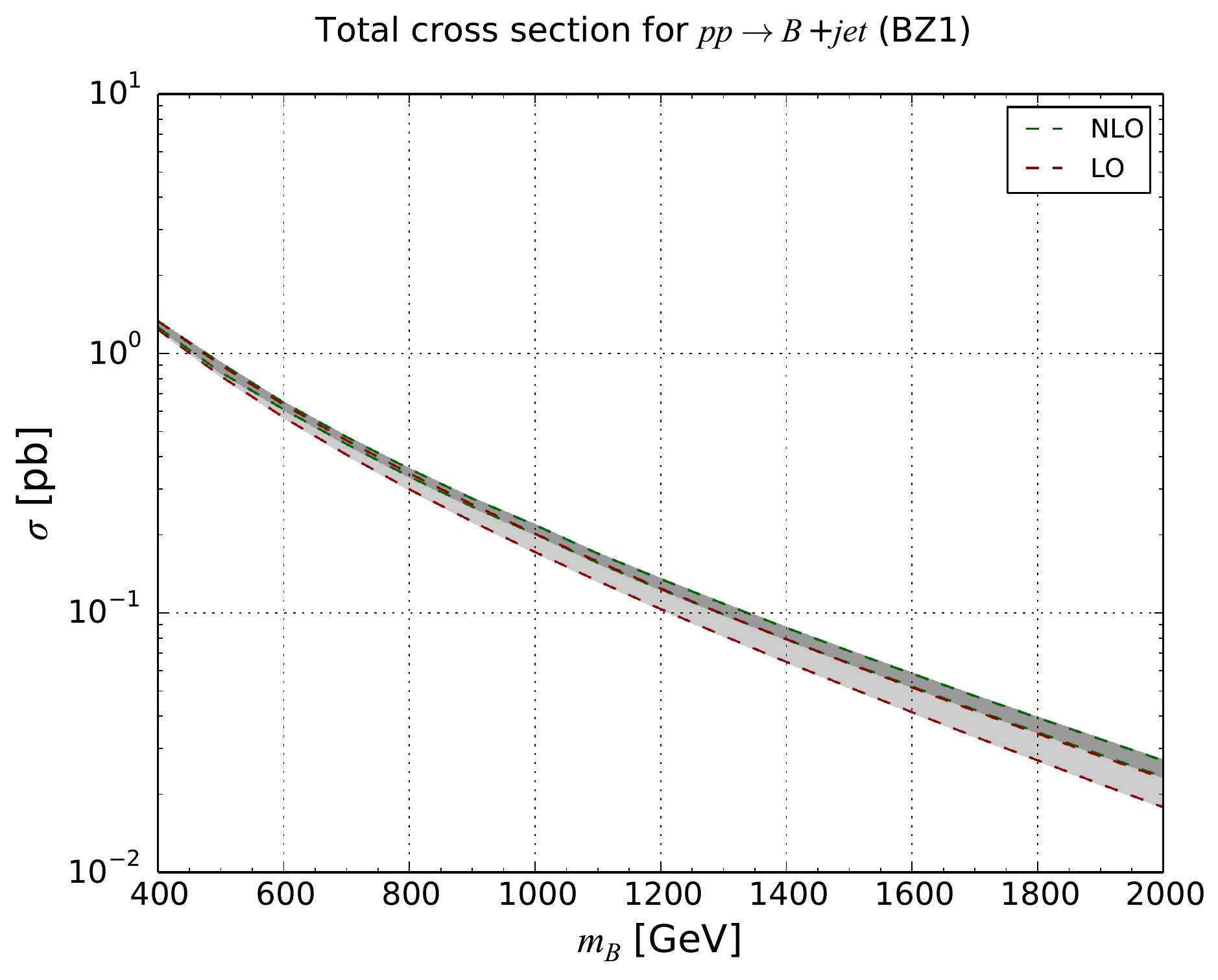}
 \includegraphics[width=0.48\textwidth]{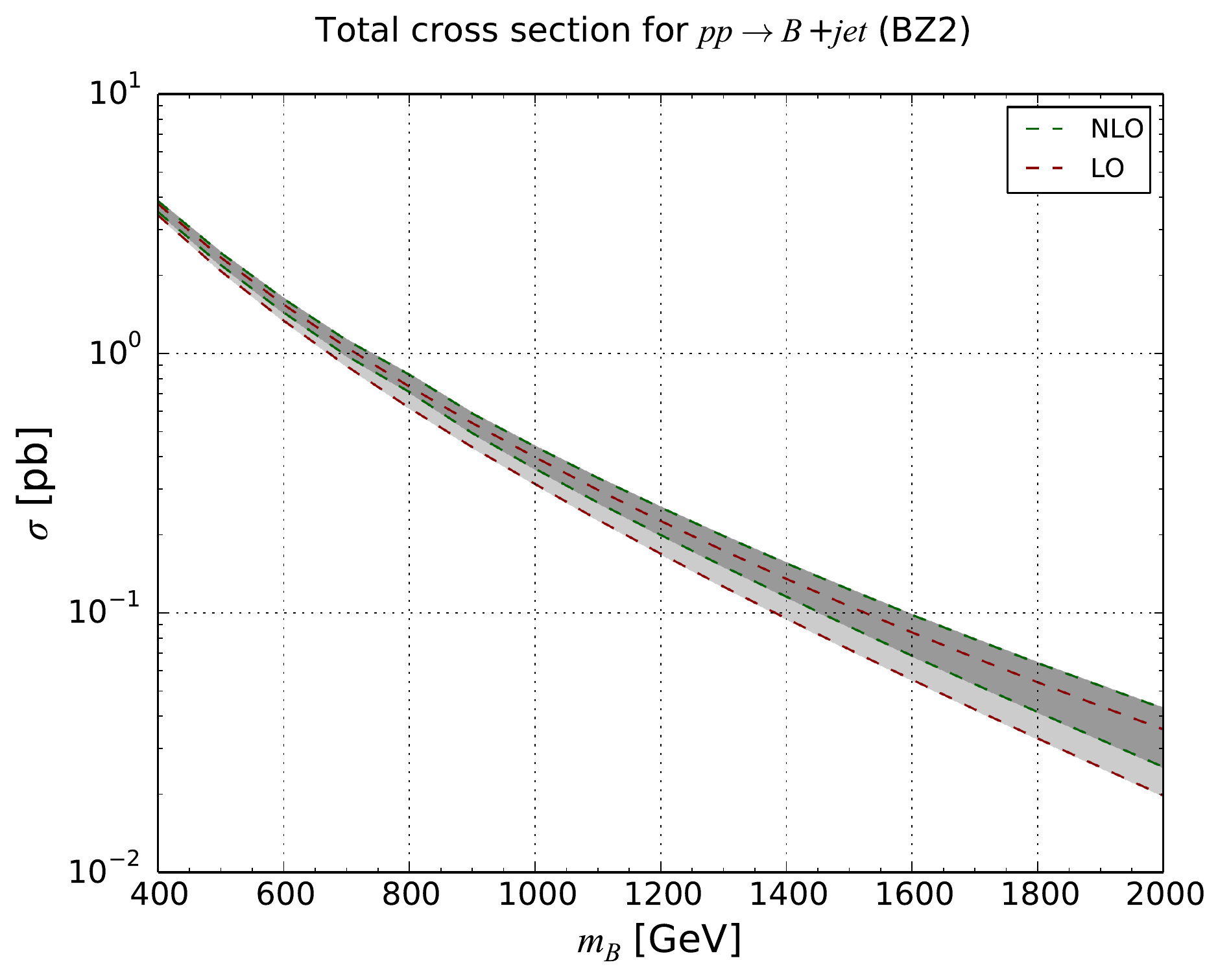}\\
 \includegraphics[width=0.48\textwidth]{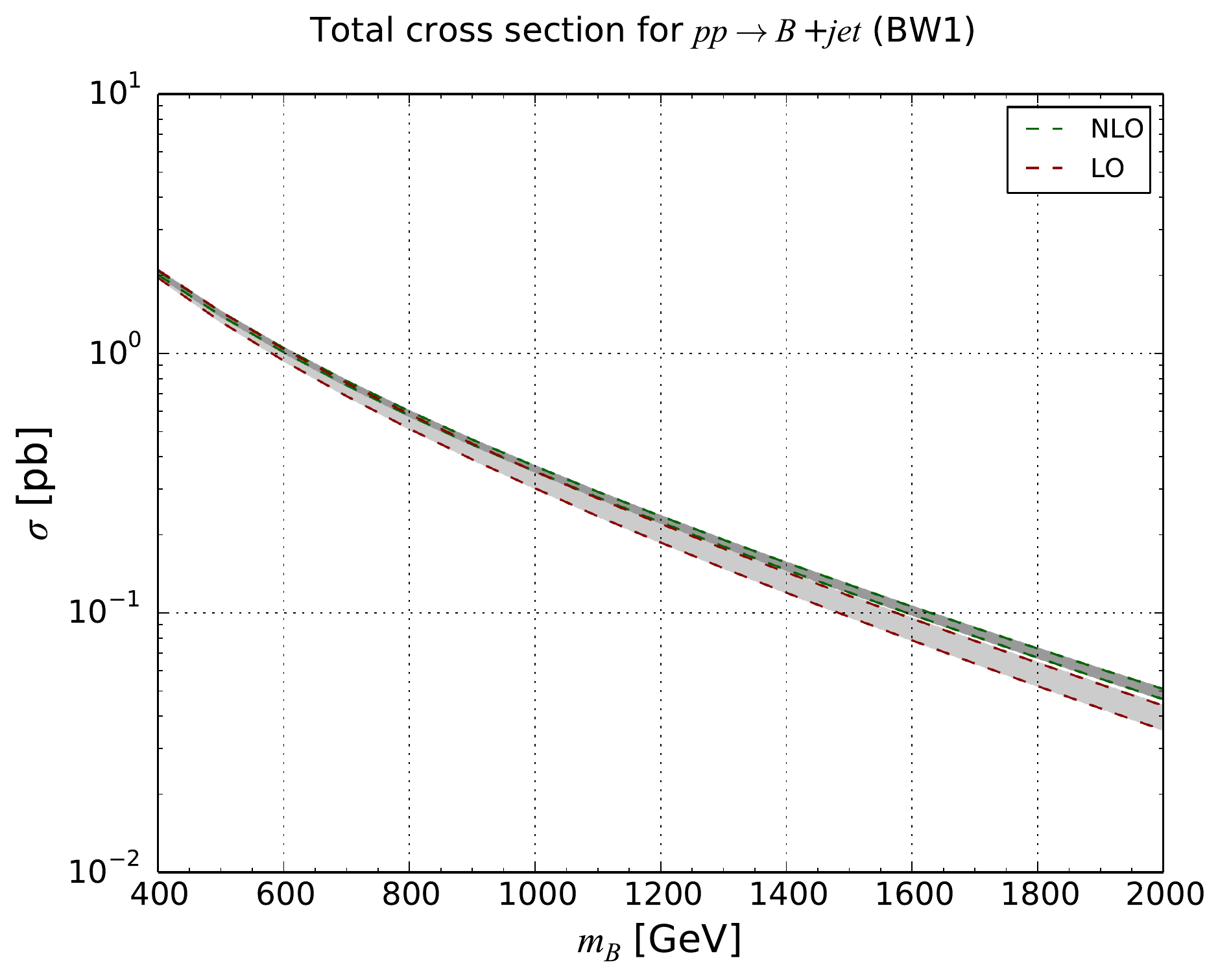}
 \includegraphics[width=0.48\textwidth]{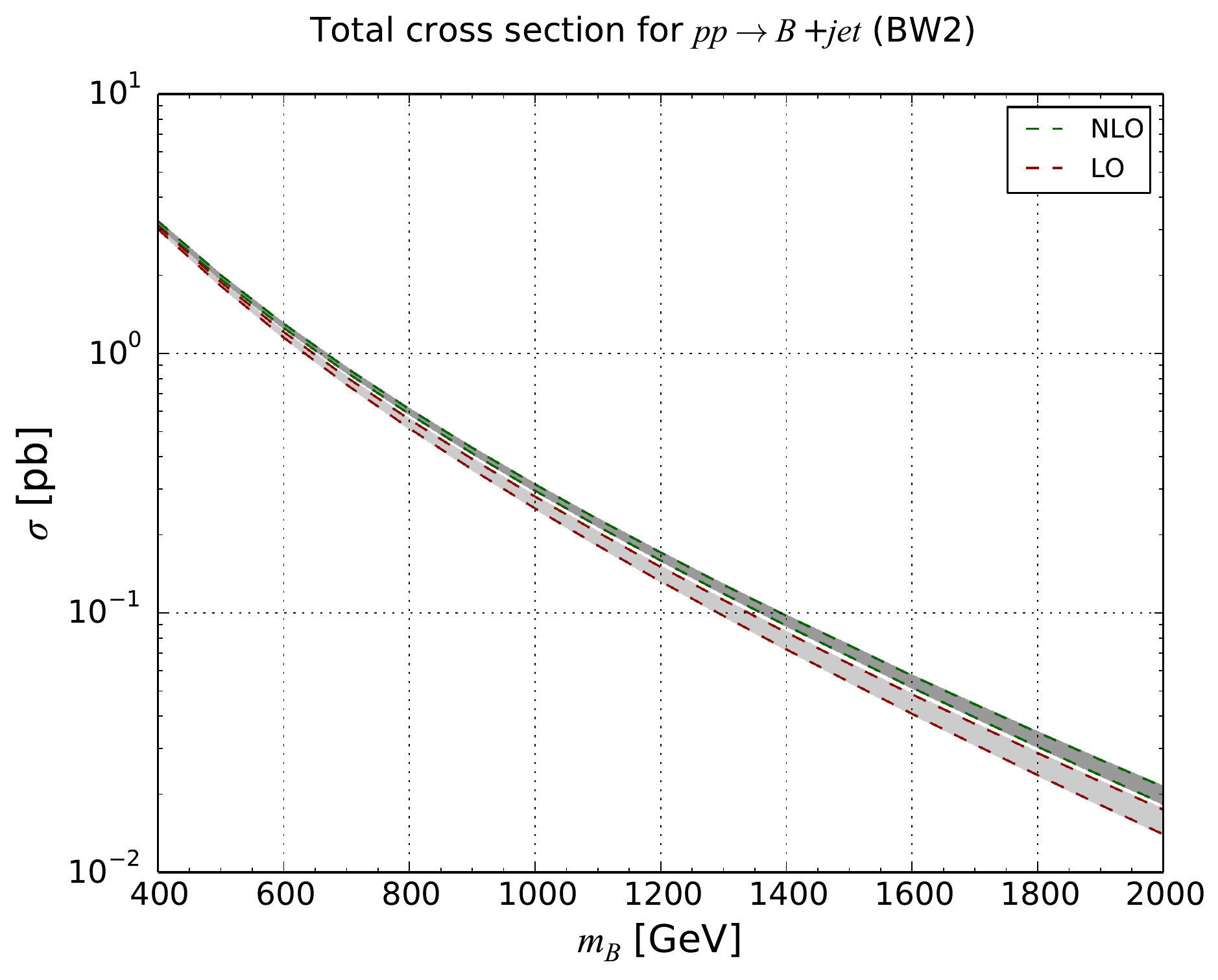}
 \caption{LO and NLO QCD inclusive cross sections for single $B$ quark
   production at the LHC with \mbox{$\sqrt{s}=13$~TeV}. The
   results are presented together with the associated theoretical
   uncertainty bands for the {\bf BZ1} (upper left), {\bf BZ2} (upper right),
   {\bf BW1} (lower left) and {\bf BW2} (lower right) scenarios.}
\label{fig:xsec_Bsingle}
\end{figure*}

In this section, we focus on vector-like quark single production in association
with jets,
\be
  p p \to Q j \ \ \text{or}\ \ \bar Q j \qquad\text{with}\qquad
    Q = T, B, X \ \ \text{or}\ \ Y.
\ee
Other single-production mechanisms exist, with, for instance, a final-state
gauge or a Higgs boson, but we refer to Section~\ref{sec:others} for the
latter. A representative set
of Feynman diagrams related to single vector-like quark production with jets is
shown on Figure~\ref{fig:graph_vlqsingle}, and
NLO cross-section calculation and event generation can be achieved
with \amc\ by typing in the program interpreter the command
\begin{verbatim}
  generate p p > tp j [QCD]
\end{verbatim}
for single $T$ production. Other processes with a different final-state
vector-like quark can be accounted for with a similar syntax, and LO event
generation only necessitates to remove the \verb+[QCD]+ tag. As for electroweak
contributions to vector-like quark pair production, mixed QCD and electroweak
loop diagrams appear at the NLO level and must be treated consistently for
getting ultraviolet-finite results. This step being not automated, we provide
scripts that steer the event generation process on the UFO model webpage.

In Figure~\ref{fig:xsec_Bsingle} and Table~\ref{tab:BJxs}, we present LO and NLO
total cross section for single $B$ quark production for the {\bf BZ1},
{\bf BZ2}, {\bf BW1} and {\bf BW2} scenarios
introduced in Section~\ref{sec:benchmarks} for which single vector-like $B$
quark production occurs via $Z$-boson or $W$-boson exchanges. This consists of
the first NLO predictions for a single vector-like quark production process.
Although all
depicted total cross sections exhibit a similar order of magnitude, a small
hierarchy is observed. It is driven by an interplay of the parton
densities that enhance mechanisms involving first generation quarks and of the
new physics coupling strengths that are much larger for vector-like quark
mixings with second generation ($\kappa_{\sss Q}=0.2$) than with
first generation quarks ($\kappa_{\sss Q}=0.07$).
For vector-like $B$-quark masses smaller than
500~GeV, single-production cross sections are slightly suppressed by a factor
of 2 or 3 with respect to the strong production of a pair of $B$ quarks
(see Appendix~\ref{app:vlqpair}), but feature a similar order of magnitude for
$m_{\sss B} \in [500, 800]$~GeV. In contrast, single-$B$ production always
dominates for heavier vector-like quarks by up to two orders of magnitude for
larger values of $m_{\sss B}$. For scenarios with smaller $\kappa_{\sss Q}$
values, the changes in the relative importance of the two production channels
can, however, be shifted towards higher masses.

\begin{figure*}
\centering
 \includegraphics[width=0.48\textwidth]{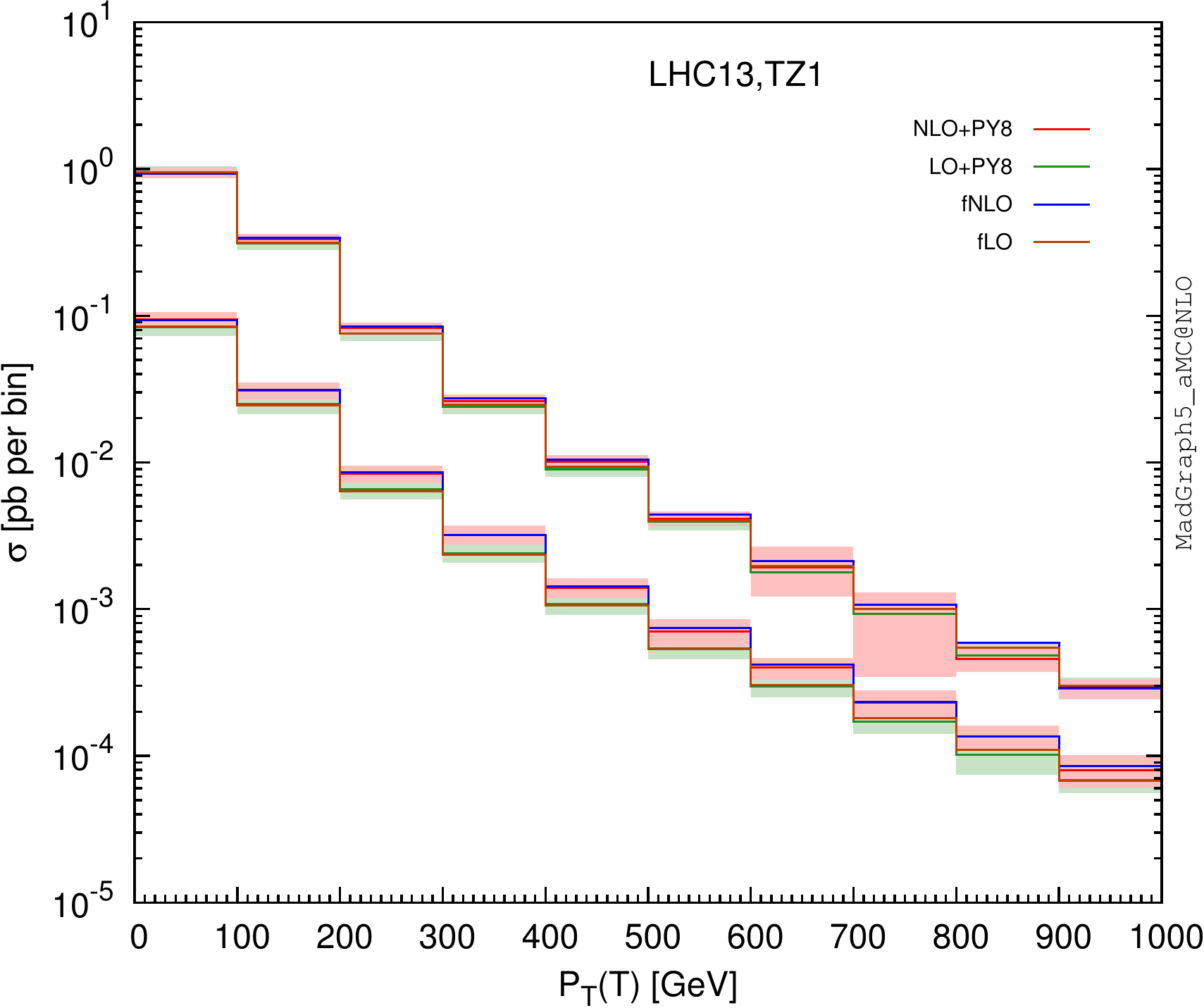}\\[.3cm]
 \includegraphics[width=0.48\textwidth]{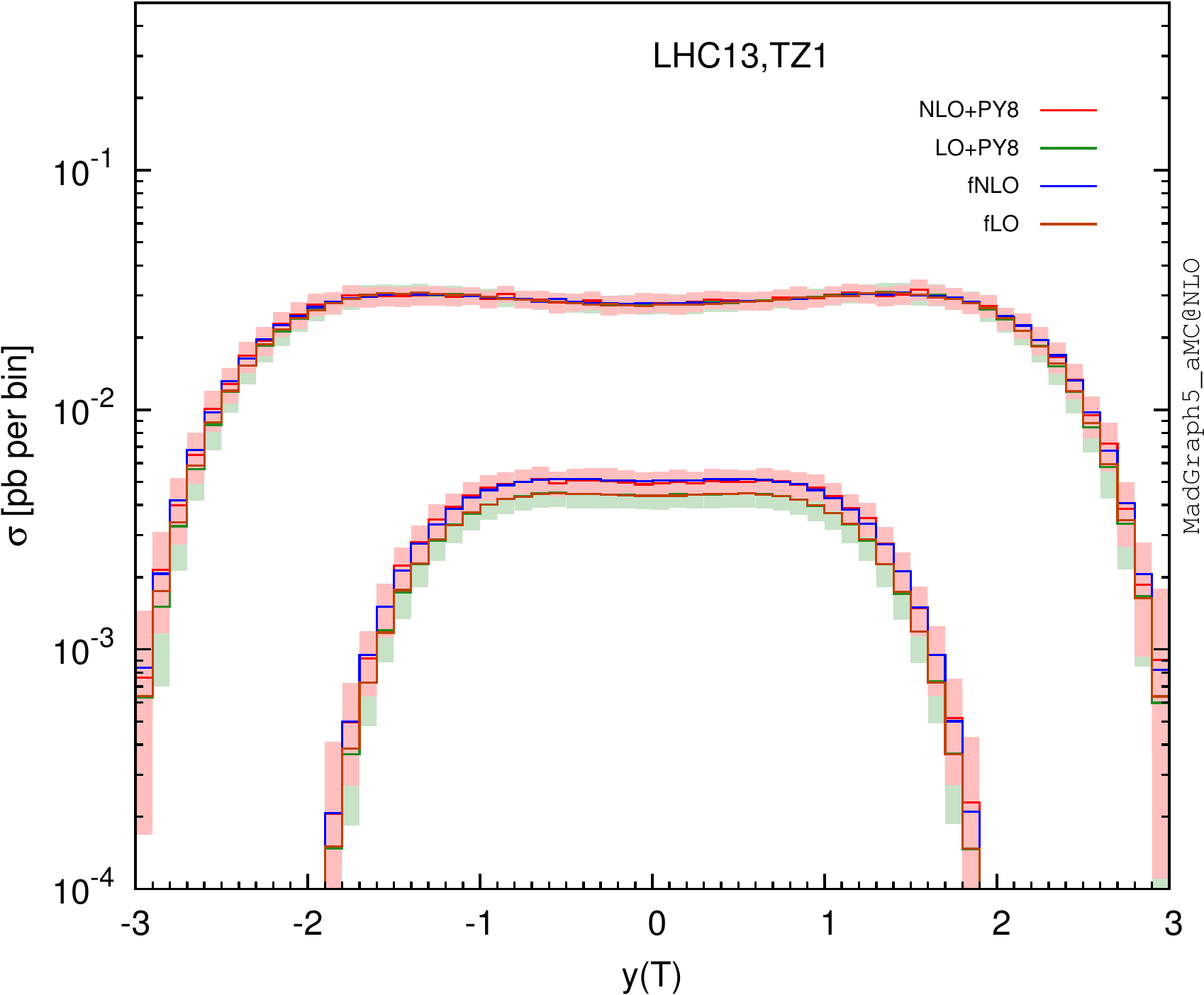}\quad
 \includegraphics[width=0.48\textwidth]{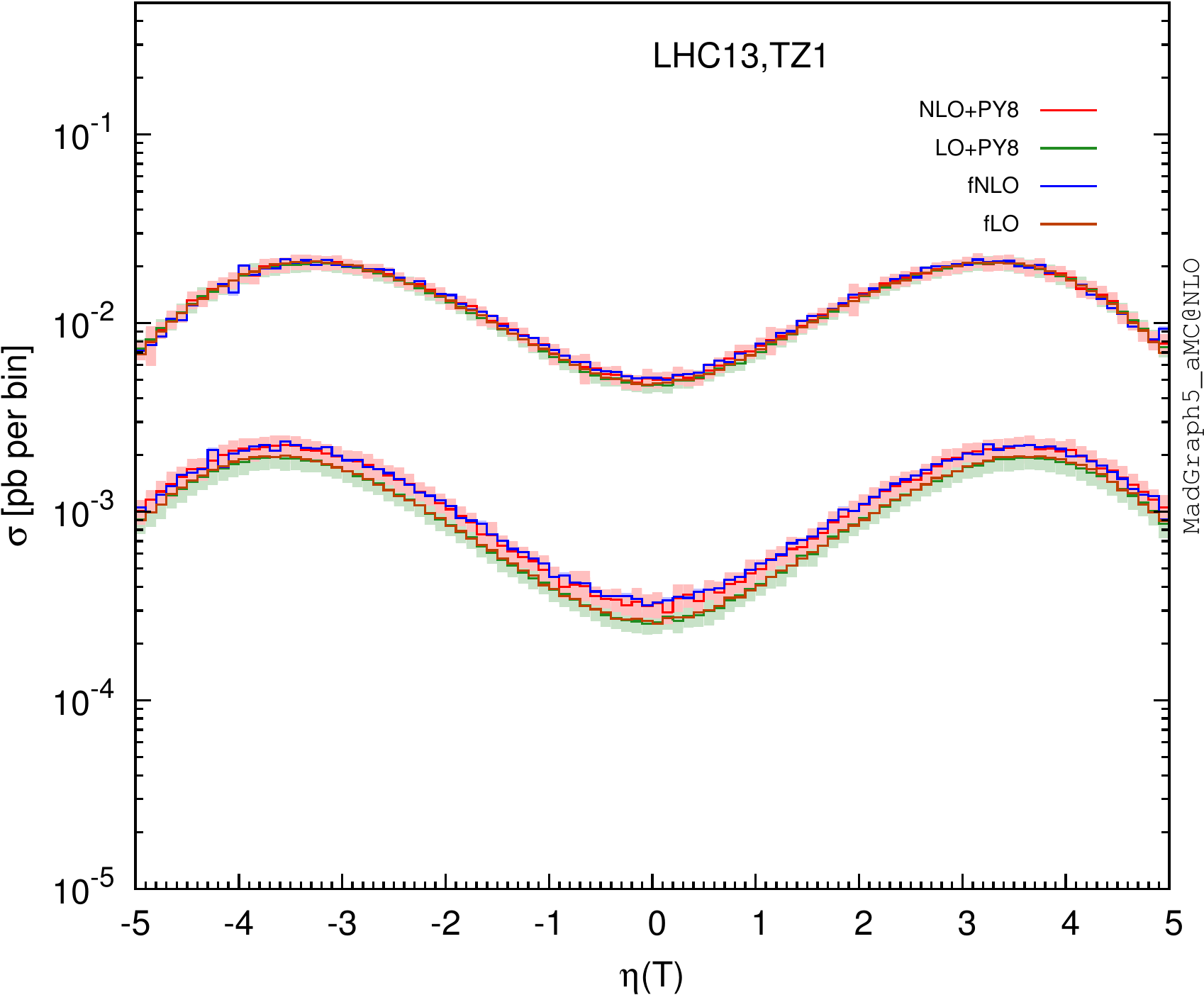}
 \caption{Differential distributions depicting the properties of a
  singly produced vector-like $T$ (or $\bar{T}$) quark. We present its
  transverse momentum (upper), rapidity (lower left) and pseudorapidity
  (lower right) spectrum and compare fixed-order predictions at the LO accuracy
  (purple curve) and NLO accuracy (blue curve), as well as prediction including
  the matching of these two calculations to parton showers (green and red
  curves for the LO and NLO cases, respectively). We have fixed the heavy quark
  mass either to 500~GeV (upper series of curves) or to 1500~GeV (lower series
  of curves).}
\label{fig:diff_singlevlq1}
\end{figure*}

In Figure~\ref{fig:diff_singlevlq1}, we turn to the study of differential
distributions related to inclusive single $T$-quark production at the LHC,
focusing on the {\bf TZ1} scenario as an illustrative benchmark point and for
a vector-like quark mass $m_{\sss T}$ fixed to 500~GeV or 1500~GeV. Event
generation and reconstruction is performed following the guidelines mentioned in
Section~\ref{sec:VLQpair}, and we present the transverse momentum $p_T(T)$,
rapidity $y(T)$ and pseudorapidity $\eta(T)$ spectrum of the vector-like
quark. We show predictions both at the fixed-order (purple and blue curves for
the LO and NLO accuracy, respectively) and after matching the results to parton
showers (green and red curves for the LO and NLO accuracy, respectively).
Theoretical uncertainties originating from scale variations and parton densities
are included for the matched predictions. In general, NLO effects only
moderately affect the shape of the different spectra but in contrast drastically
reduce the theoretical scale uncertainties and allow for a better prediction of
the spectrum normalisation. Similarly to the pair-production
case, matching to parton showers only mildly impacts fixed-order predictions
for the properties of the produced heavy quarks. Regardless of $m_{\sss T}$, the
transverse-momentum distribution of singly produced $T$ quarks exhibits a
typical steeply falling behaviour with increasing values of $p_T(T)$ (with a peak
at low $p_T$, which is invisible due to the binning choice), and the vector-like
quark is quite forward by virtue of the process topology, with $|\eta|>2$ in
average.

\begin{figure*}
\centering
  \includegraphics[width=0.48\textwidth]{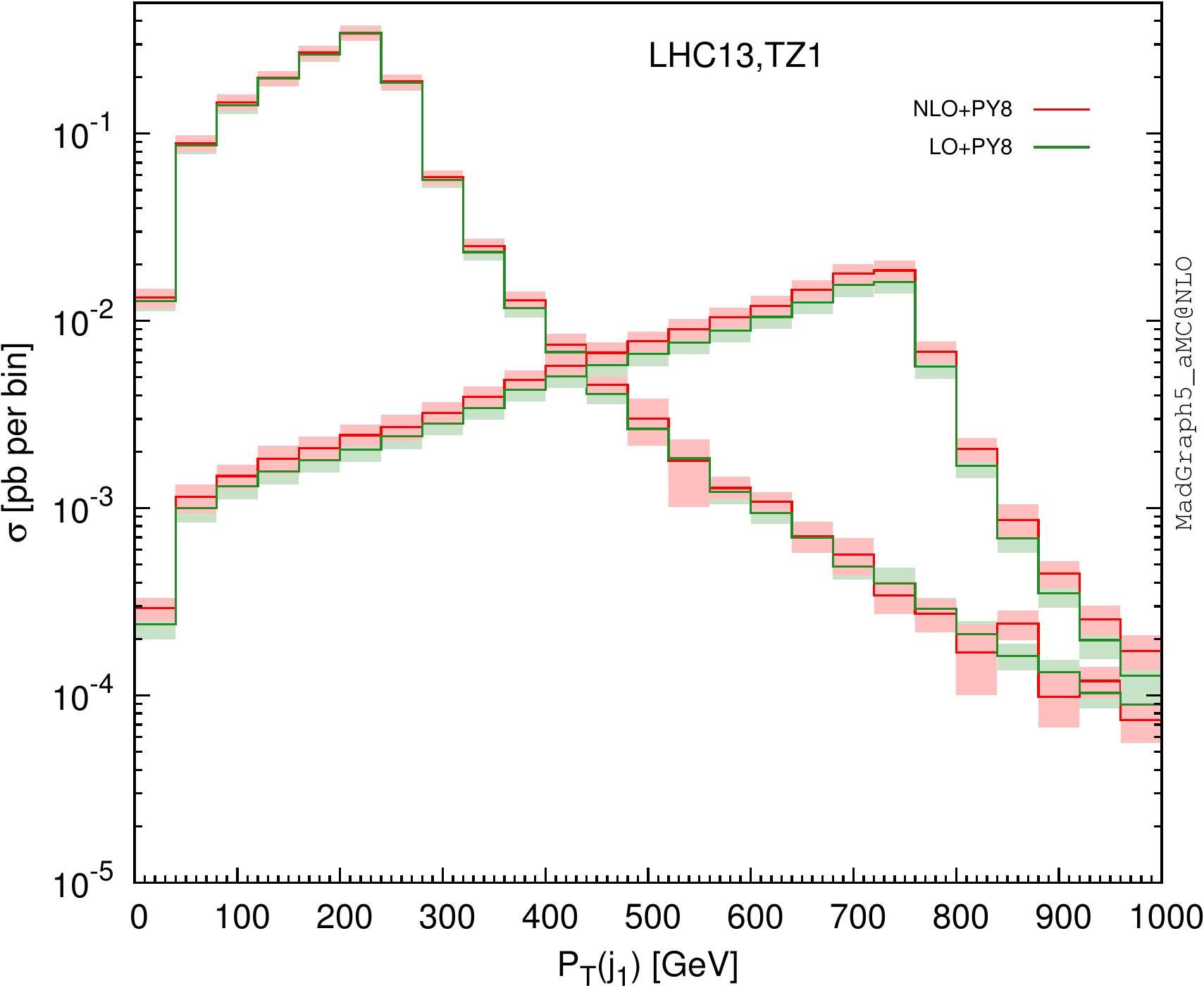}\quad
 \includegraphics[width=0.48\textwidth]{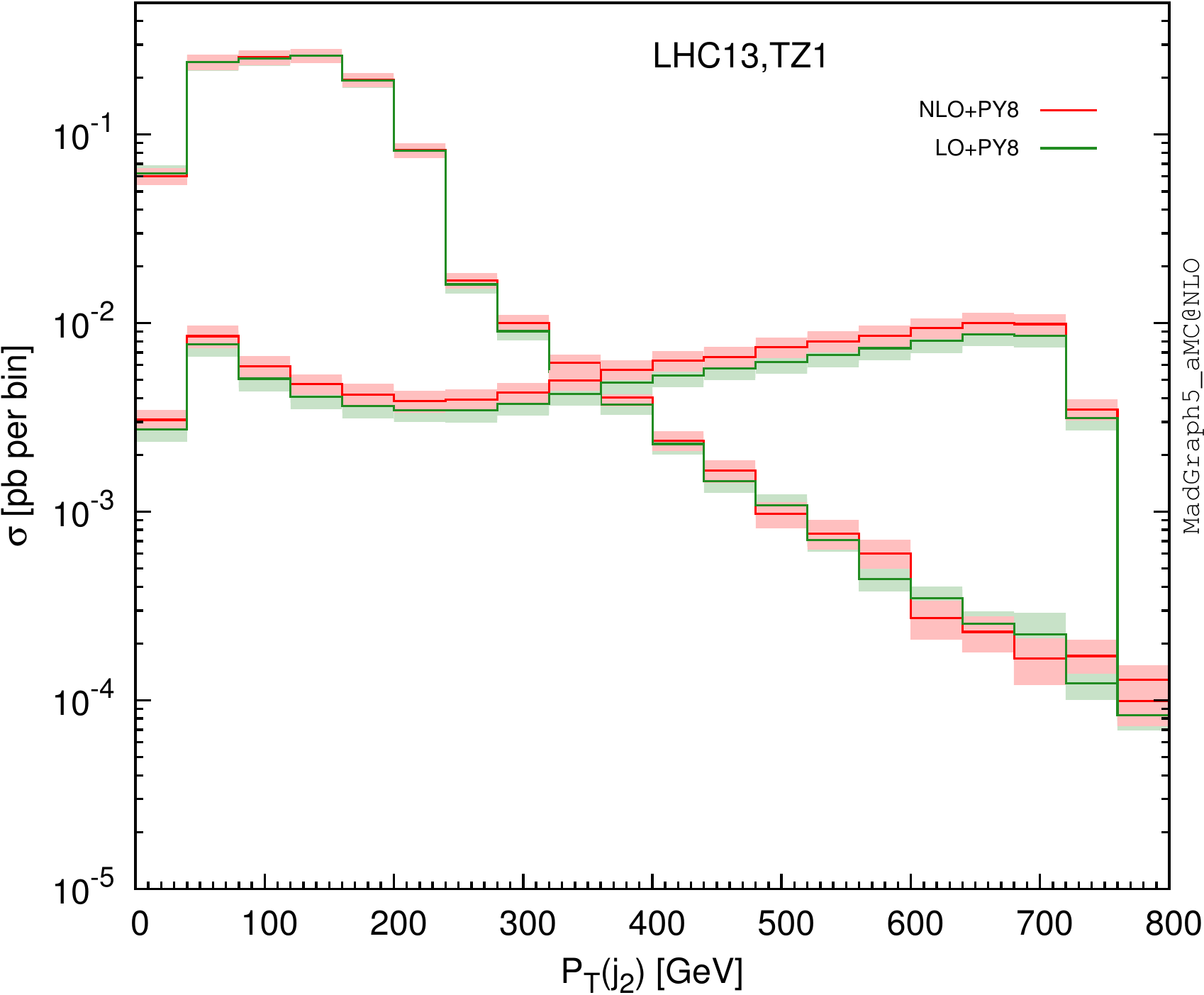}\\
 \includegraphics[width=0.48\textwidth]{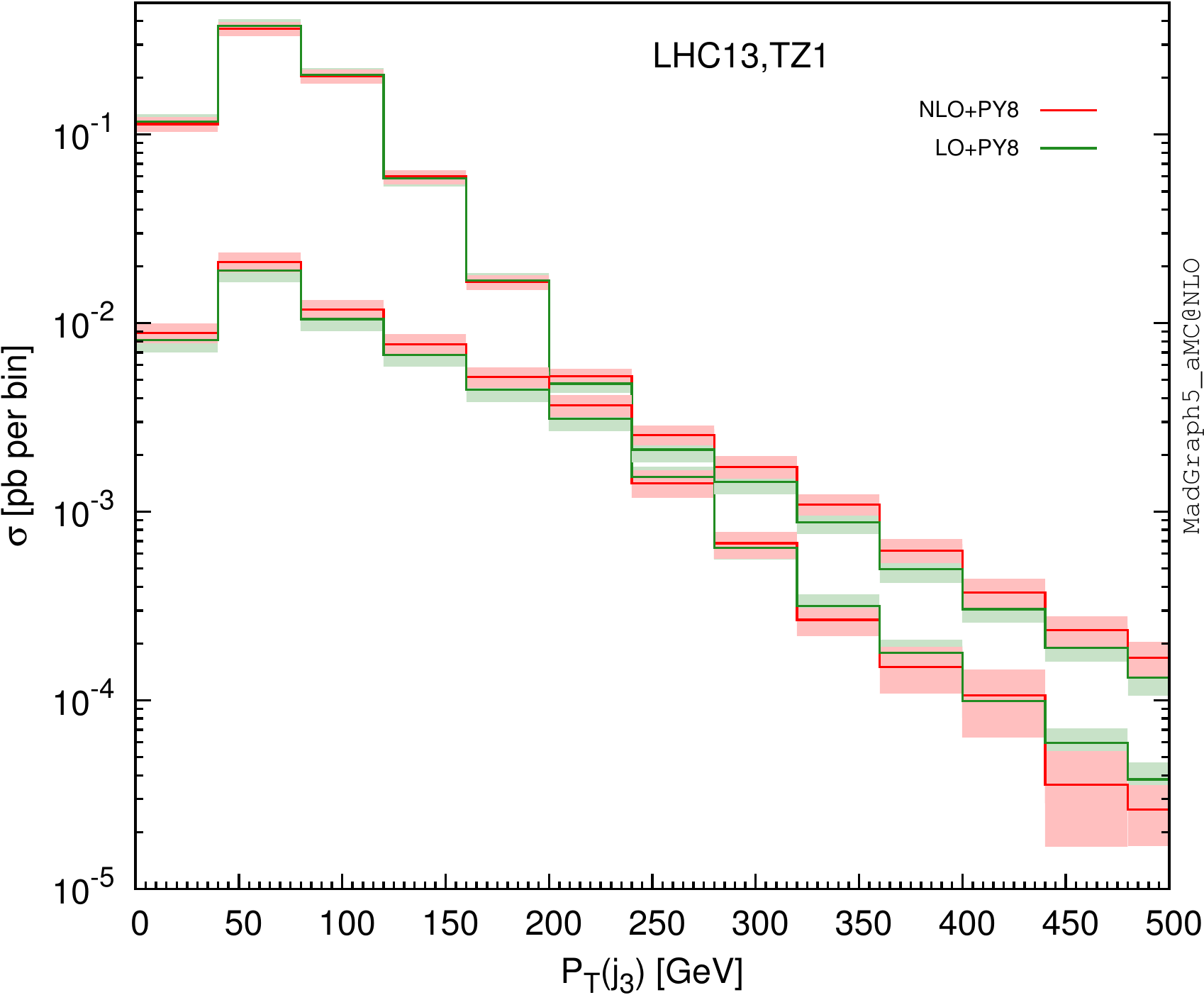}\quad
 \includegraphics[width=0.48\textwidth]{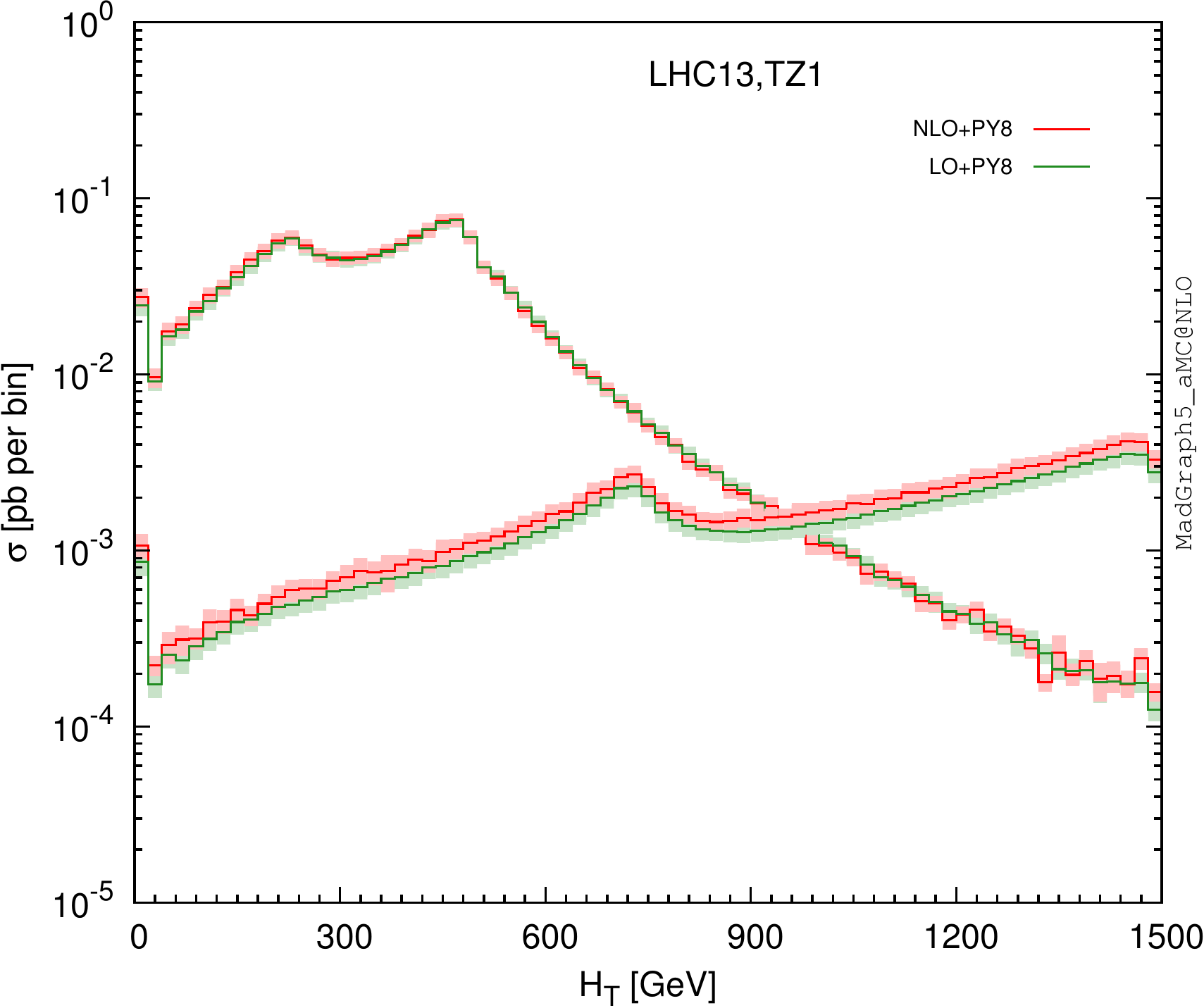}\\
 \caption{Differential distributions depicting the properties of the decay
  products of a singly produced vector-like $T$ (or $\bar{T}$) quark. We present
  the transverse-momentum distributions of the three leading jets, as well as
  the one in the $H_T$ variable defined as the scalar sum of all jet and
  isolated lepton transverse momenta. We compare LO (green) and NLO (red)
  predictions after matching the fixed-order calculations to parton showers.
  We have fixed the heavy quark mass either to 500~GeV or to 1500~GeV.}
\label{fig:diff_singlevlq2}
\end{figure*}

In Figure~\ref{fig:diff_singlevlq2}, we study the properties of the decay
products of the heavy quarks that consist of a $Z$-boson and an up quark in the
{\bf TZ1} scenario. We first present
the transverse-momentum spectrum of the three leading jets with a pseudorapidity
satisfying $|\eta|<2.5$ and a transverse-momentum larger than 30~GeV. Being
inclusive in the $Z$-boson decay, the jets can originate either from the heavy
$T$-quark decay, from the $Z$-boson decay or from initial- or final-state
radiation. Focusing on the leading jet $p_T$ distribution, we observe that it
peaks at half the $T$-quark mass, which shows that the leading jet is often
issued from the heavy quark decay. In contrast, the second jet
transverse-momentum distribution exhibit a plateau extending up to half the
$T$-mass, which we conclude that it could alternatively originate either
directly from the $T$-quark decay, or from the $Z$-boson decay. The third jet
finally shows a different behaviour, the distribution peaking this time at a
lower $p_T$ value, so that it is likely to be connected to the hard process.

In addition to the leading jet transverse-momentum distributions, we also
present the distribution in the $H_T$ variable defined as the scalar sum of the
trans\-ver\-se mo\-men\-tum of the final-state jets and isolated leptons, the latter
being only considered if their pseudorapidity fulfils $\eta|<2.4$ and their
transverse momentum $p_T>30$~GeV. Moreover, we require the leptons to be well
separated from any jet, imposing the angular distance $\Delta R$ to be larger
than 0.5. The $H_T$ variable exhibits a non-trivial structure which peaks
both at the $T$-quark mass $m_{\sss T}$ and at $m_{\sss T}/2$, the second peak
arising from cases where the $Z$-boson decays invisibly.

In all cases, the shapes turn out to be only slightly affected by the NLO
effects and the uncertainties are drastically reduced.

\subsection{Other processes impacted by the presence of vector-like quarks}
\label{sec:others}

\begin{figure*}
\centering
 \includegraphics[width=0.23\textwidth]{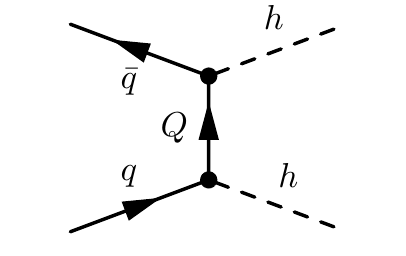}\
 \includegraphics[width=0.23\textwidth]{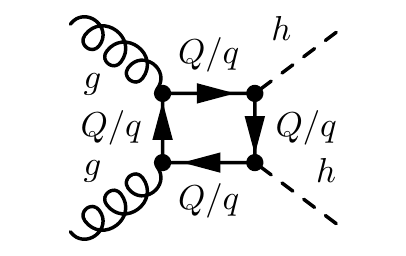}\qquad
 \includegraphics[width=0.23\textwidth]{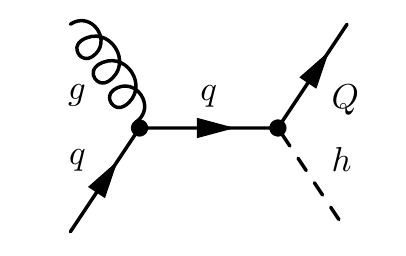}\
 \includegraphics[width=0.23\textwidth]{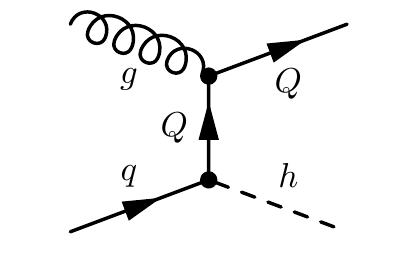}
 \caption{Representative Feynman diagram for Higgs-boson pair production that
  depend on vector-like quark exchange (two leftmost diagrams)
  and for single vector-like quark production in association with a Higgs boson
  (two rightmost diagrams). Similar diagrams exist when Higgs bosons are replaced by weak
  gauge bosons. With the second diagram, we include LO
  loop-induced contributions to Higgs-boson pair production that can only be
  calculated at the LO accuracy with \amc.}
\label{fig:graph_hh}
\end{figure*}

\begin{table*}
 \centering
 \renewcommand{\arraystretch}{1.40}
 \setlength{\tabcolsep}{12pt}
 \begin{tabular}{cc|cc}
   $m_{\sss T}$~[GeV] & Scenario & $\sigma_{\rm LO}$ [pb] & $\sigma_{\rm NLO}$ [pb]\\
   \hline
   $400$ & {\footnotesize \bf TH1} & $(1.254\ 10^{-4}){}^{+3.3\%}_{-3.3\%}{}^{+2.1\%}_{-2.1\%}$ &
     $(1.883\ 10^{-4}){}^{+4.8\%}_{-4.1\%}{}^{+1.8\%}_{-1.8\%}$\\   \hdashline
   $800$ & {\footnotesize \bf TH1} & $(1.043\ 10^{-4}){}^{+5.9\%}_{-5.3\%}{}^{+2.3\%}_{-2.3\%}$ &
     $(1.815\ 10^{-4}){}^{+7.0\%}_{-5.9\%}{}^{+1.8\%}_{-1.8\%}$\\   \hdashline
   $1200$ & {\footnotesize \bf TH1} & $(8.130\ 10^{-5}){}^{+7.3\%}_{-6.5\%}{}^{+2.7\%}_{-2.7\%}$ &
     $(1.714\ 10^{-4}){}^{+8.9\%}_{-7.5\%}{}^{+1.7\%}_{-1.7\%}$\\   \hdashline
   $1600$ & {\footnotesize \bf TH1} & $(6.092\ 10^{-5}){}^{+8.4\%}_{-7.3\%}{}^{+3.1\%}_{-3.1\%}$ &
     $(1.600\ 10^{-4}){}^{+10.8\%}_{-9.0\%}{}^{+1.6\%}_{-1.6\%}$\\   \hdashline
   $2000$ & {\footnotesize \bf TH1} & $(4.519\ 10^{-5}){}^{+9.1\%}_{-7.9\%}{}^{+3.6\%}_{-3.6\%}$ &
     $(1.513\ 10^{-4}){}^{+12.4\%}_{-10.2\%}{}^{+1.6\%}_{-1.6\%}$\\
\end{tabular}\\[.5cm]

 \begin{tabular}{cc|cc}
   $m_{\sss T}$~[GeV] & Scenario & $\sigma_{\rm LO}$ [pb] & $\sigma_{\rm NLO}$ [pb]\\
   \hline
   $400$ & {\footnotesize \bf TZ1} & $(1.017\ 10^{1}){}^{+4.7\%}_{-5.9\%}{}^{+1.5\%}_{-1.5\%}$ &
     $(1.314\ 10^{1}){}^{+3.3\%}_{-3.9\%}{}^{+1.5\%}_{-1.5\%}$\\   \hdashline
   $800$ & {\footnotesize \bf TZ1} & $(1.019\ 10^{1}){}^{+4.7\%}_{-5.9\%}{}^{+1.5\%}_{-1.5\%}$ &
     $(1.312\ 10^{1}){}^{+3.3\%}_{-3.8\%}{}^{+1.5\%}_{-1.5\%}$\\   \hdashline
   $1200$ & {\footnotesize \bf TZ1} & $(1.020\ 10^{1}){}^{+4.7\%}_{-5.8\%}{}^{+1.5\%}_{-1.5\%}$ &
     $(1.312\ 10^{1}){}^{+3.3\%}_{-3.9\%}{}^{+1.5\%}_{-1.5\%}$\\   \hdashline
   $1600$ & {\footnotesize \bf TZ1} & $(1.020\ 10^{1}){}^{+4.7\%}_{-5.9\%}{}^{+1.5\%}_{-1.5\%}$ &
     $(1.313\ 10^{1}){}^{+3.2\%}_{-3.8\%}{}^{+1.5\%}_{-1.5\%}$\\   \hdashline
   $2000$ & {\footnotesize \bf TZ1} & $(1.020\ 10^{1}){}^{+4.7\%}_{-5.9\%}{}^{+1.5\%}_{-1.5\%}$ &
     $(1.313\ 10^{1}){}^{+3.3\%}_{-3.8\%}{}^{+1.5\%}_{-1.5\%}$\\  \end{tabular}
  \renewcommand{\arraystretch}{1.0}
  \caption{\small \label{tab:HHxs}LO and NLO QCD inclusive cross sections for
   Higgs-boson (upper) and $Z$-boson (lower) pair production at the LHC, running
   at a center-of-mass energy of
   $\sqrt{s}=13$~TeV. The results are shown together with the associated scale
   and PDF relative uncertainties in the context of the {\bf TH1} (upper) and
   {\bf TZ1} (lower) class of benchmark scenarios.}
\end{table*}

Vector-like quarks can also affect Standard Model processes due
to additional Feynman diagram contributions featuring a virtual heavy quark. For
instance, the production of a pair of Higgs bosons could proceed via the
new physics diagrams shown in Figure~\ref{fig:graph_hh}, similar diagrams
existing for the diboson case. \amc\ can be used
for event generation at the NLO accuracy, once topologies featuring intermediate
heavy quark resonances are treated accordingly. This is achieved with the
command
\begin{verbatim}
  define res =  tp tp~ bp bp~ x x~ y y~ t t~
  generate p p > h h $$ tp tp~ bp bp~ [QCD]
  generate p p > v v $$ res [QCD]
\end{verbatim}
for di-Higgs and diboson production, respectively, using the \verb+$$+ symbol
to remove any possible intermediate resonance. In the last case, the pair of
symbols \texttt{v v} stands either for {\tt w+ w-} or {\tt z z} according to the
weak boson under consideration. LO event generation can finally be achieved by
removing the \verb+[QCD]+ tag.

Additional single vector-like quark production processes where the heavy quark
is produced in association with a Standard Model boson can be considered as
extra mechanisms useful for seeking for vector-like quarks (diagrams
shown in the rightmost part of Figure~\ref{fig:graph_hh}). Such processes can
be simulated with \amc, by typing in the commands
\begin{verbatim}
  generate    p p > tp h  $$ tp~ [QCD]
  add process p p > tp~ h $$ tp [QCD]
\end{verbatim}
for $T/\bar T$ production, as an example. Once again, the {\tt \$\$} symbol is
used in
order to avoid intermediate resonances. $VQ$ production can be undertaken
similarly.

\begin{figure*}
\centering
 \includegraphics[width=0.48\textwidth]{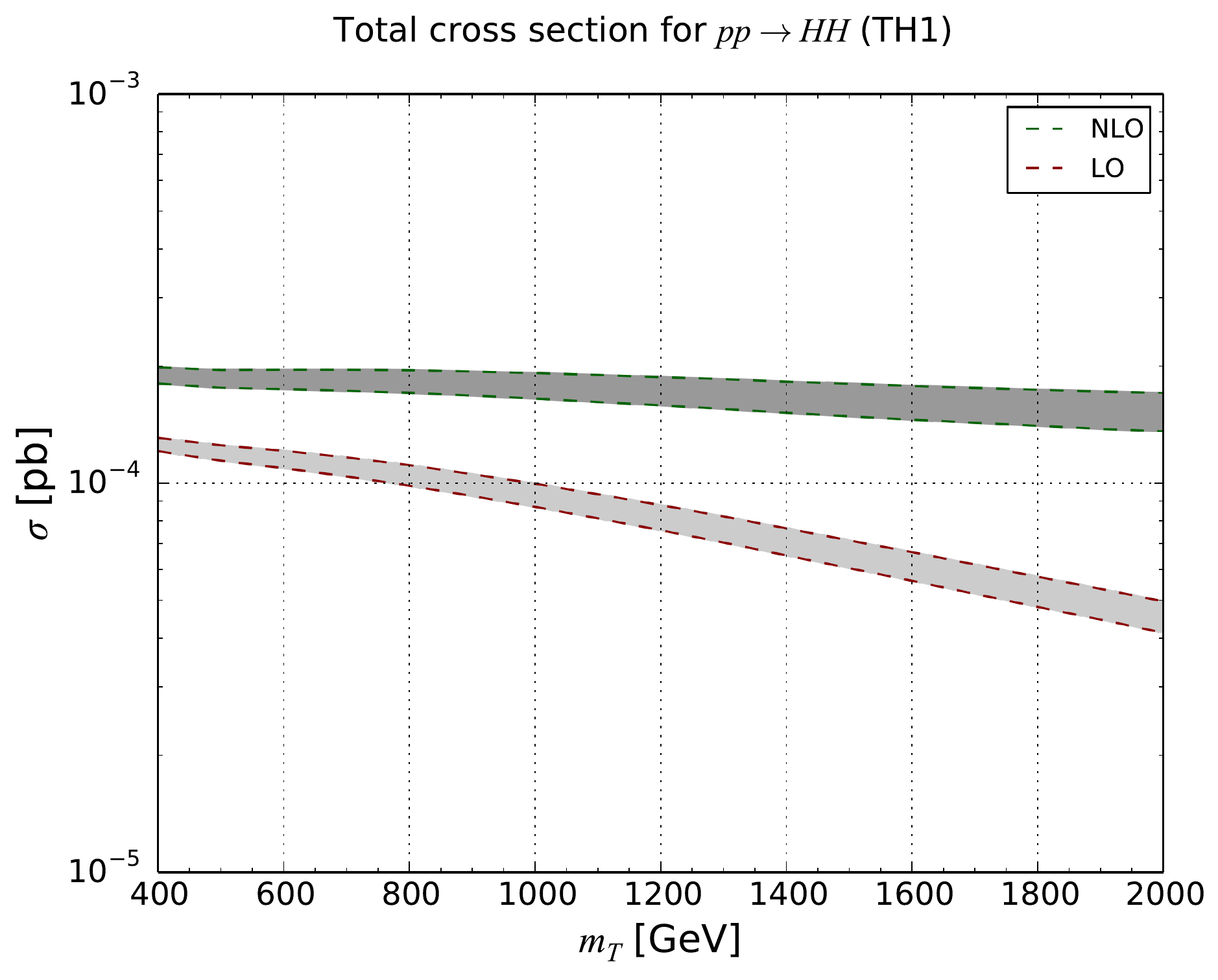}
 \includegraphics[width=0.48\textwidth]{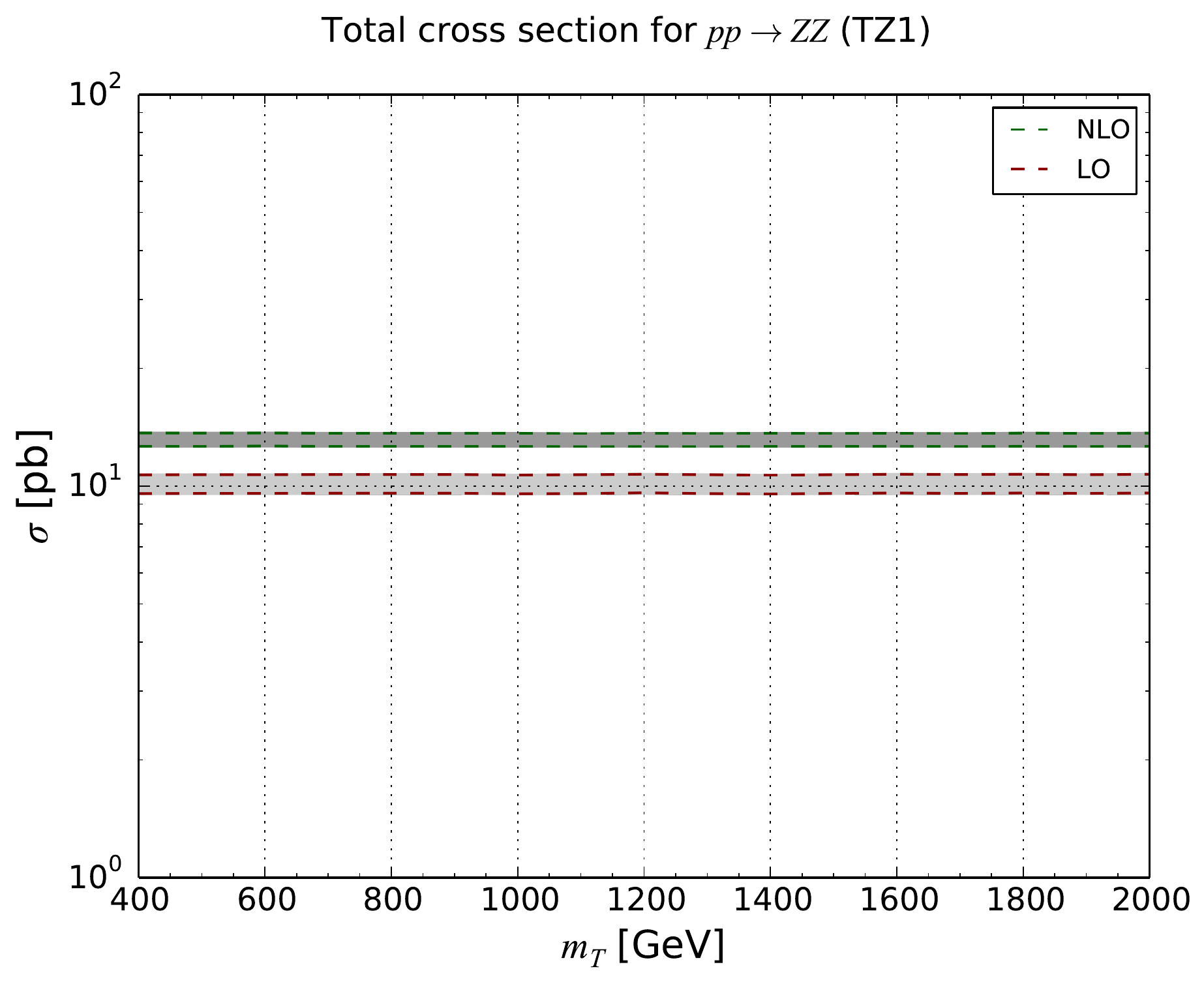}
 \caption{LO and NLO QCD inclusive cross sections for Higgs-boson (left) and
  $Z$-boson (right) pair production at the LHC with \mbox{$\sqrt{s}=13$~TeV}.
  The results are presented together with the associated theoretical
  uncertainty bands for the {\bf TH1} (left) and {\bf TZ1} (right) scenarios.}
\label{fig:sigma_bosons}
\end{figure*}

As an illustrative example, we show in Table~\ref{tab:HHxs} and
Figure~\ref{fig:sigma_bosons} total rates at LO
and NLO for Higgs-boson (upper) and Z-boson (lower) pair production,
respectively, for scenarios featuring a $T$ quark interacting with the first
generation of Standard Model quarks and either the Higgs boson or the $Z$-boson.
We observe gigantic $K$-factors in the case of di-Higgs production. This
enhancement is connected to an interplay of two effects. Turning back to the
observations made in Section~\ref{sec:VLQpair}, the vector-like quark coupling
to light quarks and a Higgs boson has strength that is proportional to the heavy
quark mass, and thus becomes stronger and stronger with increasing vector-like
quark masses. In addition, a new channel where the initial state is comprised of
a gluon and a quark opens up at NLO. This component of the NLO cross section
turns out
to dominate due to the large gluon density in the proton. As a result, the total
cross section for vector-like-quark-mediated di-Higgs production is more or less
constant with the heavy quark mass at the NLO QCD accuracy, which contrasts with
the LO case.

\section{Conclusions} \label{sec:conclusions}
We have modified a previously introduced model-in\-de\-pen\-dent parameterisation
suitable for the study of the vector-like quark phenomenology at the LHC so that
it is now suitable for NLO calculations in QCD matched to parton showers within
the \amc\ framework. We have illustrated its usage in the context of vector-like
quark pair production and vector-like quark single production in association
either with a jet or with a weak or Higgs boson.
For all showcased processes, we have considered QCD and electroweak diagram
contributions and investigated NLO and parton-shower effects on the
normalisation and shapes of the associated kinematical distributions.

We have found that NLO $K$-factors are important, both globally (at the
total-rate level) and at the differential distribution level and could hence
potentially impact limits currently extracted from vector-like quark search
results of the ATLAS and CMS collaborations. We have in particular noticed the
existence of potentially huge $K$-factors for new physics scenarios involving
the coupling of a heavy vector-like quark to first generation quarks and a Higgs
boson due to new production channels that open at the NLO accuracy. This
motivates further investigations, in particular to assess how an experimental
analysis including detector effects could benefit from the gain in cross section
stemming from the new topologies that dominate at NLO and to determine the
impact on the current vector-like quark limits and LHC discovery potential.

\section*{Acknowledgements}
The authors thank the organisers of the {\it PhysTeV Les Houches}
workshop where this work has been initiated and are grateful to
G.~Cacciapaglia, H.~Cai, A.~Carvalho, A.~Deandrea,
T.~Flacke, S.~Frixione, F.~Maltoni, D.~Majumder
and M.~Mangano for fruitful discussions. BF and HSS
acknowledge partial support by
the ERC grant 291377 \textit{LHCtheory: Theoretical predictions and analyses of
LHC physics: advancing the precision frontier}, the Research Executive Agency of
the European Union under Grant Agreement PITN-GA-2012-315877 (MCNet) and the
Theory-LHC-France initiative of the CNRS (INP/IN2P3).

\appendix

\section{Total cross sections for the production of a pair of $B$, $X$ or $Y$
quarks}
\label{app:vlqpair}

\begin{figure*}
\centering
 \includegraphics[width=0.48\textwidth]{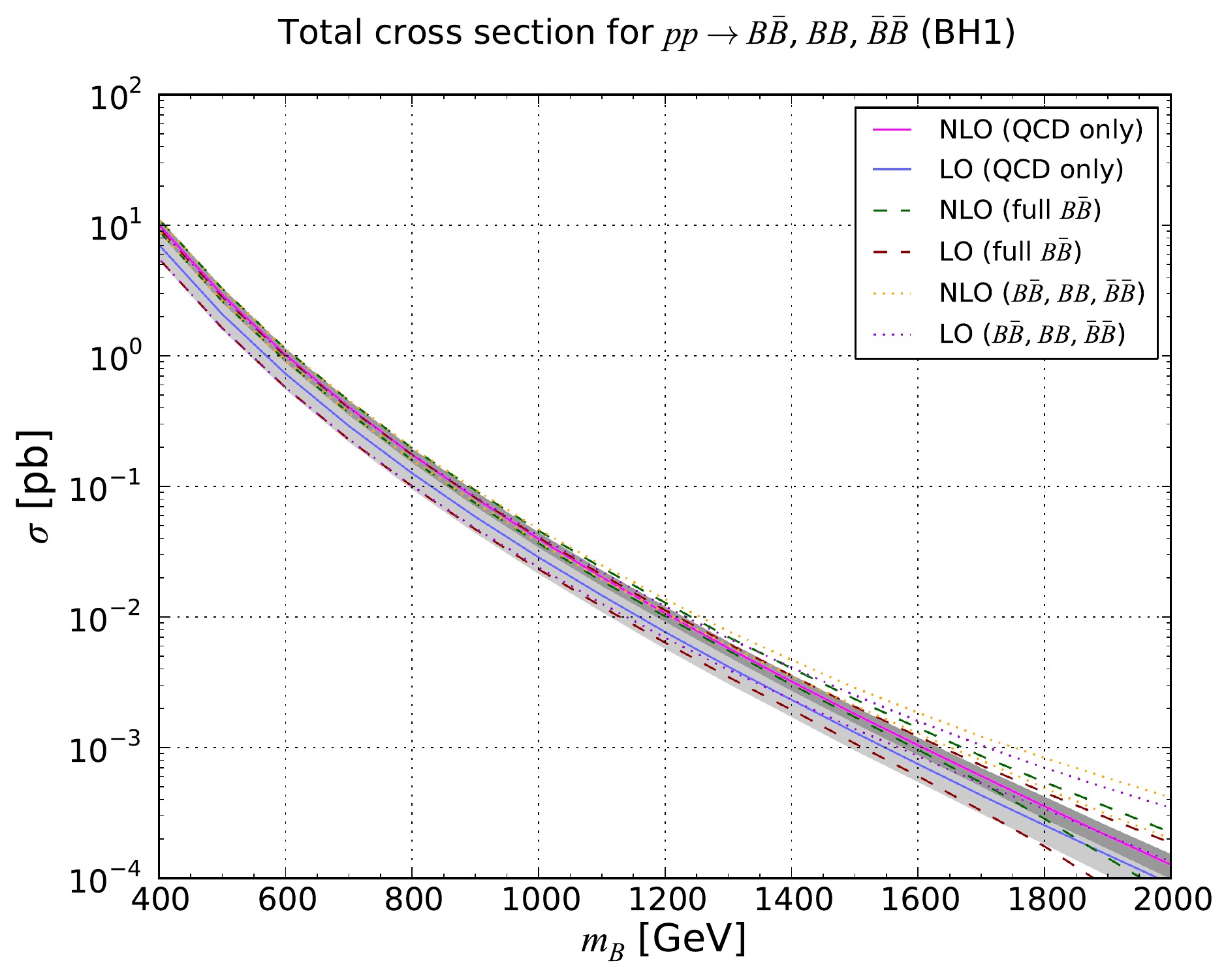}
 \includegraphics[width=0.48\textwidth]{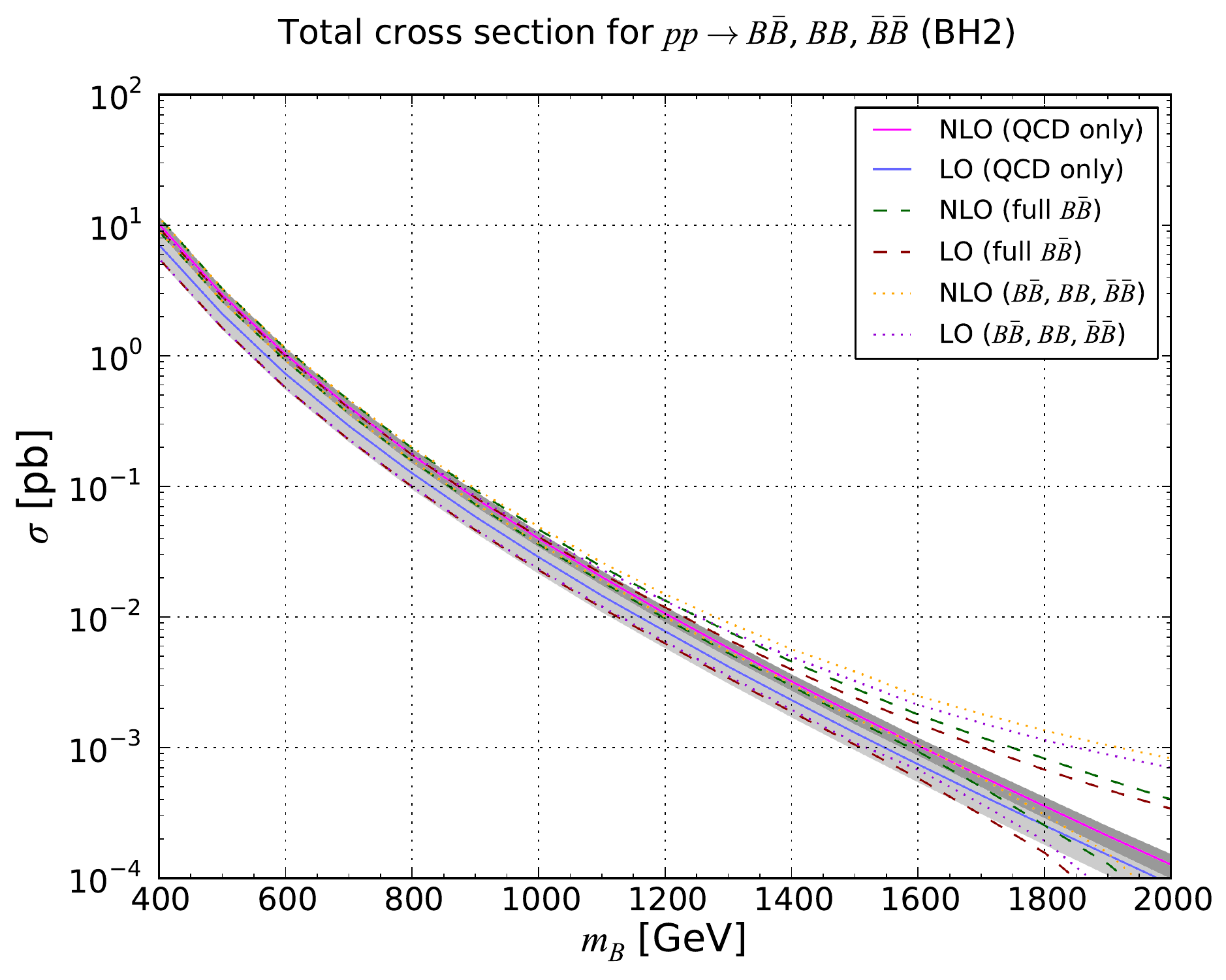}
 \caption{LO and NLO QCD inclusive cross sections for
   $B$ pair production at the LHC, for \mbox{$\sqrt{s}=13$~TeV}. The QCD
   contribution results are presented together with the associated theoretical
   uncertainty bands, and we indicate the shifts in the bands that are
   induced by weak diagram contributions in a scenario in which they are
   non-negligible.}
\label{fig:xsec_Bpair}
\end{figure*}

\begin{table*}
 \centering
 \renewcommand{\arraystretch}{1.40}
 \setlength{\tabcolsep}{12pt}
 \begin{tabular}{cc|cc}
   $m_{\sss B}$~[GeV] & Scenario & $\sigma_{\rm LO}$ [pb] & $\sigma_{\rm NLO}$ [pb]\\
   \hline
   \multirow{2}{*}{$400$} & {\footnotesize \bf BH1 } & $(7.025\ 10^{0}){}^{+31.0\%}_{-23.3\%}{}^{+1.5\%}_{-3.2\%}$ &
     $(9.995\ 10^{0}){}^{+7.6\%}_{-11.5\%}{}^{+1.6\%}_{-3.1\%}$\\
   & {\footnotesize \bf BH2 } & $(7.047\ 10^{0}){}^{+31.2\%}_{-23.1\%}{}^{+1.7\%}_{-3.0\%}$ &
     $(1.011\ 10^{1}){}^{+9.8\%}_{-12.3\%}{}^{+2.0\%}_{-3.1\%}$\\
   \hdashline
   \multirow{2}{*}{$800$} & {\footnotesize \bf BH1 } & $(1.241\ 10^{-1}){}^{+26.2\%}_{-27.5\%}{}^{+-1.8\%}_{-9.1\%}$ &
     $(1.715\ 10^{-1}){}^{+2.7\%}_{-15.5\%}{}^{+-1.1\%}_{-9.0\%}$\\
   & {\footnotesize \bf BH2 } & $(1.282\ 10^{-1}){}^{+29.2\%}_{-25.0\%}{}^{+1.8\%}_{-7.2\%}$ &
     $(1.749\ 10^{-1}){}^{+5.6\%}_{-13.1\%}{}^{+2.0\%}_{-7.1\%}$\\
   \hdashline
   \multirow{2}{*}{$1200$} & {\footnotesize \bf BH1 } & $(7.903\ 10^{-3}){}^{+16.2\%}_{-32.4\%}{}^{+-7.9\%}_{-16.9\%}$ &
     $(1.088\ 10^{-2}){}^{+-4.8\%}_{-21.5\%}{}^{+-7.0\%}_{-16.9\%}$\\
   & {\footnotesize \bf BH2 } & $(8.883\ 10^{-3}){}^{+23.0\%}_{-26.1\%}{}^{+6.7\%}_{-18.3\%}$ &
     $(1.189\ 10^{-2}){}^{+2.1\%}_{-15.4\%}{}^{+5.6\%}_{-16.7\%}$\\
   \hdashline
   \multirow{2}{*}{$1600$} & {\footnotesize \bf BH1 } & $(1.016\ 10^{-3}){}^{+10.5\%}_{-31.3\%}{}^{+-6.8\%}_{-21.1\%}$ &
     $(1.356\ 10^{-3}){}^{+-7.5\%}_{-22.3\%}{}^{+-7.6\%}_{-20.5\%}$\\
   & {\footnotesize \bf BH2 } & $(1.255\ 10^{-3}){}^{+14.8\%}_{-24.3\%}{}^{+25.7\%}_{-41.5\%}$ &
     $(1.643\ 10^{-3}){}^{+-1.7\%}_{-15.7\%}{}^{+22.3\%}_{-38.1\%}$\\
   \hdashline
   \multirow{2}{*}{$2000$}  & {\footnotesize \bf BH1 } & $(2.138\ 10^{-4}){}^{+13.1\%}_{-24.3\%}{}^{+7.3\%}_{-24.0\%}$ &
     $(2.817\ 10^{-4}){}^{+-3.1\%}_{-16.0\%}{}^{+5.3\%}_{-22.7\%}$\\
   & {\footnotesize \bf BH2 } & $(3.262\ 10^{-4}){}^{+11.0\%}_{-19.4\%}{}^{+78.6\%}_{-91.1\%}$ &
     $(4.133\ 10^{-4}){}^{+-2.1\%}_{-12.2\%}{}^{+69.7\%}_{-82.8\%}$\\
  \end{tabular}
  \renewcommand{\arraystretch}{1.0}
  \caption{\small \label{tab:Bxs}LO and NLO QCD inclusive cross sections for
   $B$ pair production at the LHC, running at a center-of-mass energy of
   $\sqrt{s}=13$~TeV. The results are shown together with the associated scale
   and PDF relative uncertainties in the context of several benchmark scenarios.
   For all non-indicated scenarios, the results match the pure QCD one given
   in Table~\ref{tab:Txs}.}
\end{table*}

In Figure~\ref{fig:xsec_Bpair} and Table~\ref{tab:Bxs}, we present LO and NLO
total cross sections related to the production of a pair of vector-like $B$
quarks for the different scenarios introduced in Section~\ref{sec:benchmarks}.
We depict the dependence of the cross sections on the vector-like quark
mass and recall that the pure QCD contributions can be found in
Table~\ref{fig:xsec_Tpair}
as they are independent of the vector-like quark nature. Results for $X$ and $Y$
pair production are not shown as non-QCD contributions are negligible for all
considered scenarios.

\section{Total cross sections for the single production of a $T$, $X$ or $Y$
quark}
\label{app:vlqsingle}
\begin{figure*}
\centering
 \includegraphics[width=0.40\textwidth]{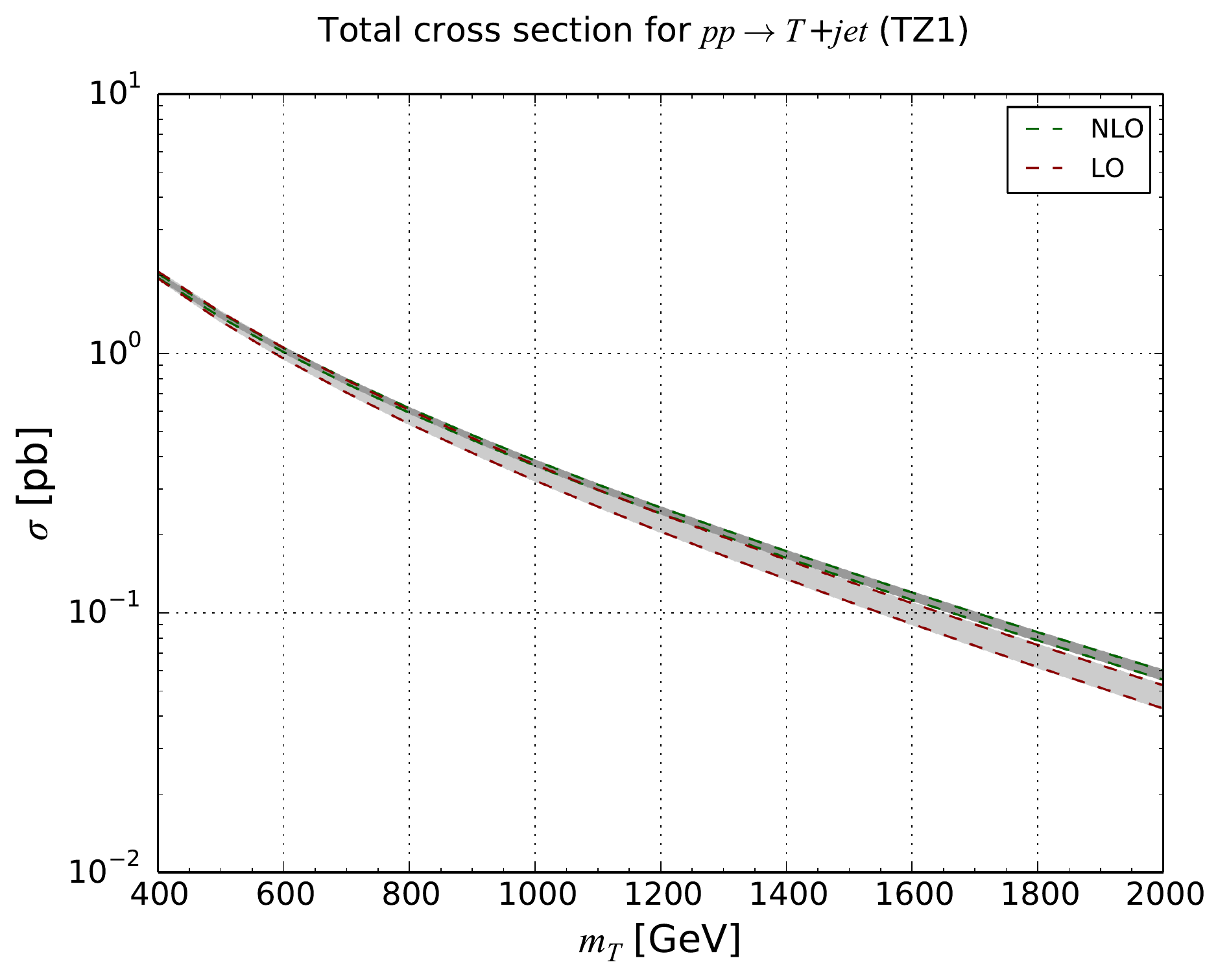}
 \includegraphics[width=0.40\textwidth]{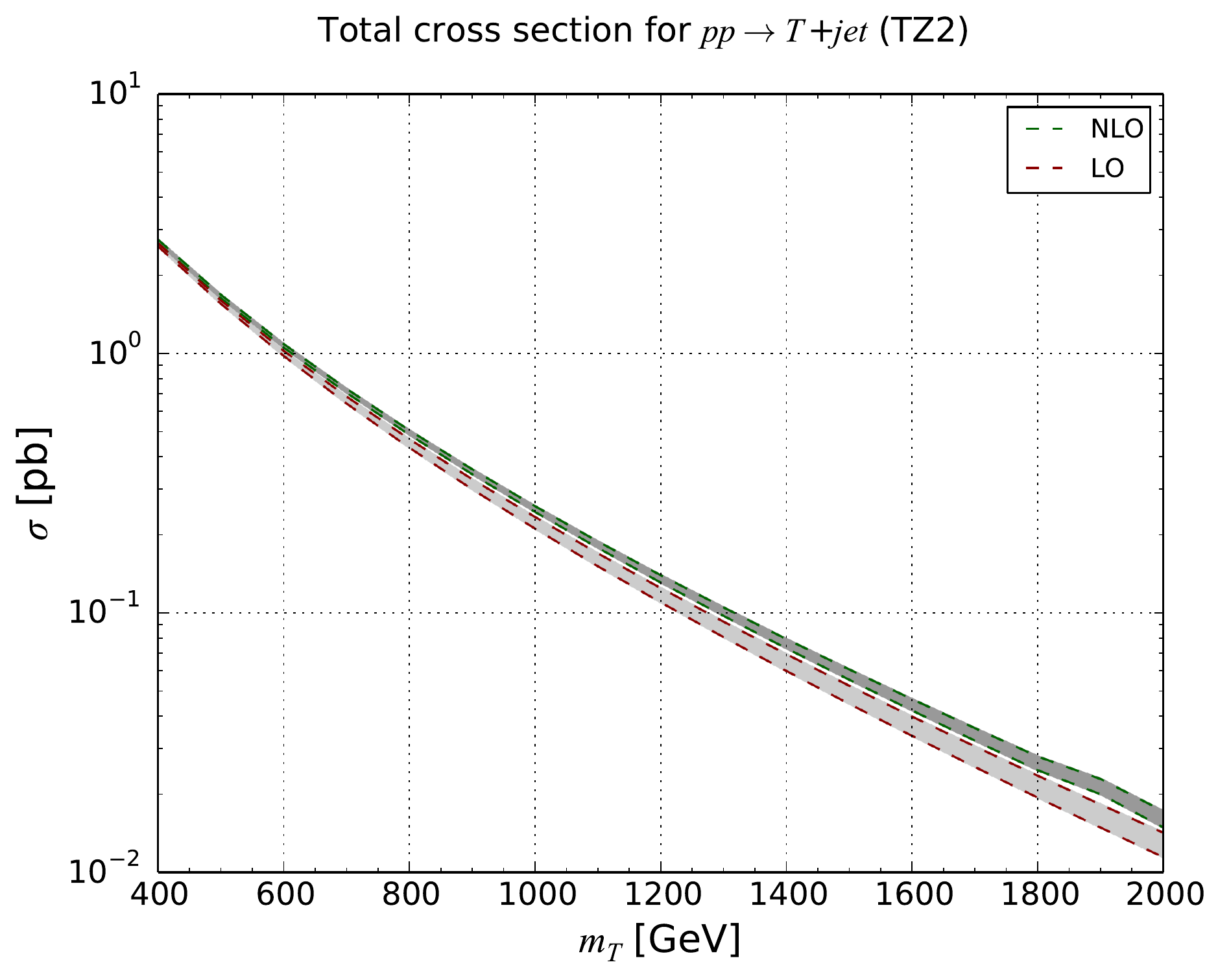}\\
 \includegraphics[width=0.40\textwidth]{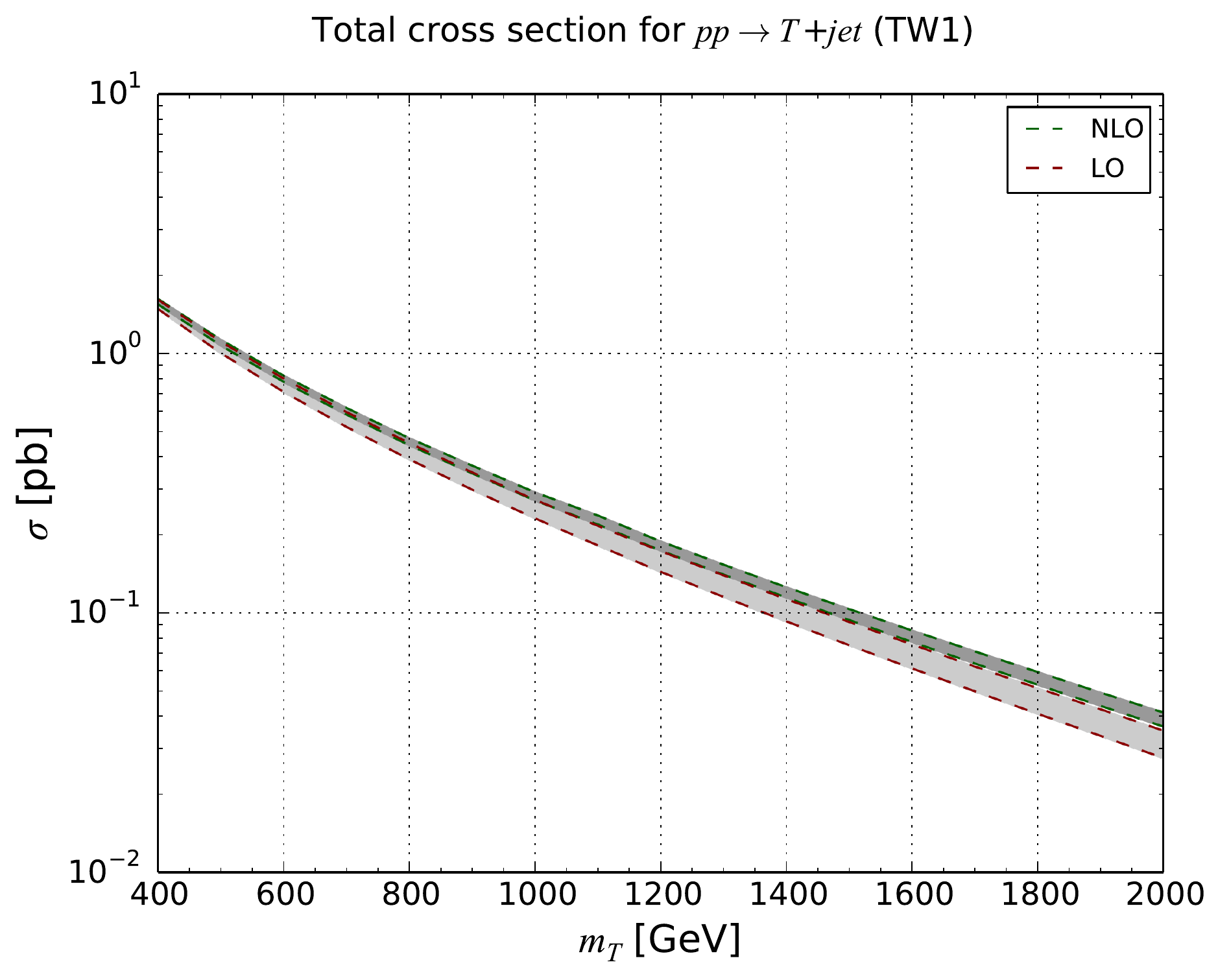}
 \includegraphics[width=0.40\textwidth]{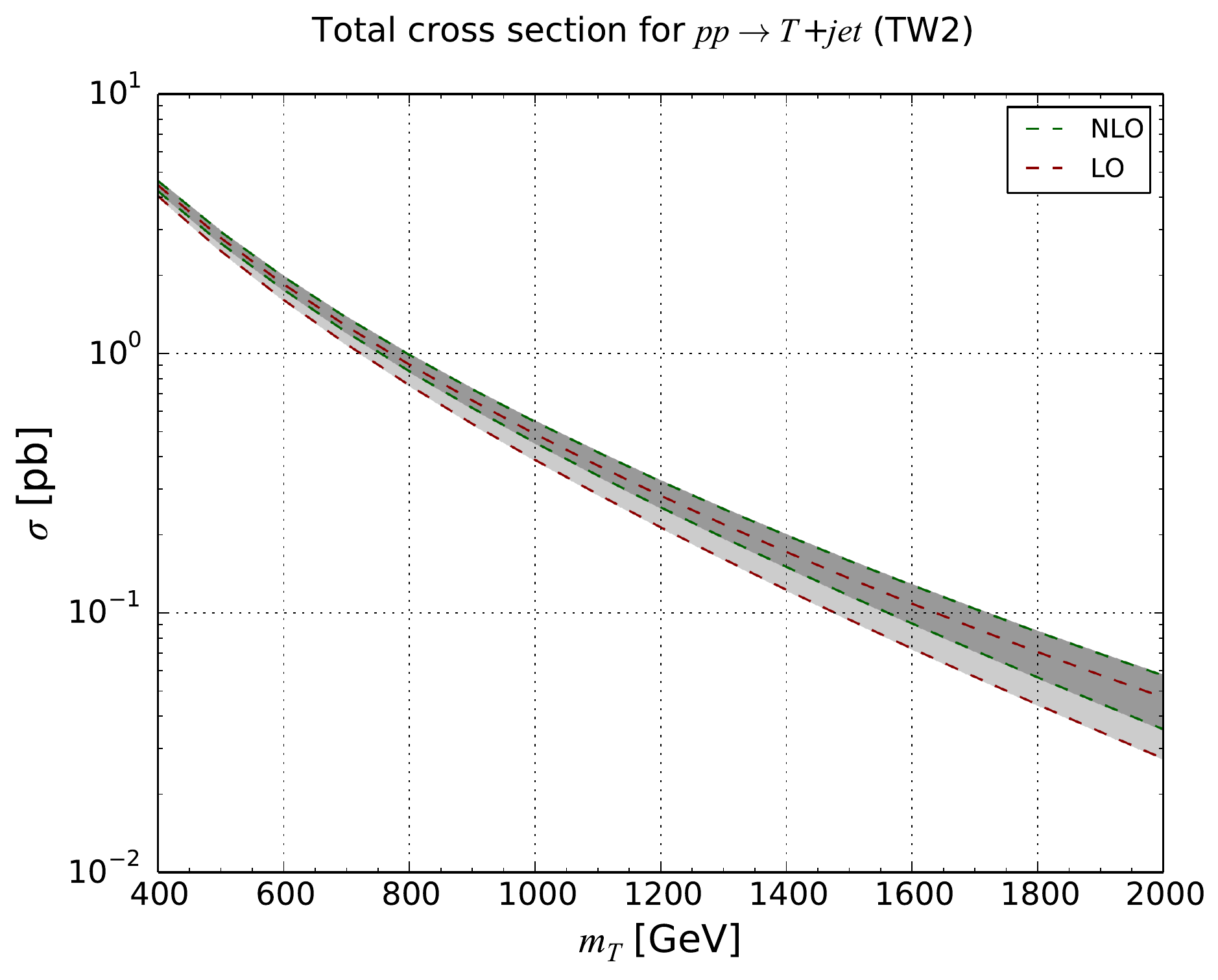}\\
 \includegraphics[width=0.40\textwidth]{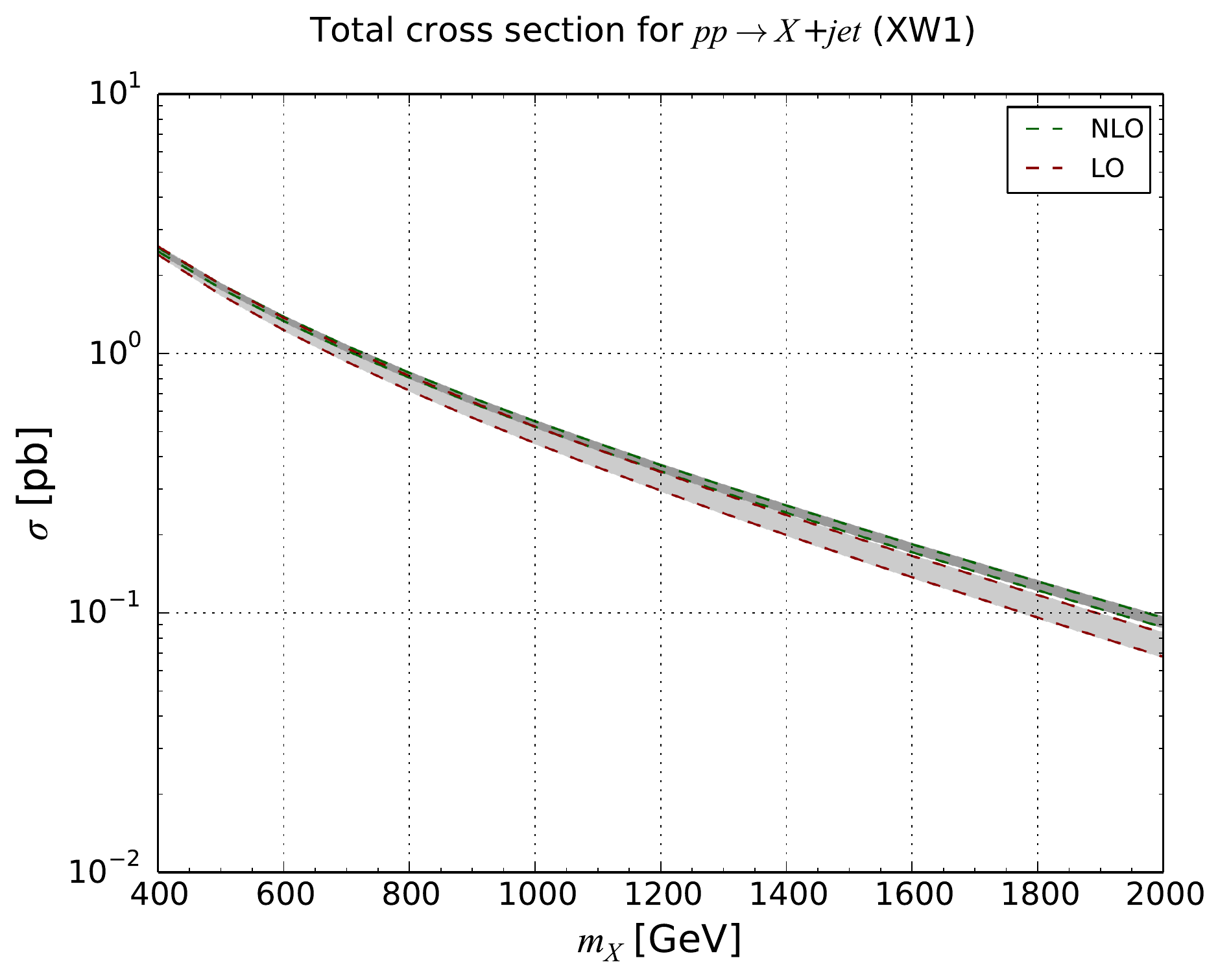}
 \includegraphics[width=0.40\textwidth]{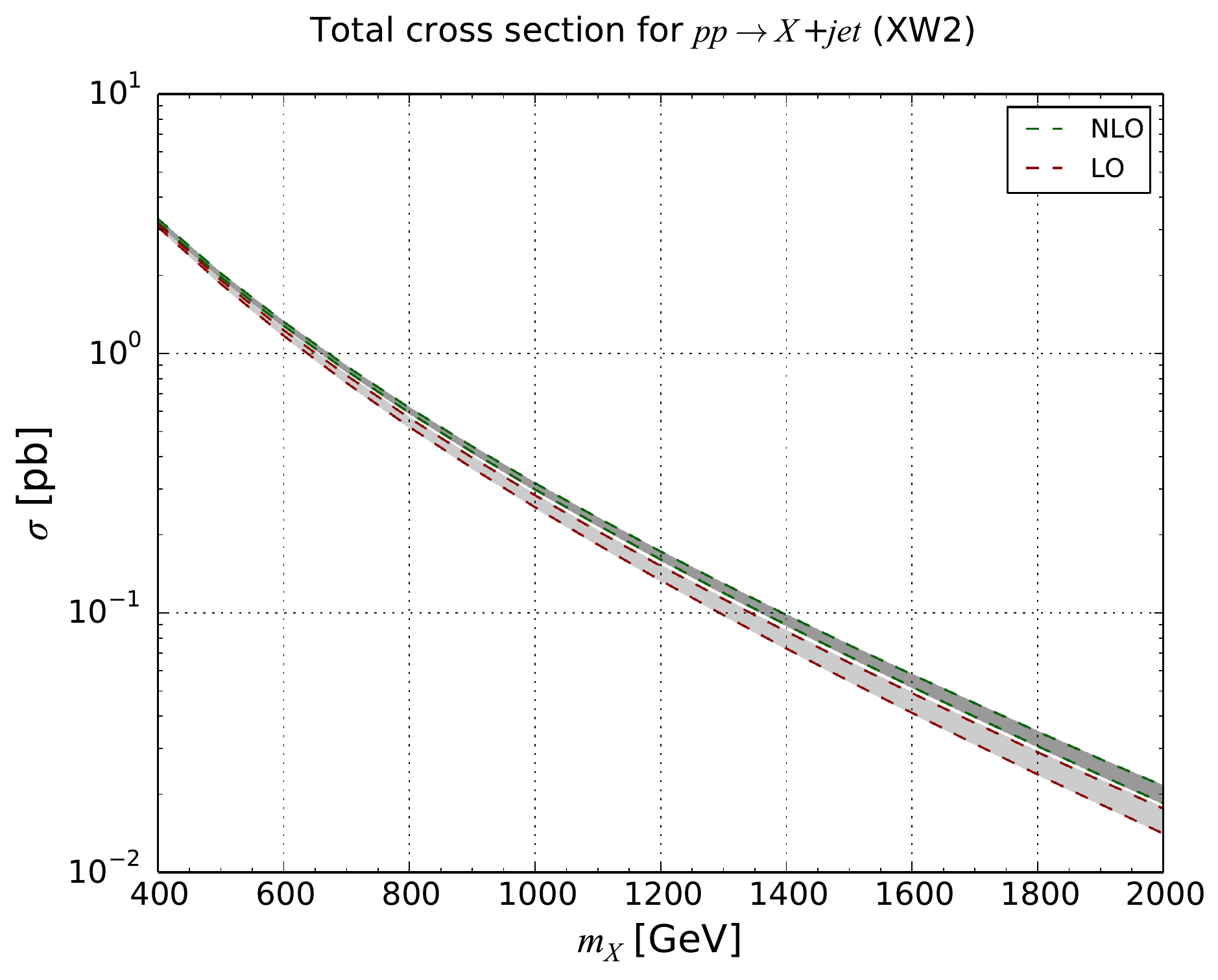}\\
 \includegraphics[width=0.40\textwidth]{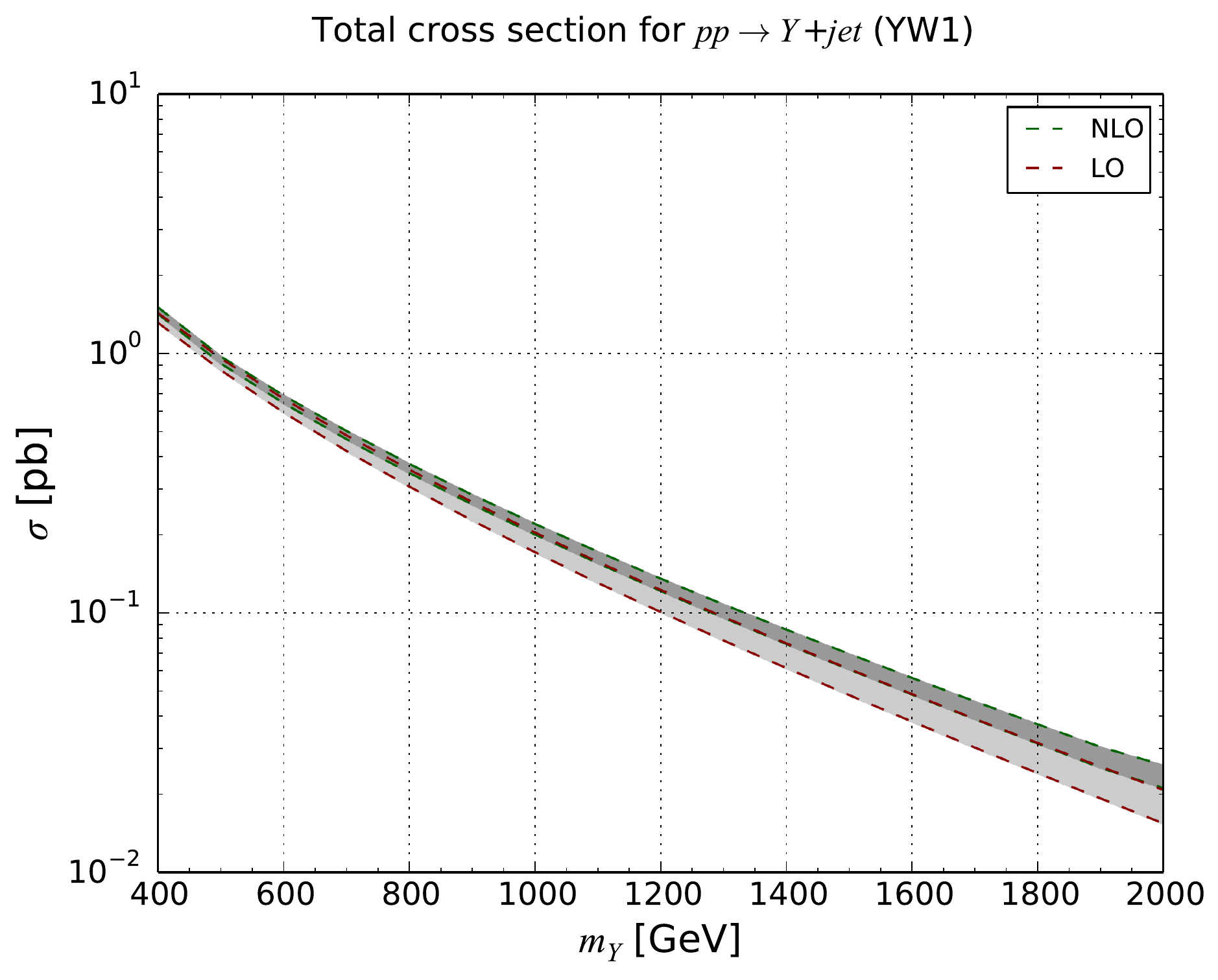}
 \includegraphics[width=0.40\textwidth]{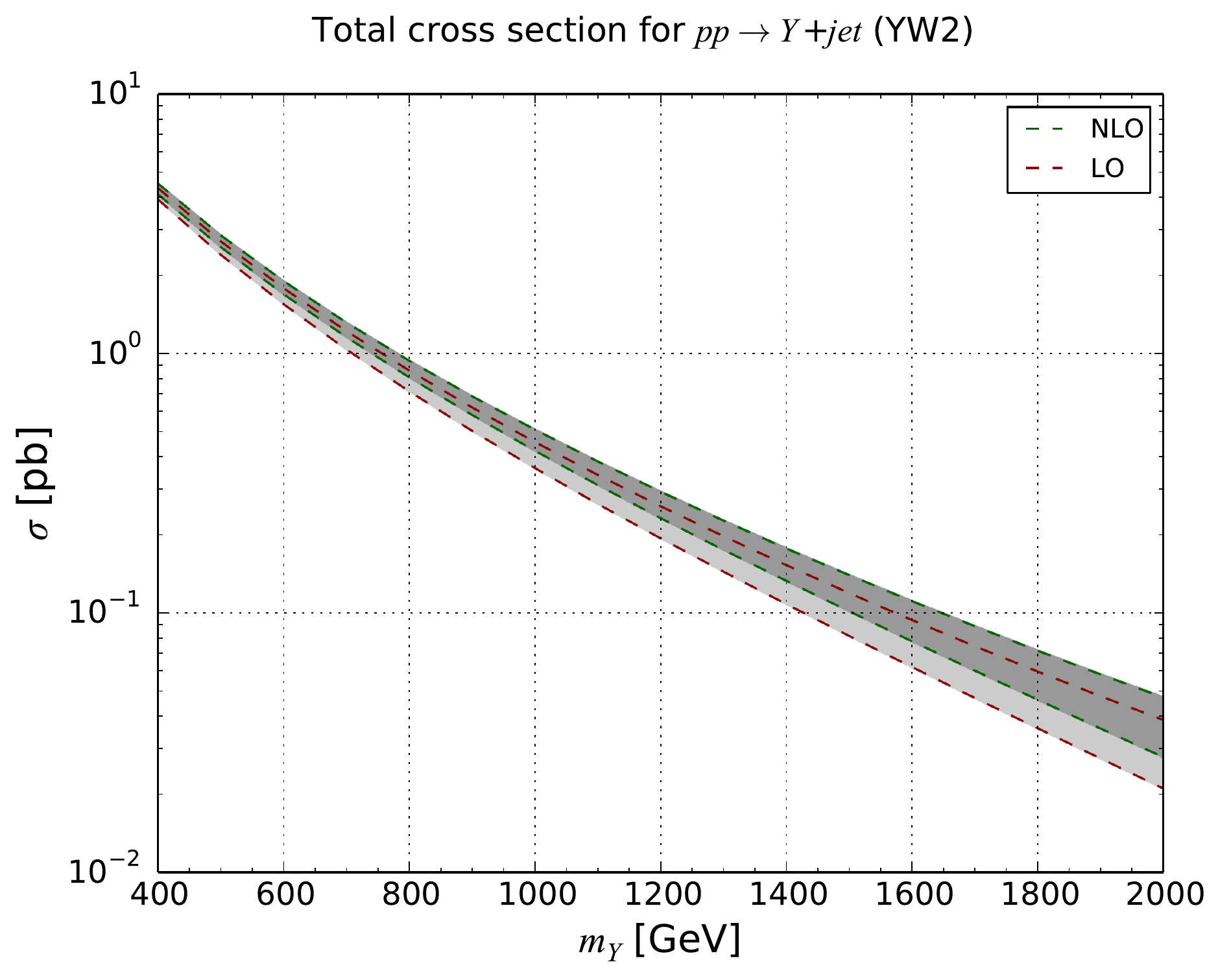}
 \caption{LO and NLO QCD inclusive cross sections for single $T$, $X$ and $Y$
   quark production at the LHC with \mbox{$\sqrt{s}=13$~TeV}, for
   various benchmark scenarios.}
\label{fig:xsec_Tsingle}
\end{figure*}

In Figure~\ref{fig:xsec_Tsingle}, Table~\ref{tab:TJxs} and Table~\ref{tab:XJxs},
we present LO and
NLO total cross sections related to the production of a single vector-like $T$,
$X$ and $Y$ quark for different scenarios introduced in
Section~\ref{sec:benchmarks}. In each case, we depict the dependence of the
cross sections on the vector-like quark mass and study the uncertainties
stemming from scale variation.

\begin{table*}
 \centering
 \renewcommand{\arraystretch}{1.40}
 \setlength{\tabcolsep}{12pt}
 \begin{tabular}{cc|cc}
   $m_{\sss T}$~[GeV] & Scenario & $\sigma_{\rm LO}$ [pb] & $\sigma_{\rm NLO}$ [pb]\\
   \hline
   \multirow{4}{*}{$400$} & {\footnotesize \bf TZ1} & $(1.995\ 10^{0}){}^{+2.6\%}_{-2.7\%}{}^{+1.6\%}_{-1.6\%}$ &
     $(1.987\ 10^{0}){}^{+0.8\%}_{-0.6\%}{}^{+1.7\%}_{-1.7\%}$\\
   & {\footnotesize \bf TZ2 } & $(2.613\ 10^{0}){}^{+0.1\%}_{-1.0\%}{}^{+1.2\%}_{-1.2\%}$ &
     $(2.685\ 10^{0}){}^{+1.1\%}_{-0.6\%}{}^{+1.2\%}_{-1.2\%}$\\
   & {\footnotesize \bf TW1 } & $(1.541\ 10^{0}){}^{+3.4\%}_{-3.3\%}{}^{+2.4\%}_{-2.4\%}$ &
     $(1.575\ 10^{0}){}^{+0.9\%}_{-0.2\%}{}^{+2.4\%}_{-2.4\%}$\\
   & {\footnotesize \bf TW2 } & $(4.229\ 10^{0}){}^{+1.1\%}_{-1.5\%}{}^{+4.5\%}_{-4.5\%}$ &
     $(4.392\ 10^{0}){}^{+1.1\%}_{-0.3\%}{}^{+4.4\%}_{-4.4\%}$\\
   \hdashline
   \multirow{4}{*}{$800$} & {\footnotesize \bf TZ1} & $(5.657\ 10^{-1}){}^{+6.2\%}_{-5.5\%}{}^{+1.8\%}_{-1.8\%}$ &
     $(6.014\ 10^{-1}){}^{+0.9\%}_{-0.9\%}{}^{+1.8\%}_{-1.8\%}$\\
   & {\footnotesize \bf TZ2 } & $(4.488\ 10^{-1}){}^{+3.2\%}_{-3.4\%}{}^{+1.8\%}_{-1.8\%}$ &
     $(4.925\ 10^{-1}){}^{+1.3\%}_{-0.6\%}{}^{+1.7\%}_{-1.7\%}$\\
   & {\footnotesize \bf TW1 } & $(4.183\ 10^{-1}){}^{+6.9\%}_{-6.1\%}{}^{+3.0\%}_{-3.0\%}$ &
     $(4.560\ 10^{-1}){}^{+1.0\%}_{-1.4\%}{}^{+2.9\%}_{-2.9\%}$\\
   & {\footnotesize \bf TW2 } & $(8.272\ 10^{-1}){}^{+5.0\%}_{-4.7\%}{}^{+7.7\%}_{-7.7\%}$ &
     $(9.175\ 10^{-1}){}^{+1.1\%}_{-1.2\%}{}^{+7.2\%}_{-7.2\%}$\\
   \hdashline
   \multirow{4}{*}{$1200$} & {\footnotesize \bf TZ1} & $(2.214\ 10^{-1}){}^{+8.2\%}_{-7.1\%}{}^{+1.9\%}_{-1.9\%}$ &
     $(2.483\ 10^{-1}){}^{+1.4\%}_{-1.9\%}{}^{+2.0\%}_{-2.0\%}$\\
   & {\footnotesize \bf TZ2 } & $(1.168\ 10^{-1}){}^{+5.6\%}_{-5.3\%}{}^{+3.2\%}_{-3.2\%}$ &
     $(1.348\ 10^{-1}){}^{+1.6\%}_{-1.6\%}{}^{+2.9\%}_{-2.9\%}$\\
   & {\footnotesize \bf TW1 } & $(1.572\ 10^{-1}){}^{+8.9\%}_{-7.6\%}{}^{+3.5\%}_{-3.5\%}$ &
     $(1.812\ 10^{-1}){}^{+1.9\%}_{-2.4\%}{}^{+3.5\%}_{-3.5\%}$\\
   & {\footnotesize \bf TW2 } & $(2.476\ 10^{-1}){}^{+7.3\%}_{-6.5\%}{}^{+12.0\%}_{-12.0\%}$ &
     $(2.878\ 10^{-1}){}^{+2.0\%}_{-2.2\%}{}^{+11.3\%}_{-11.3\%}$\\
   \hdashline
   \multirow{4}{*}{$1600$} & {\footnotesize \bf TZ1} & $(9.897\ 10^{-2}){}^{+9.6\%}_{-8.3\%}{}^{+2.2\%}_{-2.2\%}$ &
     $(1.163\ 10^{-1}){}^{+2.1\%}_{-2.7\%}{}^{+2.1\%}_{-2.1\%}$\\
   & {\footnotesize \bf TZ2 } & $(3.657\ 10^{-2}){}^{+7.3\%}_{-6.7\%}{}^{+4.9\%}_{-4.9\%}$ &
     $(4.417\ 10^{-2}){}^{+2.3\%}_{-2.4\%}{}^{+4.5\%}_{-4.5\%}$\\
   & {\footnotesize \bf TW1 } & $(6.774\ 10^{-2}){}^{+10.3\%}_{-8.8\%}{}^{+4.2\%}_{-4.2\%}$ &
     $(8.151\ 10^{-2}){}^{+2.8\%}_{-3.3\%}{}^{+4.0\%}_{-4.0\%}$\\
   & {\footnotesize \bf TW2 } & $(9.040\ 10^{-2}){}^{+9.0\%}_{-7.8\%}{}^{+17.7\%}_{-17.7\%}$ &
     $(1.095\ 10^{-1}){}^{+2.6\%}_{-3.0\%}{}^{+16.6\%}_{-16.6\%}$\\
   \hdashline
   \multirow{4}{*}{$2000$} & {\footnotesize \bf TZ1} & $(4.721\ 10^{-2}){}^{+10.9\%}_{-9.2\%}{}^{+2.4\%}_{-2.4\%}$ &
     $(5.771\ 10^{-2}){}^{+2.9\%}_{-3.5\%}{}^{+2.4\%}_{-2.4\%}$\\
   & {\footnotesize \bf TZ2 } & $(1.277\ 10^{-2}){}^{+8.7\%}_{-7.8\%}{}^{+7.0\%}_{-7.0\%}$ &
     $(1.600\ 10^{-2}){}^{+3.0\%}_{-3.1\%}{}^{+6.6\%}_{-6.6\%}$\\
   & {\footnotesize \bf TW1 } & $(3.105\ 10^{-2}){}^{+11.5\%}_{-9.7\%}{}^{+5.0\%}_{-5.0\%}$ &
     $(3.899\ 10^{-2}){}^{+3.5\%}_{-4.0\%}{}^{+4.7\%}_{-4.7\%}$\\
   & {\footnotesize \bf TW2 } & $(3.725\ 10^{-2}){}^{+10.1\%}_{-8.7\%}{}^{+24.7\%}_{-24.7\%}$ &
     $(4.653\ 10^{-2}){}^{+3.2\%}_{-3.6\%}{}^{+23.1\%}_{-23.1\%}$\\
  \end{tabular}
  \renewcommand{\arraystretch}{1.0}
  \caption{\small \label{tab:TJxs}LO and NLO QCD inclusive cross sections for
   single $T$ production at the LHC, running at a center-of-mass energy of
   $\sqrt{s}=13$~TeV. The results are shown together with the associated scale
   and PDF relative uncertainties in the context of several benchmark scenarios.}
\end{table*}

\begin{table*}
 \centering
 \renewcommand{\arraystretch}{1.40}
 \setlength{\tabcolsep}{12pt}
 \begin{tabular}{cc|cc}
   $m_{\sss Q}$~[GeV] & Scenario & $\sigma_{\rm LO}$ [pb] & $\sigma_{\rm NLO}$ [pb]\\
   \hline
   \multirow{4}{*}{$400$} & {\footnotesize \bf XW1} & $(2.483\ 10^{0}){}^{+3.1\%}_{-3.1\%}{}^{+1.8\%}_{-1.8\%}$ &
     $(2.501\ 10^{0}){}^{+0.8\%}_{-0.4\%}{}^{+1.8\%}_{-1.8\%}$\\
   & {\footnotesize \bf XW2 } & $(3.102\ 10^{0}){}^{+0.0\%}_{-0.7\%}{}^{+1.3\%}_{-1.3\%}$ &
     $(3.217\ 10^{0}){}^{+1.2\%}_{-0.6\%}{}^{+1.3\%}_{-1.3\%}$\\
   & {\footnotesize \bf YW1} & $(1.361\ 10^{0}){}^{+3.3\%}_{-3.2\%}{}^{+2.8\%}_{-2.8\%}$ &
     $(1.452\ 10^{0}){}^{+1.2\%}_{-0.4\%}{}^{+2.8\%}_{-2.8\%}$\\
   & {\footnotesize \bf YW2 } & $(4.114\ 10^{0}){}^{+1.1\%}_{-1.5\%}{}^{+4.7\%}_{-4.7\%}$ &
     $(4.278\ 10^{0}){}^{+1.1\%}_{-0.3\%}{}^{+4.6\%}_{-4.6\%}$\\
   \hdashline
   \multirow{4}{*}{$800$} & {\footnotesize \bf XW1} & $(7.639\ 10^{-1}){}^{+6.4\%}_{-5.7\%}{}^{+2.0\%}_{-2.0\%}$ &
     $(8.226\ 10^{-1}){}^{+0.9\%}_{-1.1\%}{}^{+2.0\%}_{-2.0\%}$\\
   & {\footnotesize \bf XW2 } & $(5.408\ 10^{-1}){}^{+3.3\%}_{-3.5\%}{}^{+2.0\%}_{-2.0\%}$ &
     $(6.002\ 10^{-1}){}^{+1.3\%}_{-0.8\%}{}^{+1.8\%}_{-1.8\%}$\\
   & {\footnotesize \bf YW1} & $(3.289\ 10^{-1}){}^{+6.9\%}_{-6.1\%}{}^{+3.8\%}_{-3.8\%}$ &
     $(3.593\ 10^{-1}){}^{+1.1\%}_{-1.4\%}{}^{+3.7\%}_{-3.7\%}$\\
   & {\footnotesize \bf YW2 } & $(7.827\ 10^{-1}){}^{+5.0\%}_{-4.7\%}{}^{+7.7\%}_{-7.7\%}$ &
     $(8.688\ 10^{-1}){}^{+1.2\%}_{-1.2\%}{}^{+7.2\%}_{-7.2\%}$\\
   \hdashline
   \multirow{4}{*}{$1200$} & {\footnotesize \bf XW1} & $(3.202\ 10^{-1}){}^{+8.3\%}_{-7.2\%}{}^{+2.2\%}_{-2.2\%}$ &
     $(3.617\ 10^{-1}){}^{+1.5\%}_{-2.1\%}{}^{+2.2\%}_{-2.2\%}$\\
   & {\footnotesize \bf XW2 } & $(1.420\ 10^{-1}){}^{+5.7\%}_{-5.4\%}{}^{+3.3\%}_{-3.3\%}$ &
     $(1.658\ 10^{-1}){}^{+1.8\%}_{-1.8\%}{}^{+3.0\%}_{-3.0\%}$\\
   & {\footnotesize \bf YW1} & $(1.108\ 10^{-1}){}^{+9.0\%}_{-7.8\%}{}^{+5.1\%}_{-5.1\%}$ &
     $(1.282\ 10^{-1}){}^{+1.9\%}_{-2.4\%}{}^{+4.8\%}_{-4.8\%}$\\
   & {\footnotesize \bf YW2 } & $(2.245\ 10^{-1}){}^{+7.3\%}_{-6.5\%}{}^{+12.3\%}_{-12.3\%}$ &
     $(2.618\ 10^{-1}){}^{+1.9\%}_{-2.2\%}{}^{+11.5\%}_{-11.5\%}$\\
   \hdashline
   \multirow{4}{*}{$1600$} & {\footnotesize \bf XW1} & $(1.503\ 10^{-1}){}^{+9.7\%}_{-8.3\%}{}^{+2.4\%}_{-2.4\%}$ &
     $(1.780\ 10^{-1}){}^{+2.3\%}_{-2.8\%}{}^{+2.4\%}_{-2.4\%}$\\
   & {\footnotesize \bf XW2 } & $(4.489\ 10^{-2}){}^{+7.4\%}_{-6.7\%}{}^{+5.1\%}_{-5.1\%}$ &
     $(5.474\ 10^{-2}){}^{+2.5\%}_{-2.6\%}{}^{+4.7\%}_{-4.7\%}$\\
    & {\footnotesize \bf YW1} & $(4.303\ 10^{-2}){}^{+10.6\%}_{-9.0\%}{}^{+6.9\%}_{-6.9\%}$ &
     $(5.228\ 10^{-2}){}^{+2.9\%}_{-3.4\%}{}^{+6.4\%}_{-6.4\%}$\\
   & {\footnotesize \bf YW2 } & $(7.740\ 10^{-2}){}^{+8.9\%}_{-7.8\%}{}^{+18.8\%}_{-18.8\%}$ &
     $(9.423\ 10^{-2}){}^{+2.7\%}_{-3.0\%}{}^{+17.5\%}_{-17.5\%}$\\   \hdashline
   \multirow{4}{*}{$2000$} & {\footnotesize \bf XW1} & $(7.495\ 10^{-2}){}^{+10.8\%}_{-9.2\%}{}^{+2.8\%}_{-2.8\%}$ &
     $(9.215\ 10^{-2}){}^{+3.0\%}_{-3.5\%}{}^{+2.7\%}_{-2.7\%}$\\
   & {\footnotesize \bf XW2 } & $(1.577\ 10^{-2}){}^{+8.8\%}_{-7.8\%}{}^{+7.2\%}_{-7.2\%}$ &
     $(1.997\ 10^{-2}){}^{+3.1\%}_{-3.2\%}{}^{+6.8\%}_{-6.8\%}$\\
   & {\footnotesize \bf YW1} & $(1.795\ 10^{-2}){}^{+11.9\%}_{-9.9\%}{}^{+9.3\%}_{-9.3\%}$ &
     $(2.351\ 10^{-2}){}^{+4.0\%}_{-4.3\%}{}^{+8.9\%}_{-8.9\%}$\\
   & {\footnotesize \bf YW2 } & $(2.968\ 10^{-2}){}^{+10.3\%}_{-8.9\%}{}^{+27.9\%}_{-27.9\%}$ &
     $(3.752\ 10^{-2}){}^{+3.5\%}_{-3.8\%}{}^{+25.8\%}_{-25.8\%}$\\
  \end{tabular}
  \renewcommand{\arraystretch}{1.0}
  \caption{\small \label{tab:XJxs}LO and NLO QCD inclusive cross sections for
   single $X$ and single $Y$ production at the LHC, running at a center-of-mass
   energy of
   $\sqrt{s}=13$~TeV. The results are shown together with the associated scale
   and PDF relative uncertainties in the context of several benchmark scenarios.
}
\end{table*}

\clearpage

\bibliographystyle{JHEP}
\bibliography{paper}

\end{document}